\documentclass[]{aa}

\input psfig.tex
\usepackage{graphicx}
\usepackage{natbib}
\usepackage{array}
\usepackage{graphics}
\usepackage{latexsym}
\usepackage{amssymb}
\usepackage{amsmath}
\usepackage{fancyhdr}
\usepackage{morefloats}
\usepackage{rotating}
\bibpunct{(}{)}{;}{a}{}{,}
\include{hyphe}

\begin{document}

\title{Formation and Evolution of the Dust in Galaxies. I.\\ The Condensation Efficiencies}

\author{L. Piovan \inst {1,2}, C. Chiosi \inst{1}, E. Merlin \inst{1}, T. Grassi \inst{1},
R. Tantalo \inst{1}, U. Buonomo \inst{1} and L. P. Cassar\`{a}
\inst{1}}

\institute{    $^1$ Department of Astronomy, Padova University,
Vicolo dell'Osservatorio 3, I-35122, Padova, Italy\\
$^2$Max-Planck-Institut f\"ur Astrophysik, Karl-Schwarzschild-Str.
1, Garching bei M\"unchen, Germany\\
\email{  {lorenzo.piovan\char64unipd.it} }   }

\date{Received: July 2011; Revised: *** ****;  Accepted: *** ***}

\abstract   {The growing interest in the high-z universe, where
strongly obscured objects are present, has determined an effort to
improve the simulations of dust formation and evolution in galaxies.
Three main basic ingredients enter the problem influencing the total
dust budget and the kind of mixture of the dust grains: the types
and amounts of dust injected by AGB stars and SN{\ae} and the
accretion and destruction processes of dust in the interstellar
medium (ISM). They govern the relative abundances of the gas and
dust components of the ISM of a galaxy.}{In this study, we focus on
the dust emitted by stars  and present a database of condensation
efficiencies for the refractory elements C, O, Mg, Si, S, Ca and Fe
in AGB stars and SN{\ae} that can be easily applied to the
traditional gaseous ejecta, in order to determine the amount and
kind of refractory elements locally embedded into dust and injected
into the ISM.} {The best theoretical recipes available nowadays in
literature to estimate the amount of dust produced by SN{\ae} and
AGB stars have been discussed and for SN{\ae} compared to the
observations to get clues on the problem. The condensation
efficiencies have been then analyzed in the context of a classical
chemical model of dust formation and evolution in the  Solar
Neighbourhood and Galactic Disk.} {Tables of condensation
coefficients are presented for (i) AGB stars at varying the
metallicity and  (ii) SN{\ae} at varying the density
n$_{\textrm{H}}$ of the ISM where the SN{a} explosions took place.
In particular, we show how the controversial CNT approximation
widely adopted  to form dust  in SN{\ae}, still gives good results
and agrees with some clues coming from the observations. A new
generation of  dust formation models in SN{\ae} is however required
to solve some contradictions that have recently emerged.} {A simple
database of condensation efficiencies is set up  to be used in
chemical models including the effect of dusts of different type and
meant to simulate real galaxies of different type going from
primordial proto-galaxies  to those currently seen in the local
universe.}

\titlerunning{Formation and evolution of the dust in galaxies}

\authorrunning{L. Piovan  et al.}
\maketitle

\section{Introduction}

Understanding and modelling the interstellar dust has recently
received a great deal of attention thanks to current observations
unveiling the existence of a high-z universe heavily obscured by
large quantities of dust
\citep{Shapley01,Carilli01,Bertoldi03,Robson04,Wang08a,Wang08b,Michalowski10a,Michalowski10b}.
Once established the presence of dust, some key questions must be
answered about the physical nature of a  dust-rich  universe. What
is the origin of these copious amounts of dust? What is the dust
composition? What is a plausible  mixture of the dust grains  able
to account for the observational properties of extinction of the
stellar light and emission in the mid and far infrared (MIR/FIR)?
Starting from the first simplified models simulating in some way the
formation and evolution of dust in galaxies
\citep{Lisenfeld98,Morgan03,Inoue03}, over the years models of
growing complexity have been presented: from the pioneer work of
\citet{Dwek98} on the MW till the recent ones by \citet{Zhukovska08}
and \citet{Piovan11a} on the Solar Neighborhood (SoNe) of the Milky
Way (MW) or the whole Galactic Disk of \citet{Piovan11b}, on
galaxies of different morphological types \citep{Calura08}, on
star-burst galaxies \citep{Gall11a}, on QSOs, LBGs and the Early
Universe
\citep{Valiante09,Valiante11,Pipino11,Dwek11,Mattsson11,Gall11b,Yamasawa11}.\\
\indent The concept of duty cycle for  the dust must be introduced
to suitably describe the formation and evolution of dust in high-z
and local galaxies, and to simultaneously infer  precious clues on
when and how galaxies formed and evolved. The cyclic history of the
interstellar dust is described in detail by \citet{Zhukovska08} and
nicely illustrated in the classical diagram by  \citet{Jones04}. In
brief,  low and intermediate AGB stars thanks to mass loss by
stellar winds, and massive stars thanks to the  Core Collapse SN{a}
explosions (CCSN{\ae}), inject refractory elements in the ISM: most
of this material is in gaseous form, but important amounts of it
condense into the so-called star-dust. Once mixed in the turbulent
ISM, star-dust grains are subjected to destructive processes that
restitute the material to the gaseous phase. The competition between
this process and the one of dust  accretion onto the so-called seeds
in dense and cold molecular clouds (MCs), determines the total
budget of dust in the ISM and the observed depletion of the
refractory elements by formation of new dust grains. In the MCs,
where dust accretes and cools down  the region, star formation takes
place generating new  stars that in turn evolve and die, thus more
and more enriching the ISM with new metals and star-dust (the
fraction of it able to survive to  local shocks). It is soon evident
even from this simple description  that some key agents must
intervene  to drive the evolution of dust. They are identified with
some grains or  grain families with given composition and
properties, the physical mechanisms of formation/accretion and
destruction of dust in  the ISM,  and finally the yields of  dust by stars.\\
\indent In this study we focus the attention on the dust emitted by
stars of different mass and metallicities during the AGB
evolutionary phase and/or the SN{a} explosion as appropriated to
their initial mass \citep{Zhukovska08}. The amount of produced
star-dust and the injection timescales are the object of a vivid
debate, largely motivated by the high-z galaxies. Indeed, it is not
clear (i) whether star-dust of SN{a} origin  is able alone to
explain the amount of dust observed in high-z objects
\citep{Gall11b,Gall11b}, (ii) up to  which redshift and how strong
is the role played by massive AGB stars \citep{Valiante09,Dwek11},
and (iii) whether the contribution by the dust accreted in the ISM
cannot be neglected \citep{Dwek09,Draine09,Mattsson11}. In this
study, we intend to thoroughly discuss what could be the best
compilation of theoretical condensation efficiencies currently
available in literature and how much of each refractory element
could locally condense in form of star-dust. To this aim we present
here a easy-to-use compilation of condensation efficiencies $\delta$
\citep{Dwek98,Calura08,Piovan11b} to be applied to the masses of
\textit{single elements} restituted by stars to the ISM. In other
words, starting from the classical compilations of the gas mass in
form of a given element ejected by each  star during its life back
into the ISM  \citep{Portinari98,vandenHoek97,Francois04}, we
provide a compilation of coefficients giving the dust-to-gas ratio
for that specific ejecta.

\renewcommand{\arraystretch}{1.1}
\begin{table*} \label{DeltaLiterature}
\footnotesize
\begin{center}
\caption[]{Prescriptions taken from  literature to model the
star-dust contribution to the ISM:}
\begin{tabular}{ccc}          \hline \hline \noalign{\smallskip} Work & AGB stars & SN{\ae} \\   \hline
\noalign{\smallskip}
\citet{Calura08}$^{1}$ & \citet{Dwek98} & \citet{Dwek98} \\
\citet{Zhukovska08}$^{2}$ & \citet{Ferrarotti06}    &  Its own $\delta^{SN}$ scheme     \\
\citet{Valiante09}$^{3}$ & \citet{Ferrarotti06}   &   \citet{Bianchi07} \\
\citet{Pipino11}$^{4}$ & \citet{Dwek98} & revised \citet{Dwek98} \\
\citet{Yamasawa11}$^{5}$ & no AGB stars & \citet{Nozawa03,Nozawa07} \\
\citet{Gall11a,Gall11b}$^{6}$  & \citet{Ferrarotti06}      & \citet{Todini01,Nozawa03,Nozawa06}     \\
\citet{Dwek11}$^{7}$  & \citet{Dwek98}      & Its own $\delta^{SN}$ scheme     \\
\hline
\end{tabular}
\end{center}
\begin{flushleft}\footnotesize
\footnotesize$^{1}${The same condensation efficiencies $\delta$
proposed by \citet{Dwek98} are adopted.}\, \footnotesize$^{2}${Low
and constant  condensation efficiencies are proposed and adopted for
SN{\ae}.}\, \footnotesize$^{3}${The original model by
\citet{Todini01} for dust formation in SN{\ae} is extended to a
wider set of initial conditions and model assumptions.}\,
\footnotesize$^{4}${Condensation efficiencies $\delta$ by
\citet{Dwek98} are lowered to match the observation of dust in
CCSN{\ae}, thus including in some way the uncertainties of the
destructive reverse shock effects.}\, \footnotesize$^{5}${Only
SN{\ae} are included as dust factories because the study is limited
to the very early universe.}\, \footnotesize$^{6}${Average
coefficients are obtained in order to study the evolution of the
total dust mass in star-bursters and QSOs.}\,
\footnotesize$^{7}${Only the average total amount of dust formed in
SN{\ae} and WR is considered.}\,
\end{flushleft}
\end{table*}
\renewcommand{\arraystretch}{1}

Many different recipes are proposed to deal with the two main
factories of star-dust (AGB stars and SN{\ae}), each of which  with
a different level of complexity  \citep{Gail09,Dwek05}: some of them
consider  only the total amount of dust that is injected and neglect
its composition (spectrum of elements), others adopt simple schemes
to follow the evolution of a group of elements and/or molecules
taken as representative of the dust in the ISM. In Table
\ref{DeltaLiterature} we list all the prescriptions we have adopted
based on the most recent models of
dust formation and destruction.\\

\indent  The plan of the paper is as follows. In Sect.
\ref{SNEyield_details} we discuss the amount of dust injected by a
SNa in the ISM, both from theoretical and observational point of
view, look at the different types of SN{\ae} producing dust,
calculate and present the condensation efficiencies for the single
elements from various sources in literature, and finally analyze the
various alternatives highlighting their merits and drawbacks. In
Sect. \ref{AGByield} we examine the production of dust by  AGB
stars. In Sect. \ref{Yieldseffect} we analyze the prescription  for
stardust we have just derived and their effects with the aid of the
classical model for the Galactic Disk and SoNe of the MW by
\citet{Piovan11b}. Although the model includes the injection of
star-dust, dust accretion and destruction in the ISM, radial flows
of matter and effects of the Bar for the innermost regions, we limit
ourselves here to examine only the effects brought about by type II
SN{\ae}, type Ia SN{\ae}, and AGB stars in  three regions of the MW
disk: an inner region, the SoNe, and an outer region. In Sect.
\ref{Discus_Concl} we summarize the results and draw some
conclusions. This paper is the first of series of three
\citep{Piovan11b,Piovan11c} dedicated to the wide subject of dust
formation/ destruction  and evolution, and its effects. Particular
attention is paid to the MW which is the ideal workbench for any
model of chemical evolution. In \citet{Piovan11b} we will present
our chemical model for the MW-SoNe with dust and formation and
evolution based upon the classical model with infall developed long
ago by \citet{Chiosi80} and ever since used by many authors. In
\citet{Piovan11c} we will apply the same model to investigate the
radial chemical properties of the MW Disk.

\section{Yields of dust by SN{\ae}} \label{SNEyield_details}

\indent It is long known  that SN{\ae} are primary sites of dust
formation. The direct evidence began with the pioneering
observations of the SN 1987A
\citep{Danziger91,Dwek92,McCray93,Bautista95,Dwek98} until the
recent and deep observations of the SN 1987A itself and other
SN{\ae}, like E0102 in SMC or Cas A in the MW. These new data
strengthening our knowledge about dust and SN{\ae}, are obtained by
means of the new generation of IR and sub-mm instruments, like
Spitzer \citep{Bouchet06,Meikle07,Rho08,Rho09a,Kotak09a,Rho09b},
Akari \citep{Sakon09}, SCUBA \citep{Dunne03} and PACS/SPIRE onboard
Herschel Space Observatory \citep{Matsuura11}.

\renewcommand{\arraystretch}{1.1}
\begin{table*}
\footnotesize
\begin{center}
\caption[]{Observational data on the  dust formation in the ejecta and remnants of CCSN{\ae} and in
the Kepler SNa, the progenitor of which is still controversial.}
\begin{tabular}{ccccc}
\hline \hline \noalign{\smallskip}
SN{\ae}$^{(}$\footnotemark[1]$^{)}$ &
Galaxy$^{(}$\footnotemark[2]$^{)}$ &Type$^{(}$\footnotemark[3]$^{)}$
&$M_{*,prog}$$^{(}$\footnotemark[4]$^{)}$
&$M_{d}$$^{(}$\footnotemark[5]$^{)}$\\
&   & $(M_{\odot})$     &$(M_{\odot})$  \\
\noalign{\smallskip} \hline \noalign{\smallskip}
SN 1987A   &LMC  &II-peculiar  &20$^{(}$\footnotemark[25]$^{)}$ & $2\cdot 10^{-4}-1.3\cdot10^{-3}$$^{(}$\footnotemark[25]$^{)}$, 0.4-0.7$^{(}$\footnotemark[27]$^{)}$ \\
SN 1999em  &NGC1637 &II-P$^{(}$\footnotemark[6]$^{)}$  &$12-14$$^{(}$\footnotemark[6]$^{)}$ & $>10^{-4}$ $^{(}$\footnotemark[6]$^{)}$ \\
SN 2003gd  &M74 &II-P$^{(}$\footnotemark[7]$^{)}$  &$8^{+4}_{-2}$
$^{(}$\footnotemark[9]$^{)}$ &
 $10^{-4}$ $^{(}$\footnotemark[7]$^{)}$ $-0.02$ $^{(}$\footnotemark[9]$^{)}$ \\
Kepler &MW
&Ia$^{(10,12)}$,Ib$^{(}$\footnotemark[11]$^{)}$,II-L$^{(}$\footnotemark[11]$^{)}$
&$8$ $^{(}$\footnotemark[10]$^{)}-$ \,$>10$
$^{(}$\footnotemark[11]$^{)}$ & $0.09$ $^{(}$\footnotemark[13]$^{)}$
$-0.14$ $^{(}$\footnotemark[13]$^{)}$; $<1.2 $$^{(}$\footnotemark[13]$^{)}$ \\
SNR1E0102.2-7219  &SMC  &Ib,Ic,II-L$^{(}$\footnotemark[17]$^{)}$
&25$^{(}$\footnotemark[14]$^{)}$ &
$8\cdot10^{-4}$$^{(}$\footnotemark[15]$^{)}$
- $0.014$$^{(}$\footnotemark[16]$^{)}$ \\
Cassiopeia A    &MW   &IIn$^{(}$\footnotemark[21]$^{)}$-IIb$^{(}$\footnotemark[20]$^{)}$
&13-30$^{(18,20)}$ & $0.02-0.054$$^{(}$\footnotemark[19]$^{)}$; $<1.0 $$^{(}$\footnotemark[26]$^{)}$ \\
SN 2005af   &NGC 4945   &II-P$^{(}$\footnotemark[22]$^{)}$  &13-35$^{(}$\footnotemark[22]$^{)}$ & $4\cdot 10^{-4}$$^{(}$\footnotemark[23]$^{)}$ \\
N 132D  &LMC   &Ib$^{(}$\footnotemark[24]$^{)}$ &30-35$^{(}$\footnotemark[24]$^{)}$ & $> 8\cdot 10^{-3}$$^{(}$\footnotemark[24]$^{)}$ \\
\hline \label{tabella1}
\end{tabular}
\end{center}
\renewcommand{\arraystretch}{1}
\begin{flushleft}
\footnotesize$^{1}${Identification name of the  SN{\ae} and
remnants}.\,\footnotesize$^{2}${Galaxy in which the supernova has
been observed}.\,\footnotesize$^{3}${Classification of the CCSN{\ae}
and thermonuclear SN{\ae} according to the observational scheme:
almost all of the tabulated objects are CCSN{\ae}, only for  Kepler
SNa the classification is still
debated}.\,\footnotesize$^{4}${Estimated mass of the progenitor in
solar masses}.\,\footnotesize$^{(5)}${Estimated mass of  dust
condensed in the remnant}.
\,\footnotesize$^{(6)}${\citet{Elmhamdi03}}.\,\footnotesize$^{(6)}${\citet{Meikle07}}.
\,\footnotesize$^{(8)}${\citet{Hendry05}}.\,\footnotesize$^{(9)}${\citet{Sugerman06}}.
\,\footnotesize$^{(10)}${\citet{Reynolds07}}.
\,\footnotesize$^{(11)}${\citet{Bandiera87}}.\,\footnotesize$^{(12)}${\citet{CassamChenai04}}.
\,\footnotesize$^{(13)}${\citet{Gomez09}. The estimated mass depends
strongly on the absorption coefficient $\kappa$. The adopted value
is the one appropriate for SNa dust according to \citet{Dunne09}.
But for different $\kappa$ the estimate could grow until $1.2
M_{\odot}$ or even more \citep{Gomez09}.
}.\,\footnotesize$^{(14)}${\citet{Sandstrom09}}.
\,\footnotesize$^{(15)}${\citet{Stanimirovic05}}.\,\footnotesize$^{(16)}${\citet{Rho09a},
but according to the estimate by \citet{Sandstrom09}, up to $0.6
M_{\odot}$ of cold dust could be present}.
\,\footnotesize$^{(17)}${\citet{Finkelstein04}}.\,\footnotesize$^{(18)}${\citet{Young06}}.
\,\footnotesize$^{(19)}${\citet{Rho08}}.\,\footnotesize$^{(20)}${\citet{Krause08}}.
\,\footnotesize$^{(21)}${\citet{Chevalier03}}.\,\footnotesize$^{(22)}${\citet{Kotak06}}.
\,\footnotesize$^{(23)}${\citet{Kotak08}}.\,\footnotesize$^{(24)}${\citet{Rho09b}}.
\,\footnotesize$^{(25)}${\citet{Ercolano07}}.\,\footnotesize$^{(26)}${\citet{Dunne09}}
\,\footnotesize$^{(27)}${\citet{Matsuura11}}.
\end{flushleft}
\end{table*}
\renewcommand{\arraystretch}{1}

\begin{figure}
\centering
\includegraphics[height=7.8cm,width=9.0cm]{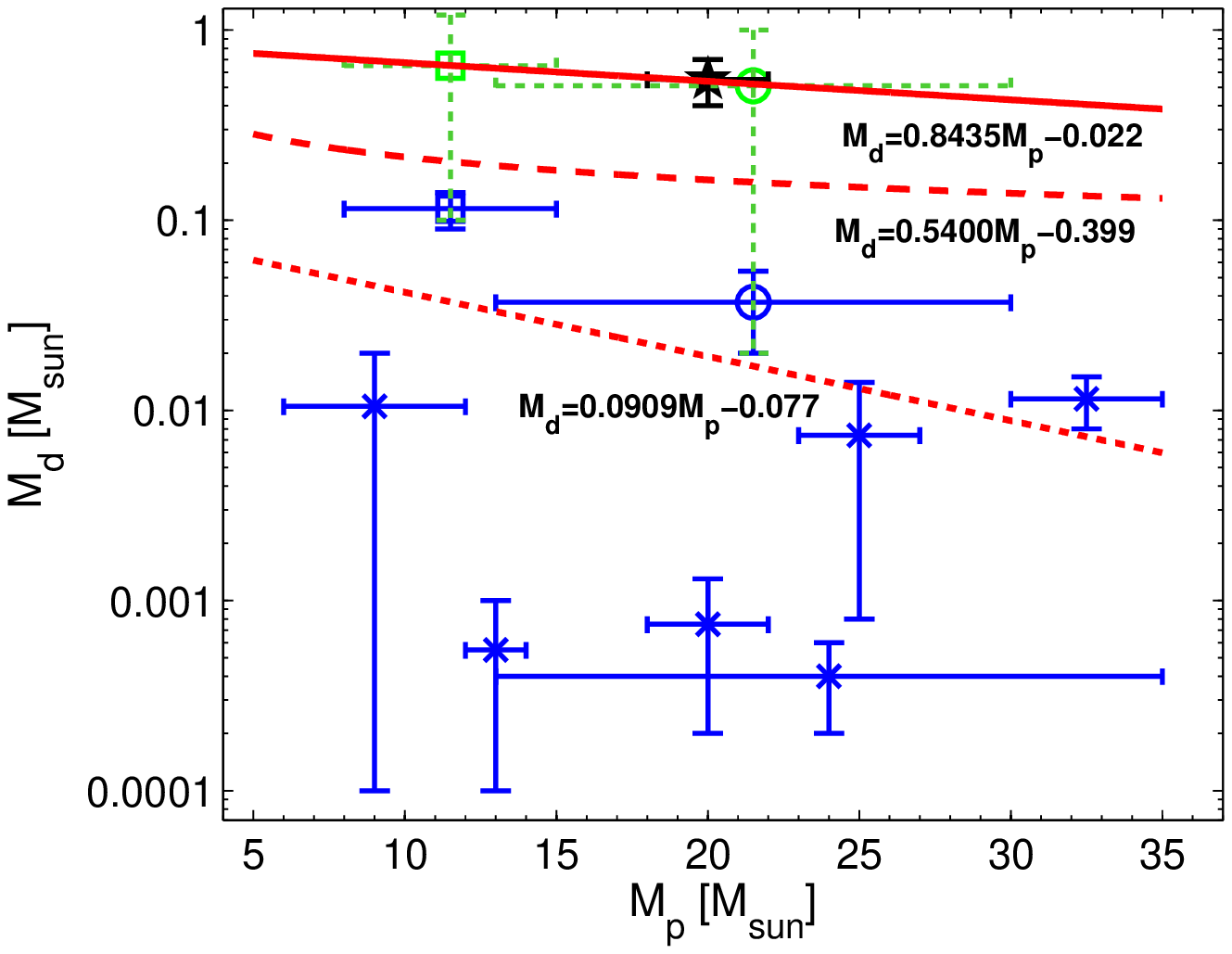}
\caption{Observational estimates of the masses of newly formed dust
M$_{\textrm{d}}$ in the CCSN{\ae} (four-points stars and solid line
error bars) as a function of the progenitor mass M$_{\textrm{p}}$,
both expressed in solar units. The recent estimates of the amount of
newly formed dust in Kepler SNa (square with dotted line error
bars), Cas A (circle with dotted line error bars) obtained by sub-mm
observations, and in SN 1987A (five-pointed star with continuous
line error bars) with the PACS/SPIRE onboard the Herschel Space
Observatory are also displayed. Three fits are shown: the dotted
line represents the best fit obtained from using only the masses of
dust determined by NIR/MIR observations; the dashed line represents
the best fit to all the data; finally, the continuous line
represents the best fit only to the amounts of dust derived from
FIR/sub-mm data.} \label{SNeYieldObserved}
\end{figure}

Given these premises, several important questions arise. How much
dust is produced by a single SNa according to the observational
data\,? What is the condensation efficiency of the different
refractory elements during the evolution of the SNa remnants
\citep{Nozawa03,Ercolano07,Cherchneff08,Zhukovska08,Calura08}, in
particular when  the effects of  forward and reverse shocks are
taken into account \citep{Nozawa06,Nozawa07,Kozasa09}\,? Do SN{\ae}
produce enough dust to significantly contribute to the obscuration
of primordial galaxies \citep{Dwek07,Nozawa08,Dwek11,Gall11b} or a
substantial amount of that dust is due to nucleation in the ISM with
SN{\ae} mainly providing  the seeds on which dust grains of the ISM
grow \citep{Dwek09,Draine09,Mattsson11}\,? Do current theoretical
models of dust formation
\citep{Todini01,Nozawa03,Schneider04,Kozasa09} agree with the
observational data
\citep{Rho08,Rho09a,Kozasa09,Rho09b,Matsuura11}\,? Finally, which
kind of SN{\ae} produce dust? We need to deal with all these
questions to build a reliable set of dust yields by SN{\ae} to be
included in  chemical models of galaxies.\\
\indent \textsf{How much dust can a single supernova inject into the
ISM\,?} Since  the early observations of the SNa 1987A, this
question has long been debated with  controversial answers. The
reasons of uncertainty can be summarized as follows: (1) The sample
of observed SN{\ae} with ongoing dust formation is small
\citep{Kozasa09} so that it is almost impossible to get some
reliable clues about the link between mass and metallicity of the
progenitor and the amount of produced dust; (2) The MIR-NIR
observations could miss the presence of a significant amount of cold
and very cold dust. Only with SCUBA-2, ALMA, and Herschel Space
Observatory  we might be able to highlight this issue
\citep{Gomez07,Rho08,Nozawa08,Dunne09,Gomez09}. The very recent
discovery of a significant amount of very cold dust grains in
 SN 1987A \citep{Matsuura11} seems to strengthen this point, thus
suggesting that, as suspected, NIR/MIR observations are not able to
trace a complete picture of the dust in SN{\ae}; (3) It is not clear
if and how much dust is embedded in a thin envelope or in thick
clumpy regions \citep{Ercolano07}, thus making quantitative
estimates highly uncertain. In some cases the assumption that the
radiation emitted by dust comes from  an optically thin region could
lead to large errors \citep{Kozasa09,Meikle07}; (4) It is always a
cumbersome affair to discriminate between contamination by
foreground dust and dust residing and forming locally in the
observed SNa \citep[see for instance the discussion  on the
foreground contamination in the case of CasA
by][]{Dunne03,Krause04,Wilson05,Rho08}. What do observations tell
us? Till now, ongoing dust formation in the ejecta of thermonuclear
type Ia SN{\ae} has not been observed \citep{Kozasa09,Draine09},
even if for instance the classification of the Kepler SNa is still
uncertain and perhaps suggesting  a type Ia SNa \citep{Gomez09}. As
nowadays, a great deal of the observational evidence of dust
formation comes from  family of type II SN{\ae} otherwise known as
CCSN{\ae}. The Only exceptions in the family of CCSN{\ae} are the
type Ic SN{\ae}, in which no dust has been revealed so far
\citep{Kozasa09}.\\ \indent In Table \ref{tabella1} we summarize the
most significant observations of dust formation in SN{\ae}, together
with the available information about the progenitor mass and the SNa
type. It is soon evident that, despite the growing number of
observed objects and the improved quality of the data with the new
IR telescopes, our current knowledge of the problem is still far
from being satisfactory. Because of the uncertainties and the poor
statistics, it is  not possible to disentangle the complex
dependence of the observed ongoing dust formation on physical
parameters like the mass and metallicity of the progenitor star  and
the density of the underlying environment where the explosion took
place. In Fig. \ref{SNeYieldObserved} we display the current
observational estimates of the  amounts of dust together with their
uncertainties as a function of the progenitor mass (the entries of
Table \ref{tabella1}). For Kepler and Cas A SN{\ae} we plot also the
estimates derived from taking  into account recent sub-mm
determinations of the cold dust contribution
\citep{Dunne09,Gomez09}. In the same way, for SN 1987A we plot the
new estimate of the dust mass derived from FIR/sub-mm observations
with PACS and SPIRE  onboard Herschel Space Observatory. Finally, we
fit our small and scattered sample of data with simple analytical
expressions. If we consider all the objects whose estimates of the
dust content is based \textit{only} upon observations of warm dust
in the NIR/MIR, the analytical fit yields about 0.006-0.05
M$_{\odot}$ of dust per SNa, depending on the progenitor mass.
Clearly, this is only a mean lower limit because we are neglecting
the cold dust emitting at longer wavelengths. As suggested by the
FIR/sub-mm data for Cas A \citep{Dunne09}, Kepler \citep{Gomez09}
and SN 1987A \citep{Matsuura11}, the contribution by cold dust could
easily increase the average estimate by one or even two orders of
magnitude, i.e. up to 0.1-0.2 M$_{\odot}$ of dust per SNa (dashed
line in Fig. \ref{SNeYieldObserved}). If we consider only the
observations taking into account FIR/sub-mm  data, we get 0.4-0.7
M$_{\odot}$ per SNa: in this case SN{\ae} would be very efficient
dust factories! However, with a sample of only  three data drawing
any conclusion would be premature.  In any case, the data on the
cold wing of the dust population clearly indicates that SN{\ae} are
not poor dust producers as  claimed by \citet{Zhukovska08}. The
issue is anyway still open. Therefore, even ignoring the other
points of uncertainty we have mentioned above, i.e. the thin layer
approximation, the poor statistics and the foreground contamination,
the sole large uncertainty on the  contribution by cold dust renders
the whole subject highly uncertain.  More sub-mm data from SCUBA-2,
ALMA, and Herschel Space Observatory are needed  to solve the
problem.

\begin{figure*}
\centerline{\hspace{-39pt}
\includegraphics[height=6cm,width=8.5truecm]{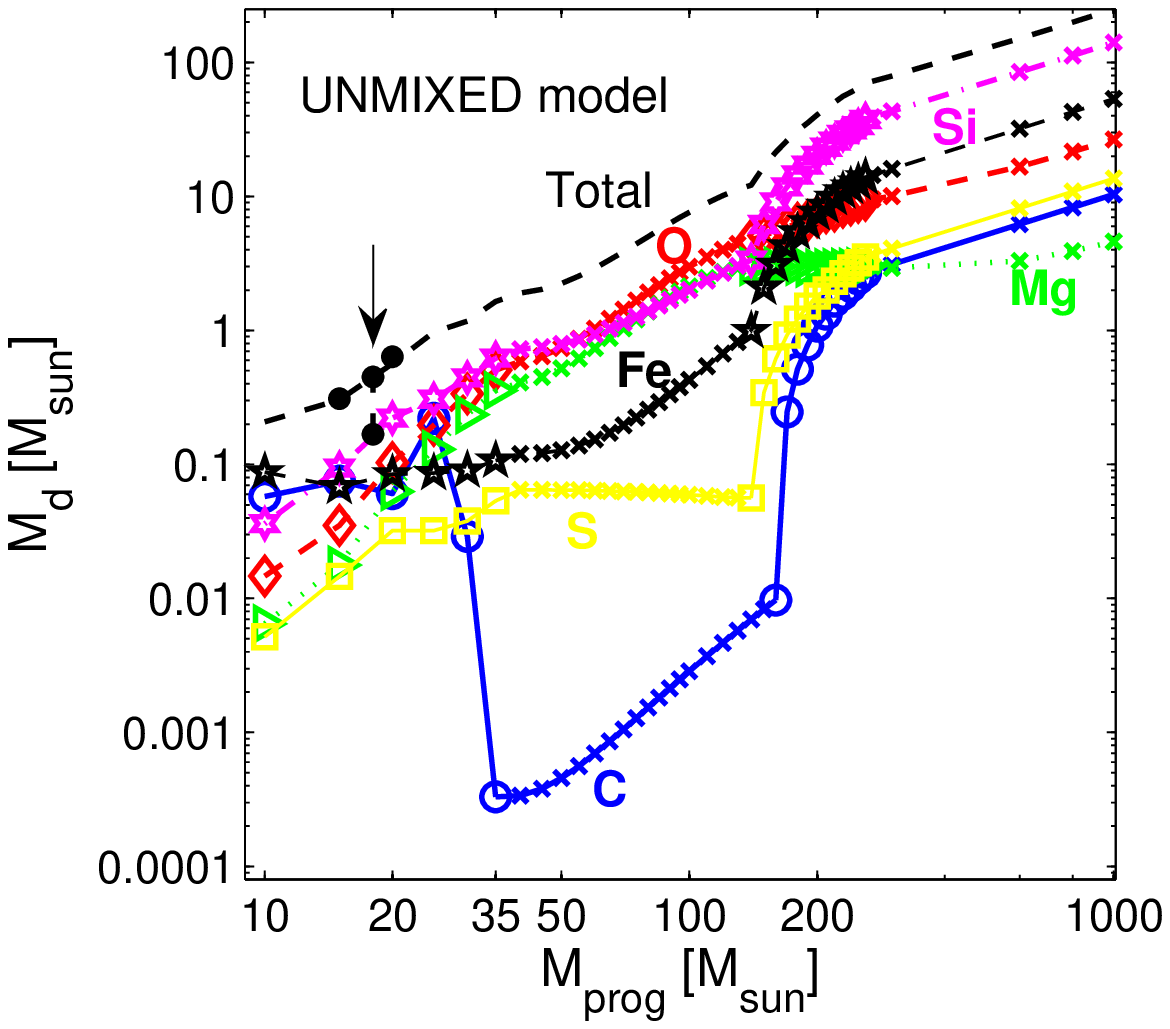}
\includegraphics[height=6cm,width=8.5truecm]{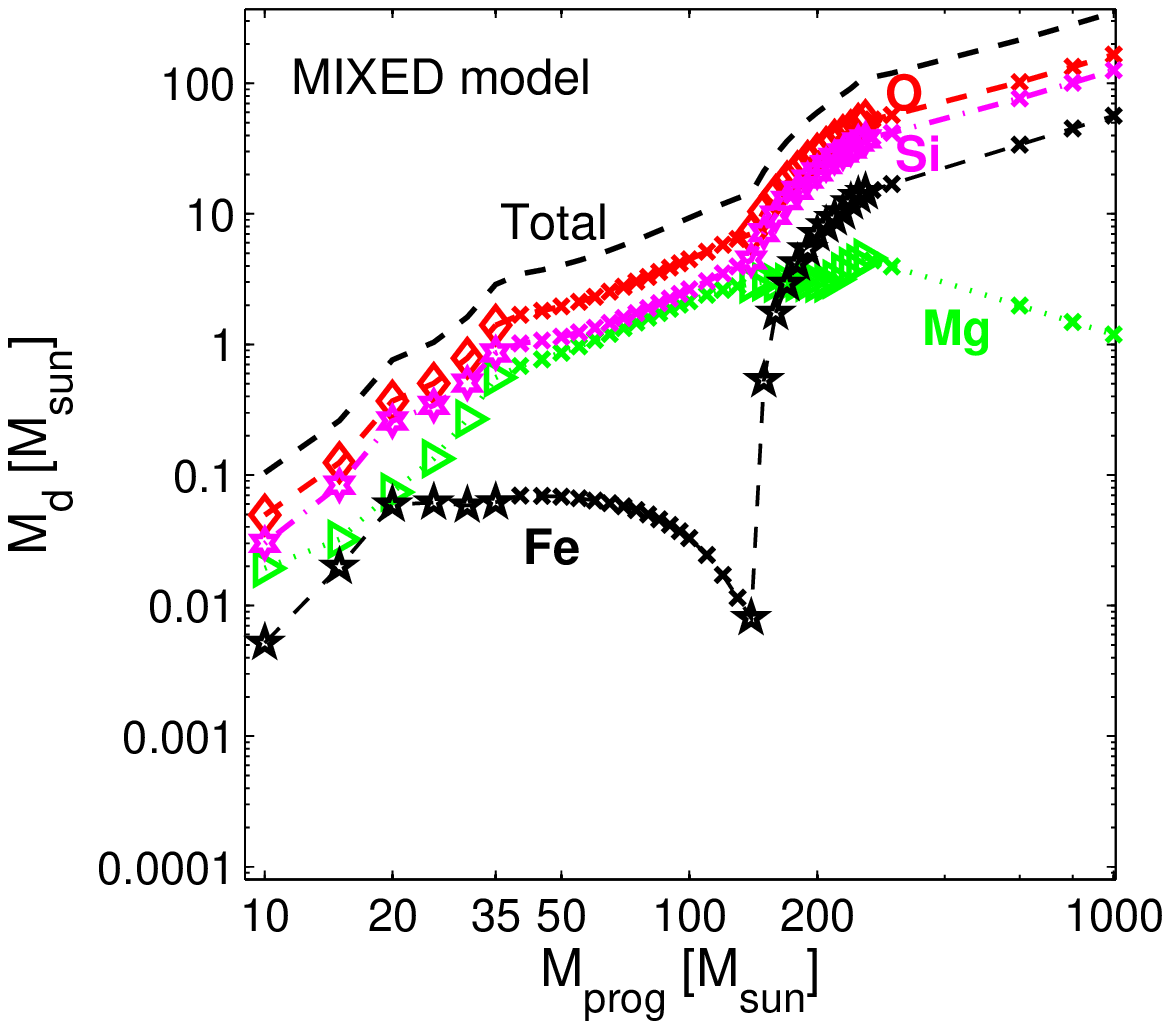}
\hspace{-39pt}} \caption{Yields of dust for C, O, Mg, Si, S and Fe,
calculated from the list of dust compounds of \citet{Nozawa03} and
the \textit{unmixed} (\textbf{Left Panel}) and \textit{mixed}
(\textbf{Right Panel}) models of ejecta as a function
 of the progenitor mass. All quantities on display are
expressed in solar masses. Small crosses represent extrapolations
from the data of \citet{Nozawa03} to other mass ranges. The legend
is as follows: C (empty circles and continuous line); O
(diamonds and dashed line); Mg (triangles and dotted line); Si
(six-pointed stars and dot-dashed line); S (squares and continuous
line) and Fe (five-pointed stars and dashed line). The dashed line
without markers represents the total amount of dust left over by the
shocks in the SNR. We also show (filled circles) the total yields of
dust  by \citet{Kozasa09} for the unmixed 15, 18 and
20M$_{\odot}$ models. The effect of a different hydrogen-rich
envelope on the amount of dust formed by a  18M$_{\odot}$ model is
also indicated by the arrow. }\label{YieldsBOTH}
\end{figure*}

\begin{figure*}
\centerline{
\includegraphics[height=6cm,width=7truecm]{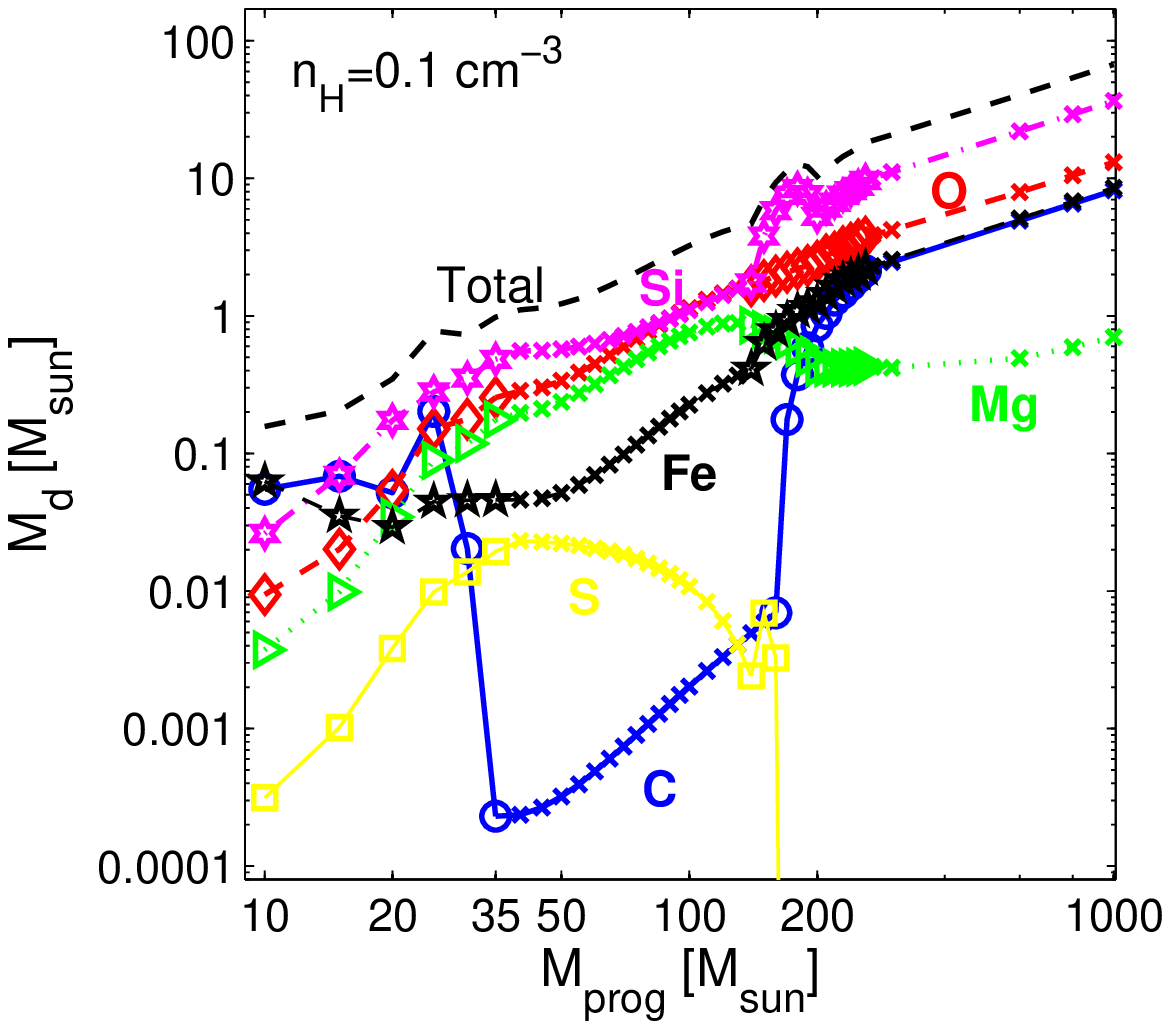}
\hspace{-35pt}
\includegraphics[height=6cm,width=7truecm]{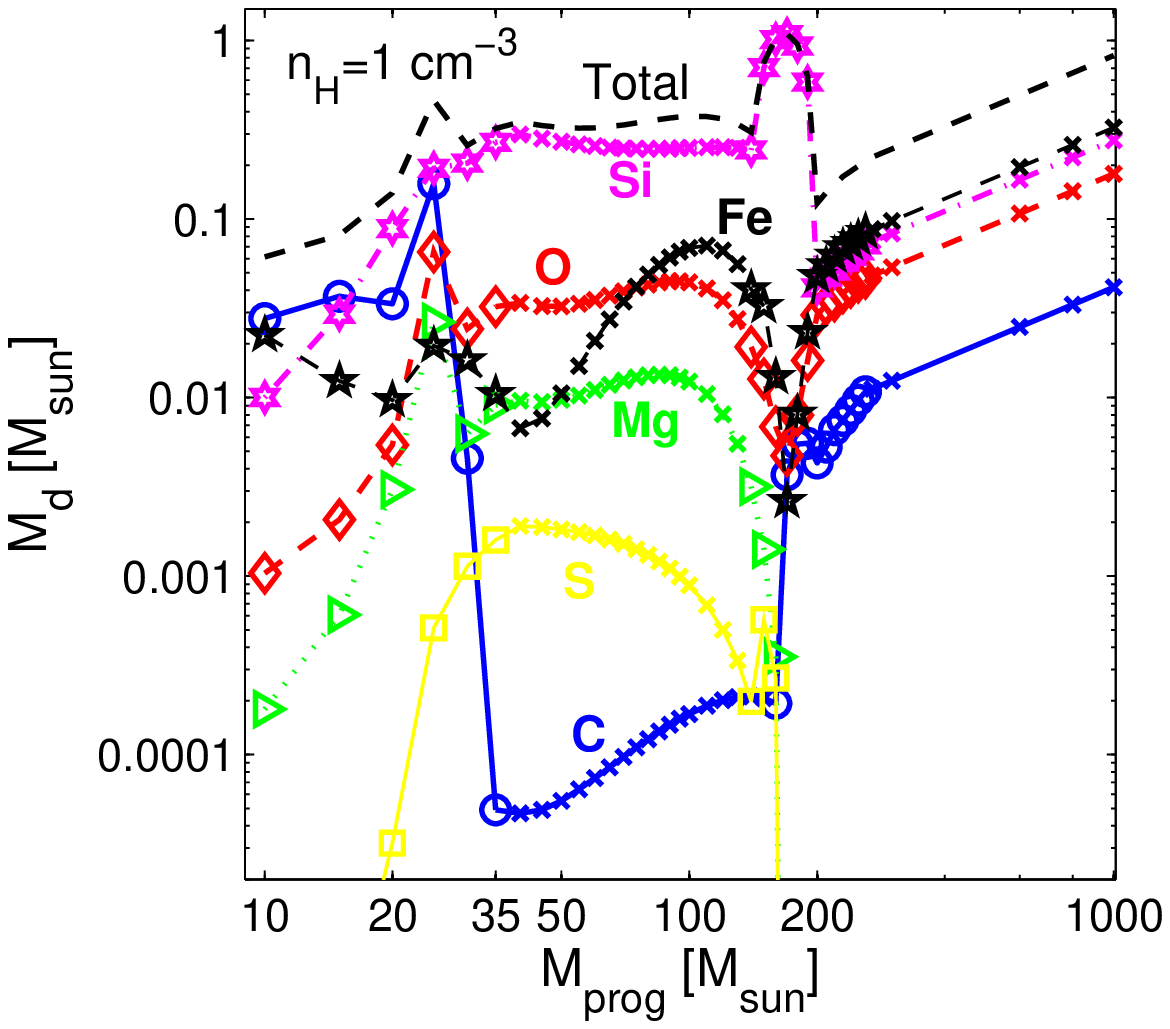}
\hspace{-35pt}
\includegraphics[height=6cm,width=7truecm]{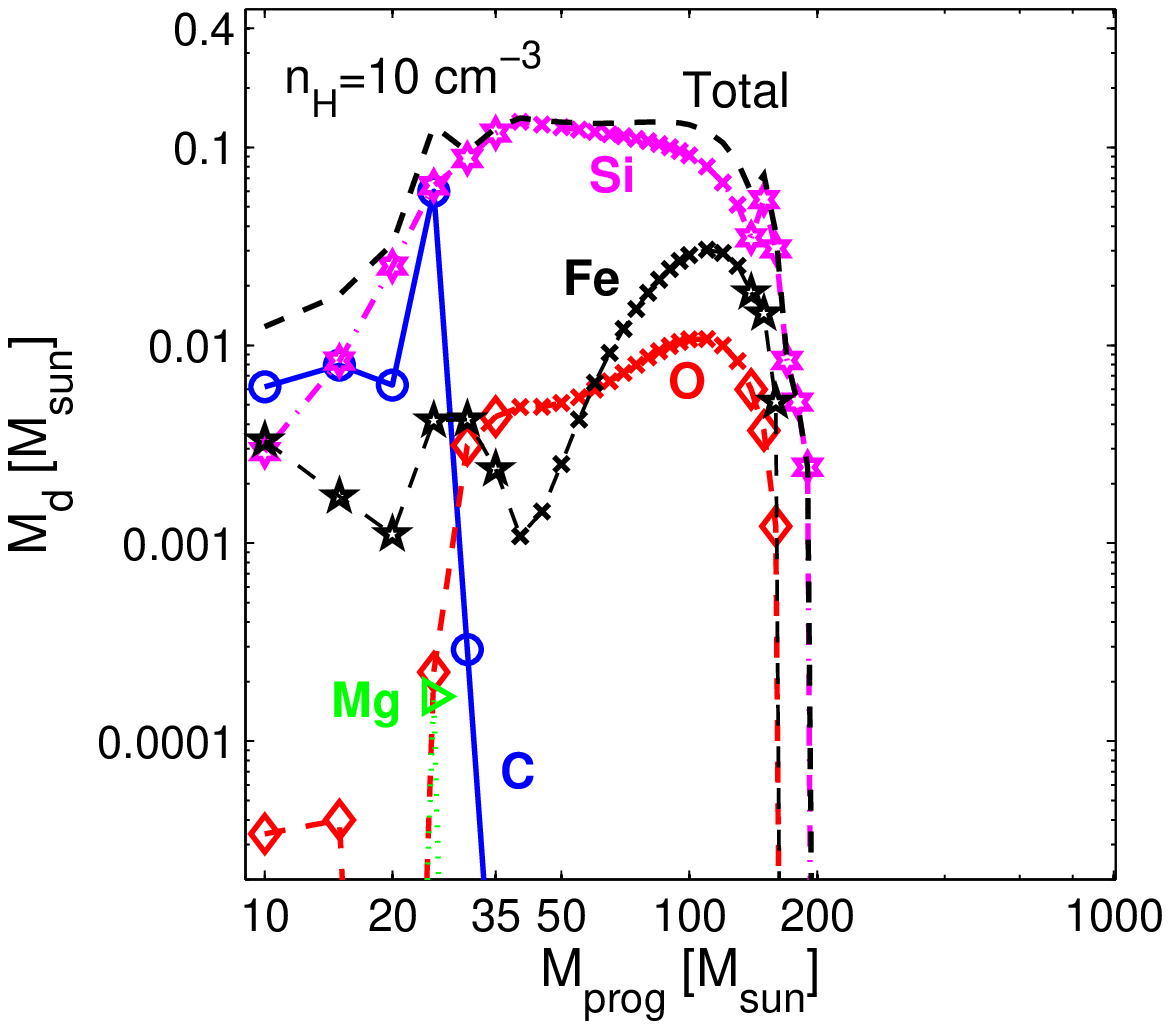}}
\caption{Masses of C, O, Mg, Si, S and Fe, hidden  in
the dust  and left over by the  reverse shocks in SNRs
  \citep{Nozawa07} as a  function of the progenitor mass, for the
\textit{unmixed} grain model of \citet{Nozawa03} and at varying the
hydrogen number density n$_{\textrm{H}}$. All quantities on display  are
expressed in solar masses. Small crosses represent extrapolations
from the dust  yields calculated by  \citet{Nozawa03} to other mass
ranges. The legend is as follows: C (empty circles and continuous
line); O (diamonds and dashed line); Mg (triangles and dotted
line); Si (six-pointed stars and dot-dashed line); S (squares
and continuous line) and Fe (five-pointed stars and dashed line).
The dashed line without markers represents the total amount of dust
survived to the shocks in the SNR. \textbf{Left Panel}: Masses of
C, O, Mg, Si, S and Fe in dust survived to reverse
shocks for n$_{\textrm{H}}=0.1 \,$cm$^{-3}$. \textbf{Middle Panel}: The same
as in left panel but for n$_{\textrm{H}}=1\,$cm$^{-3}$. \textbf{Right Panel}:
The same as in left panel but for
n$_{\textrm{H}}=10\,$cm$^{-3}$.}\label{UnmixedDestroyed}
\end{figure*}

\begin{figure*}
\centerline{
\includegraphics[height=6cm,width=7truecm]{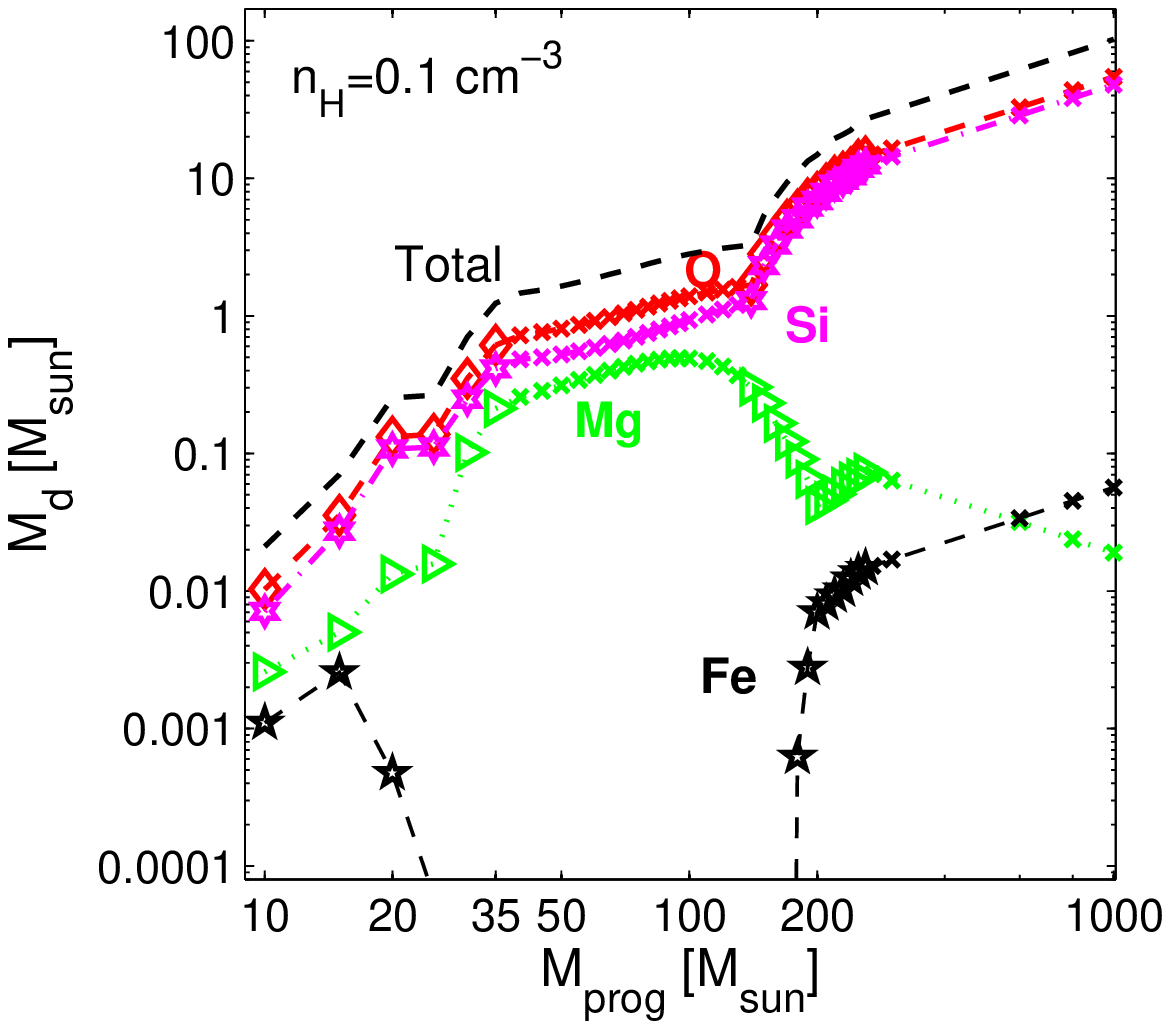}
\hspace{-35pt}
\includegraphics[height=6cm,width=7truecm]{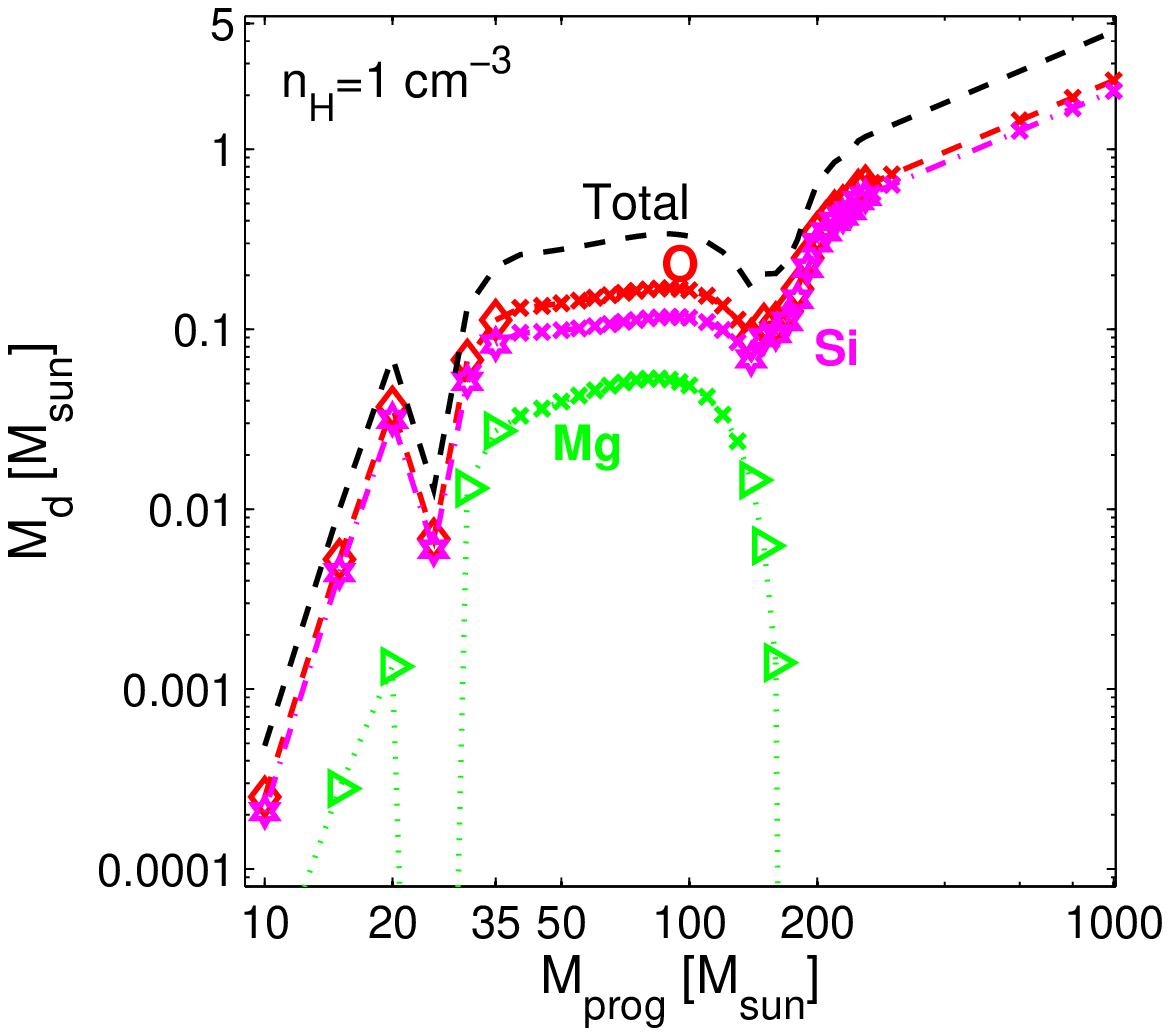}
\hspace{-35pt}
\includegraphics[height=6cm,width=7truecm]{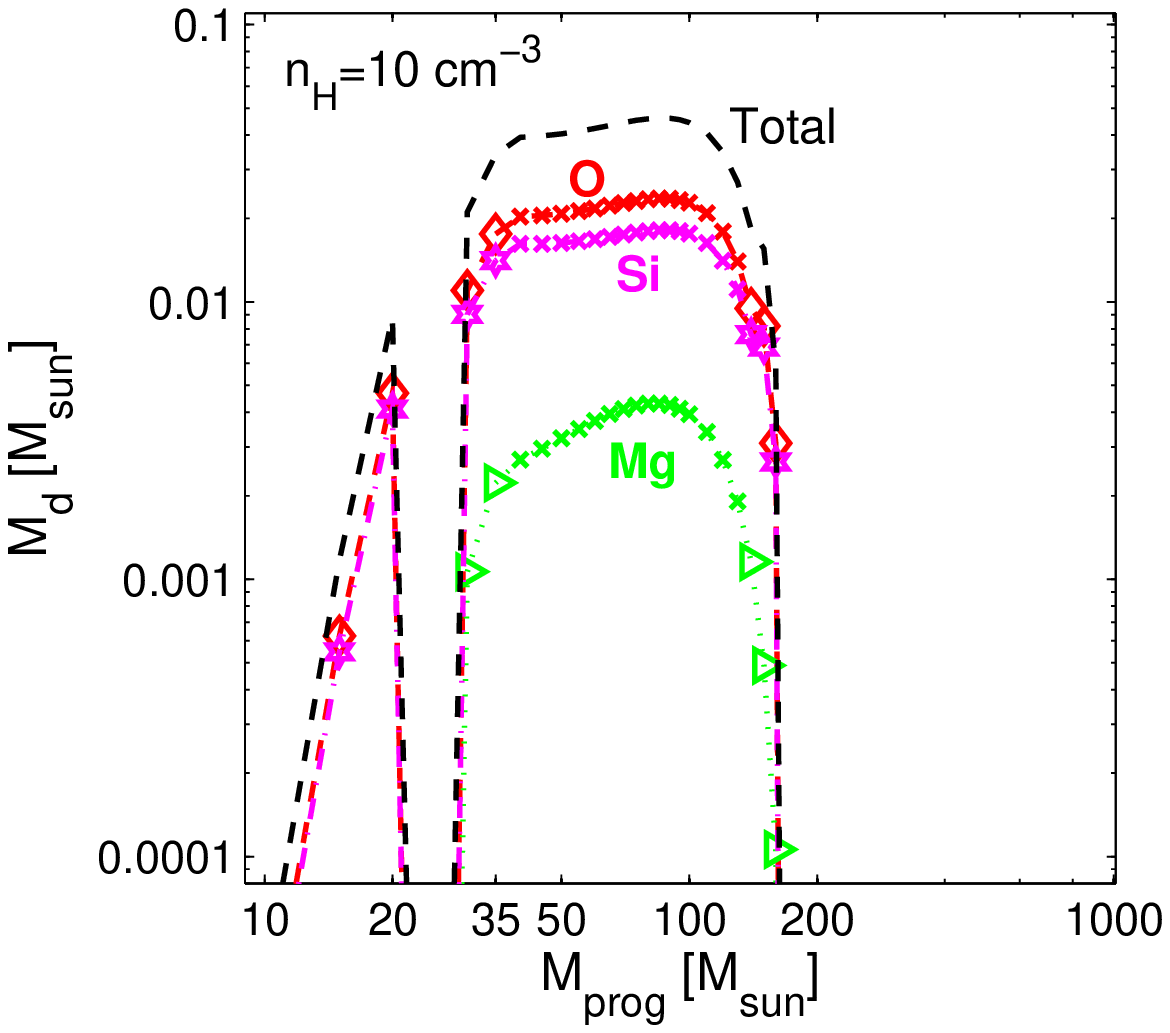}}
\caption{The same as in Fig. \ref{UnmixedDestroyed} but   for the
\textit{mixed} grain model by \citet{Nozawa03}. The meaning of all
the symbols is the same as in Fig. \ref{UnmixedDestroyed}.
\textbf{Left Panel}: The masses of O, Mg, Si and Fe in dust
that survived to the reverse shocks for n$_{\textrm{H}}=0.1\,$cm$^{-3}$.
\textbf{Middle Panel}: The same as in left panel but for
n$_{\textrm{H}}=1\,$cm$^{-3}$. \textbf{Right Panel}: The same as in left
panel but for n$_{\textrm{H}}=10\,$cm$^{-3}$.} \label{MixedDestroyed}
\end{figure*}

\begin{figure*}
\centerline{\hspace{-39pt}
\includegraphics[height=6.5cm,width=7.5truecm]{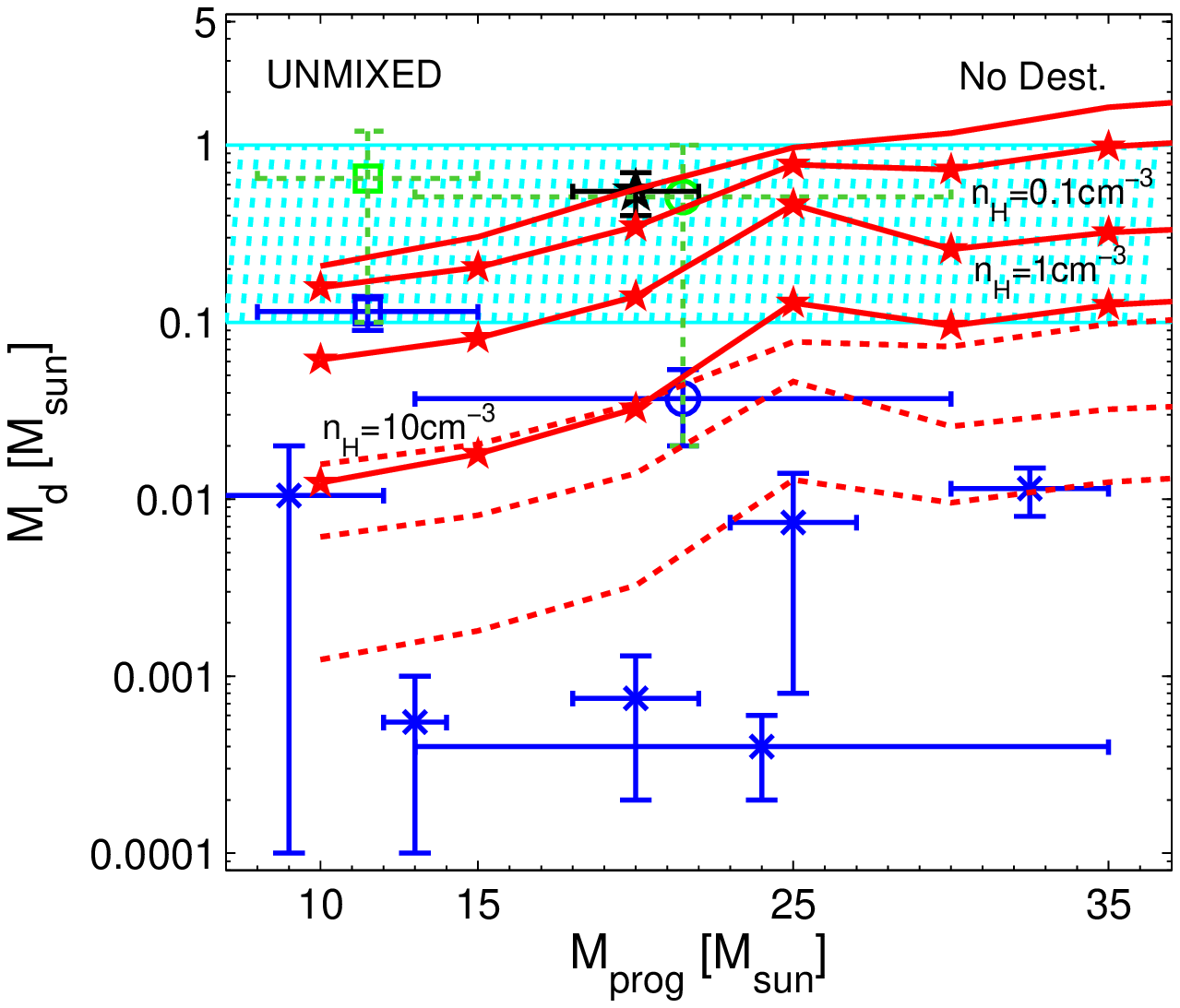}
\includegraphics[height=6.5cm,width=7.5truecm]{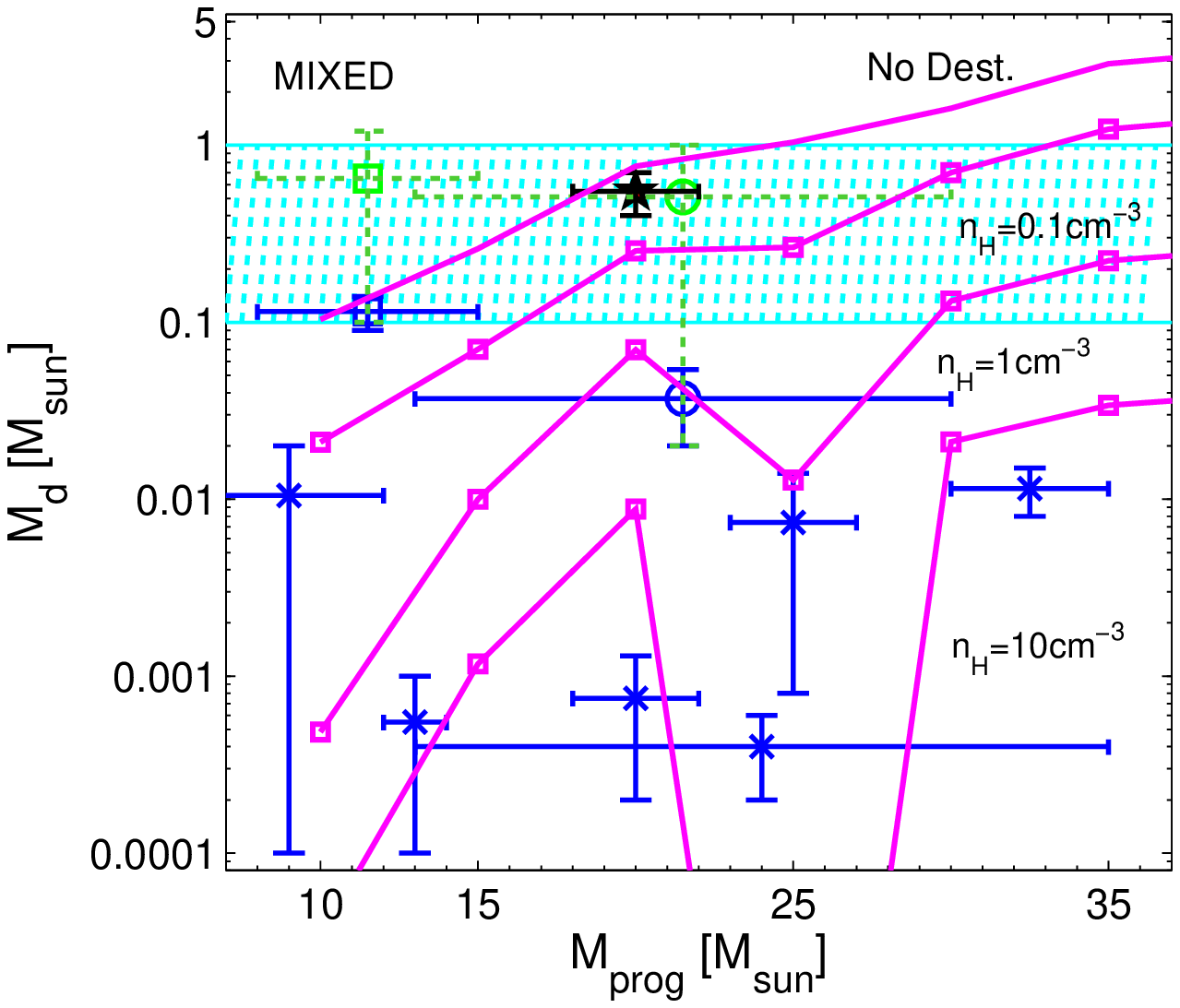}
\hspace{-39pt}} \caption{Comparison between theoretical models and
observational data. The amount of dust surviving destruction is
shown for three ambient number densities
n$_{\textrm{H}}=0.1$\,cm$^{-3}$, n$_{\textrm{H}}=1$\,cm$^{-3}$ and
n$_{\textrm{H}}=10$\,cm$^{-3}$. Also, the original undestroyed
yields by \citet{Nozawa03} are displayed. The crosses, circles,
squares and five-pointed star represent observational data from
Table \ref{tabella1} relative to freshly formed dust in SNRs, as in
Fig. \ref{SNeYieldObserved}, with the same meaning of the symbols.
The hatched area in both panels represent the amount of dust per SNa
needed to explain the obscured high-z quasars, according to the
estimates by \citet{Dwek09}. \textbf{Left panel} Theory vs.
observation for the \textit{unmixed} model. The solid  line without
marks shows the undestroyed yields. The continuous lines from top to
bottom show the yields at increasing n$_{\textrm{H}}$. Dotted lines
represent the three n$_{\textrm{H}}$ re-scaled by a factor of 10.
\textbf{Right panel} Theory vs. observation for the \textit{mixed}
model. The continuous lines have the same meaning as in the left
panel. }\label{TheoryVSObservations}
\end{figure*}

\begin{figure*}
\centerline{
\includegraphics[height=6cm,width=7truecm]{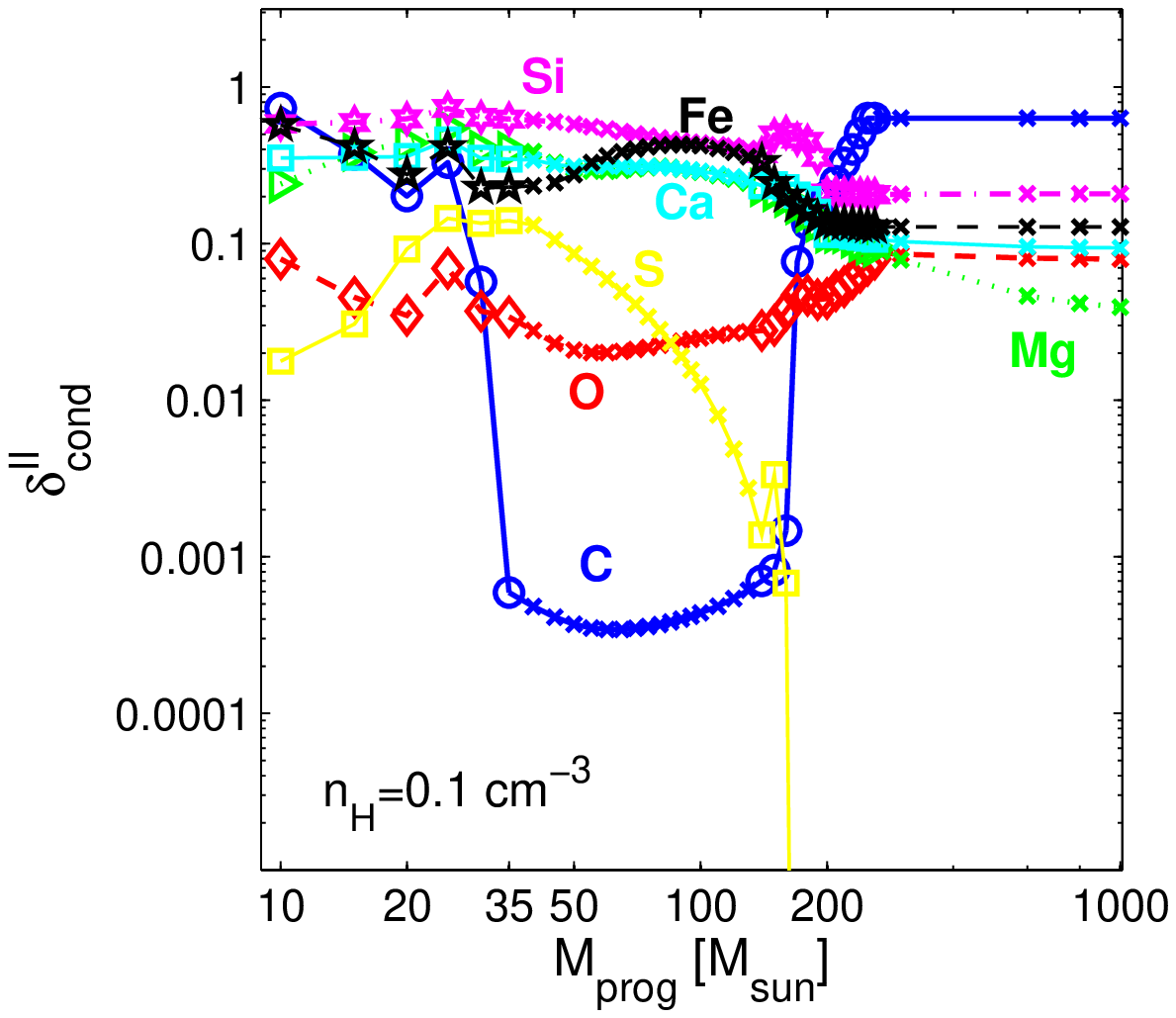}
\hspace{-35pt}
\includegraphics[height=6cm,width=7truecm]{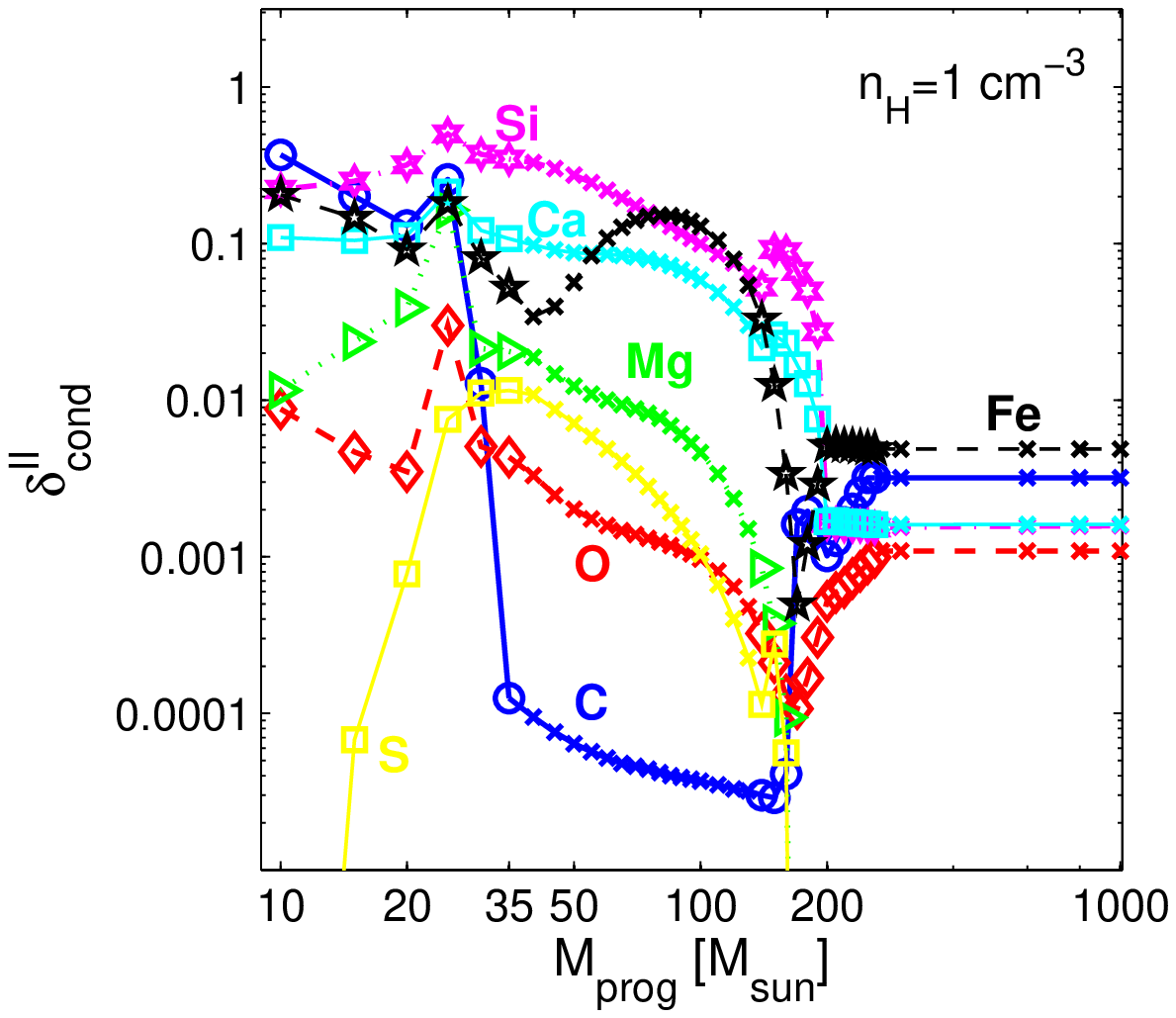}
\hspace{-35pt}
\includegraphics[height=6cm,width=7truecm]{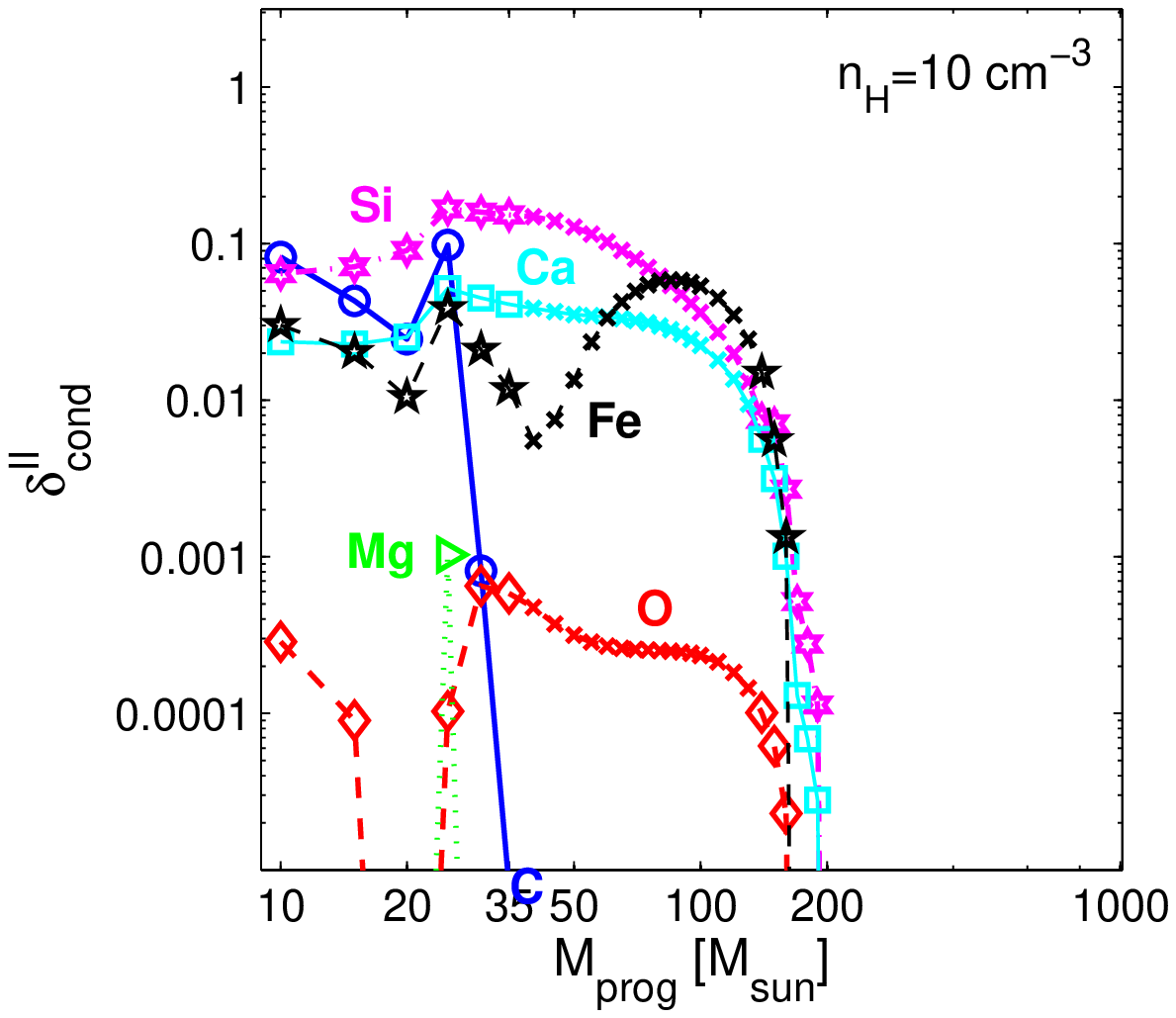}}
\caption{Condensation efficiencies of the elements C, O, Mg,
Si, Ca, S and Fe in SNRs as a function of the progenitor
mass, according to the \textit{unmixed} grain models of
\citet{Nozawa03,Nozawa07} and at varying the hydrogen number density
n$_{H}$. The small crosses represent extrapolations of the yields of
dust by \citet{Nozawa03} to other mass ranges. We plot: C (empty
circles and continuous line); O (diamonds and dashed line); Mg
(triangles and dotted line); Si (six-pointed stars and dot-dashed
line); S (squares and continuous line); Ca (squares and
continuous line) and Fe (five-pointed stars and dashed line).
\textbf{Left Panel}: Condensation efficiencies of C, O, Mg,
Si, S and Fe in dust survived to reverse shocks in a medium
with n$_{\textrm{H}}=0.1 \,$cm$^{-3}$. \textbf{Middle Panel}: The same as in
left panel but for n$_{\textrm{H}}=1\,$cm$^{-3}$. \textbf{Right Panel}: The
same as in left panel, only for n$_{\textrm{H}}=10\,$cm$^{-3}$.}
\label{UnmixedDelta}
\end{figure*}

\begin{figure*}
\centerline{
\includegraphics[height=6cm,width=7truecm]{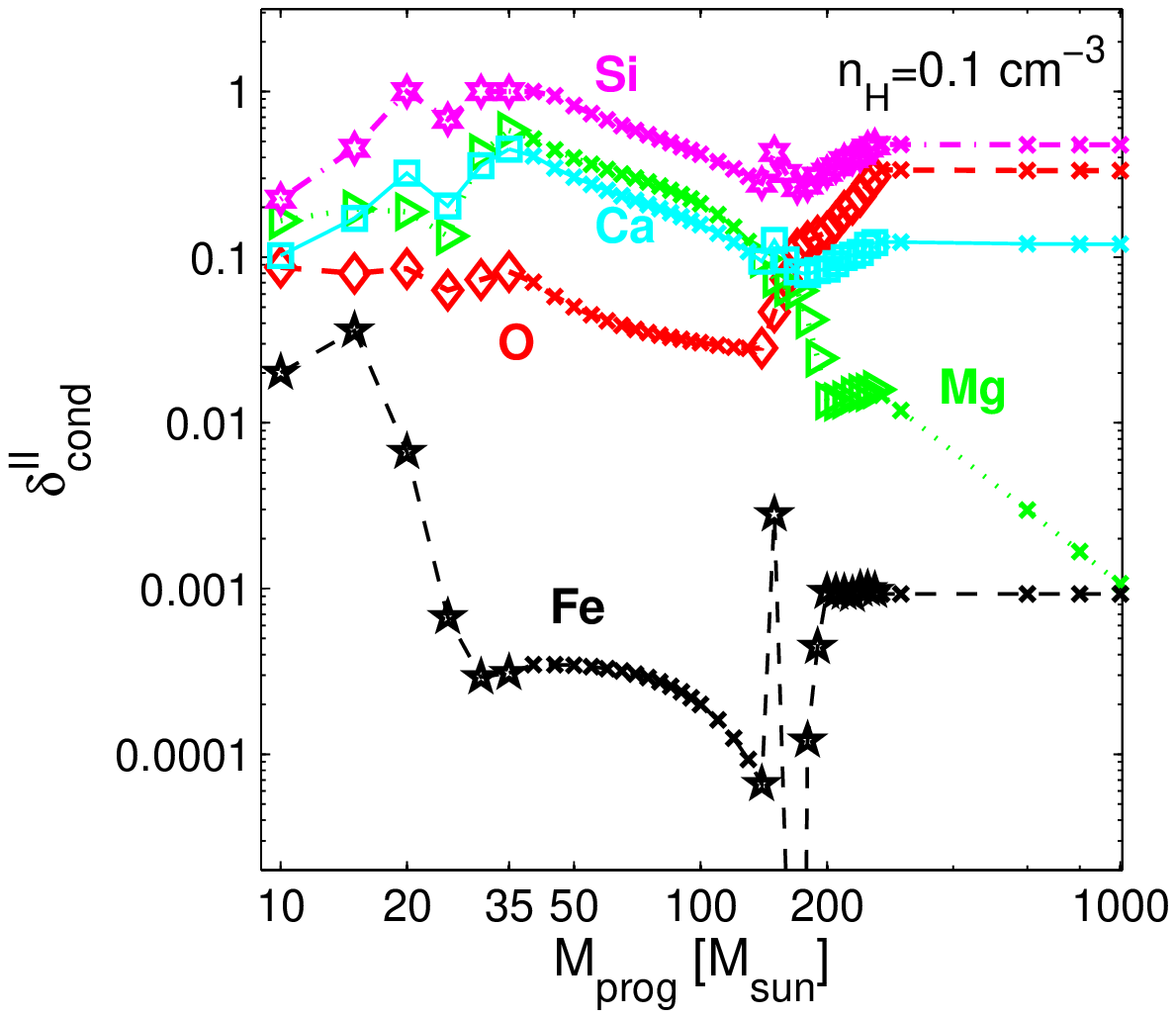}
\hspace{-35pt}
\includegraphics[height=6cm,width=7truecm]{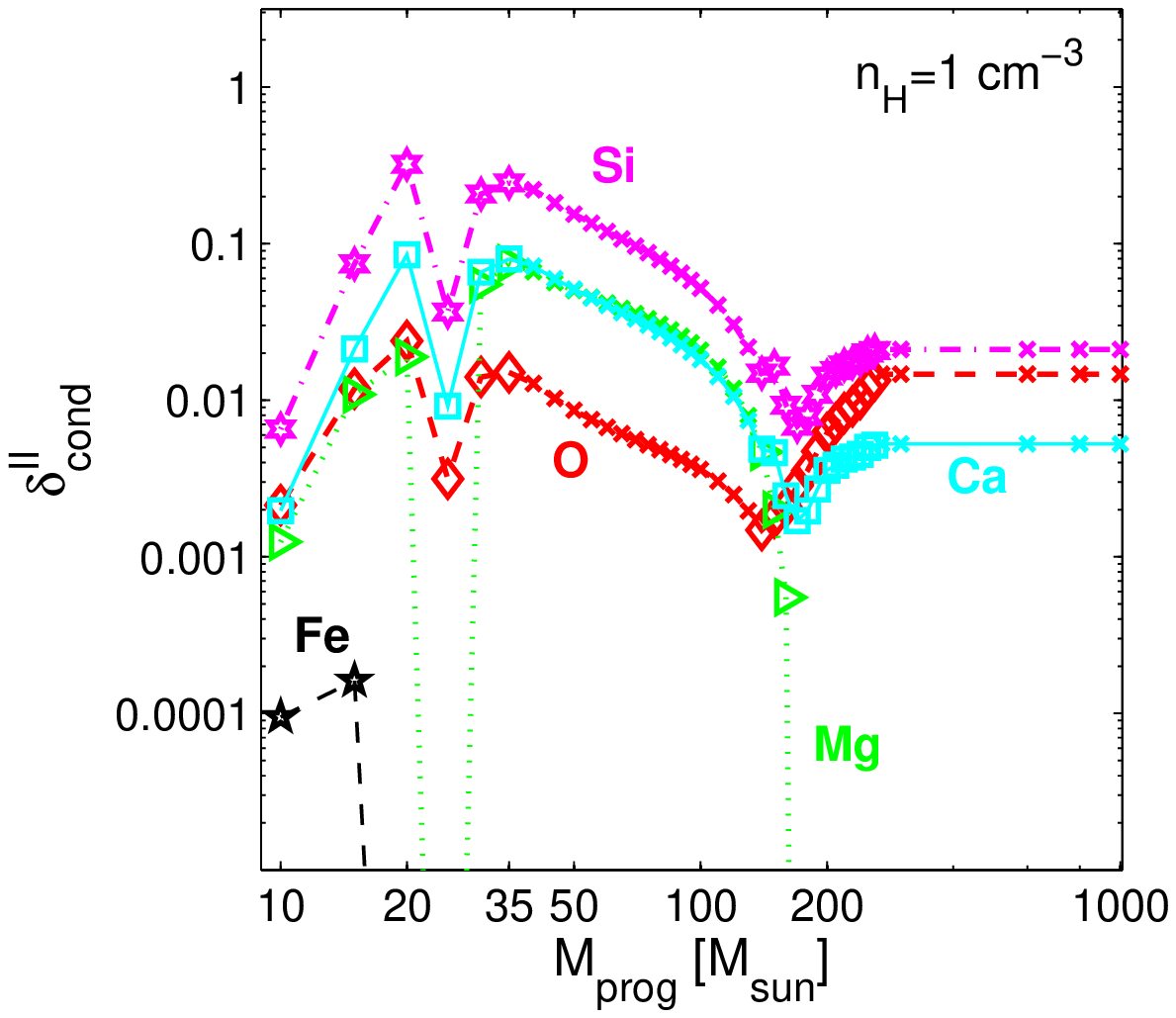}
\hspace{-35pt}
\includegraphics[height=6cm,width=7truecm]{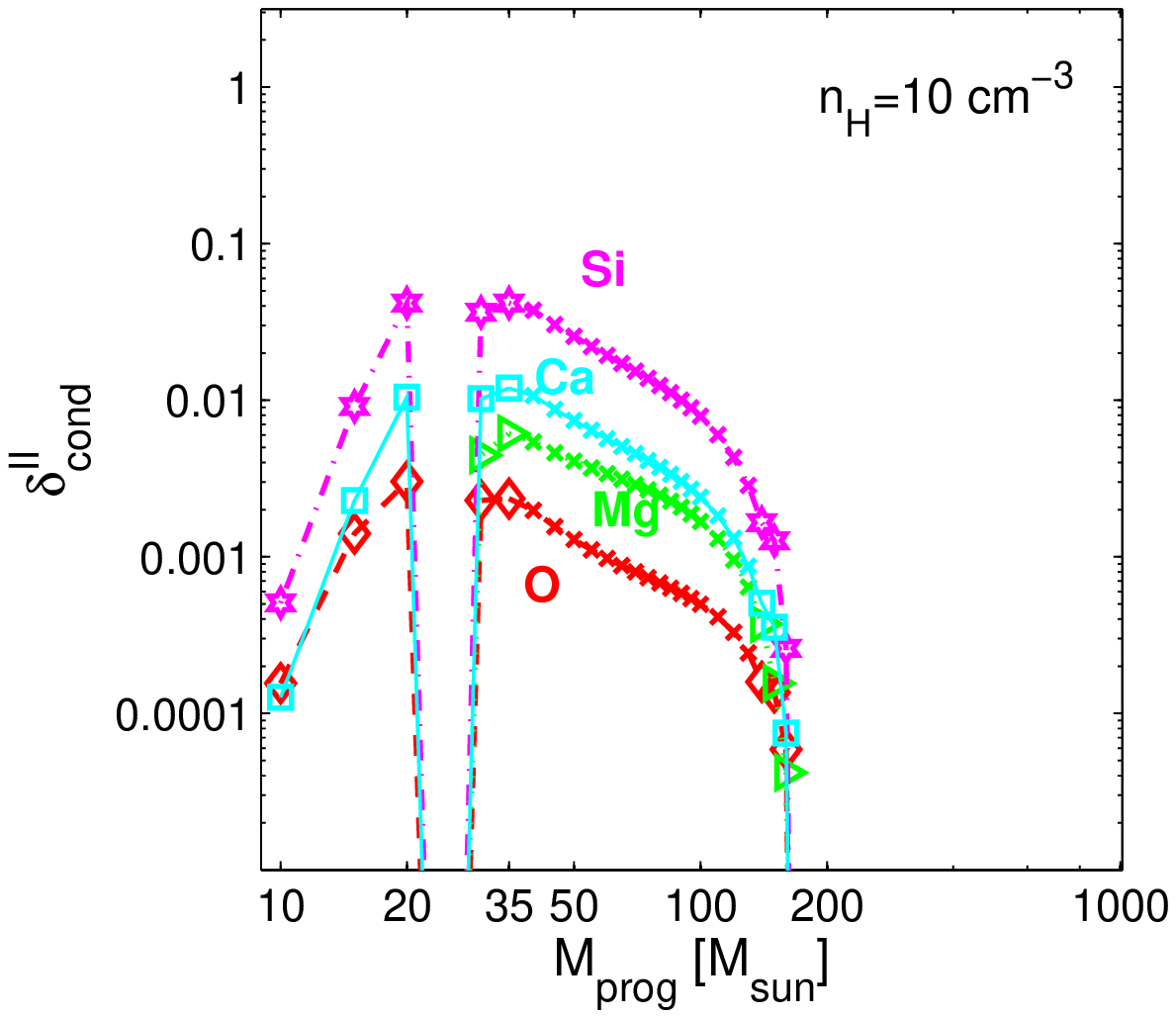}}
\caption{The same as in Fig. \ref{UnmixedDelta} but for the
\textit{mixed} grain model of \citet{Nozawa03,Nozawa07}. The meaning
of all the symbols is the same as in Fig. \ref{UnmixedDelta}.
\textbf{Left Panel}: Condensation efficiencies of O, Mg, Si,
Ca and Fe for n$_{\textrm{H}}=0.1\,$cm$^{-3}$. \textbf{Middle Panel}: The
same as in left panel but for n$_{\textrm{H}}=1\,$cm$^{-3}$. \textbf{Right
Panel}: The same as in left panel but for n$_{\textrm{H}}=10\,$cm$^{-3}$.}
\label{MixedDelta}
\end{figure*}

\textsf{How the empirical data compare with the theoretical models?}
Until now, an handful of studies have tried to theoretically model
dust \citep{Todini01,Nozawa03,Schneider04,Kozasa09} and molecules
\citep{Cherchneff08,Cherchneff09a,Cherchneff10} formation in
SN{\ae}, coupling a more or less refined classical nucleation theory
(CNT) or kinetic theory with models of SN{\ae} explosions able to
follow for hundred of days the evolution of the expanding envelope.
Even if some of these studies have been dedicated to  Population III
SN{\ae}, their results and conclusions can be applied to SN{\ae}
with progenitors of different metallicities, even with super-solar
values. Indeed, according to \citet{Todini01} and \citet{Nozawa03},
dust formation in the ejecta is almost insensitive to the
metallicity of the progenitor stars. The processes of dust
destruction and cooling in the surrounding ISM are also scarcely
dependent on the ISM metallicity \citep{Nozawa07,Nozawa08}. The most
complete compilation of dust yields are, even if limited to  Pop III
SN{\ae}, by \citet{Nozawa03}. In brief, they modelled the formation
of  dust in CCSN{\ae} from 13 to 30 M$_{\odot}$ and Pair-Instability
SN{\ae} (PISN{\ae}) from 170 to 200 M$_{\odot}$ for both unmixed and
mixed He cores  and including a wide range of dust compounds. From
their database we derived  the mass of each element  embedded in the
dust components. More details of it are given in Appendix A. In Fig.
\ref{YieldsBOTH} we show the yields of dust for each element and the
total yield, for both unmixed (left panel) and mixed (right panel)
cases. Since the progenitor masses in the \citet{Nozawa03} grid do
not cover the whole range of possible values some
interpolation/extrapolation of the data have been applied. Owing to
the coarse coverage of large mass intervals, the
interpolation/extrapolation procedure may be affected by large
uncertainties. In general, in the case of unmixed cores, many dusty
compounds form, in particular the carbon and sulphur dust, that do
not form in the mixed case. In this latter, oxygen atoms are more
abundant than carbon atoms and only silicates and oxides form
\citep{Nozawa08}. The general trend of all the elements is quite
regular, with just some exceptions, like carbon (in the unmixed
model) and iron (in the mixed model): the yields grow at growing
mass of the progenitor.\\
\textsf{Compared with the observational data, are the theoretical
yields satisfactory?}  Before comparing theory and observations, we
have taken into account the dynamical evolution of the dust and its
destruction in SNRs, in particular due to the passage of the reverse
shock \citep{Nozawa07,Bianchi07}. Basing on previous studies by
\citet{Nozawa03} and \citet{Nozawa06}, \citet{Nozawa07} calculated
the dust yields and sizes of dust grains surviving destruction.
Starting from the yields  described in Appendix A and multiplying
them for the destruction coefficients \citep{Nozawa07}, we derive
the new yields as a functions of the ambient numerical density
n$_{\textrm{H}}$ that are presented in  Figs. \ref{UnmixedDestroyed}
(unmixed model) and \ref{MixedDestroyed} (mixed model).  Here, we
show the amount of each element embedded into dust grains and
finally injected into the ISM without being destroyed in the SNR
evolution as a function of the ambient gas number density
$\textrm{n}_{\textrm{H}}$. In both mixed and unmixed cases, the
higher the ambient density $\textrm{n}_{\textrm{H}}$, the higher is
the amount of dust destroyed and the smaller the yields. Some
elements, like S or Mg in the unmixed case or Fe in the mixed one,
are completely destroyed in high density environments.\\
\indent Finally, in Fig. \ref{TheoryVSObservations} we compare the
theoretical yields with the observational data, for both the unmixed
and mixed models. The yields based on the unmixed models marginally
agree with the observational data obtained from the MIR observations
of SNRs. To get a satisfactory agreement with the MIR estimates of
the dust content we would need to re-scale the  yields from unmixed
models by at least a factor of 10 (see as Fig.
\ref{TheoryVSObservations}). However, these yields  much better
agree with the estimates of the dust content  in Kepler, Cas A
(dotted crosses) and 1987A (continuous black cross) SN{\ae} , once
the contribution by cold dust is included
\citep{Gomez09,Dunne09,Matsuura11}. These more recent data increase
the dust production by  SN{\ae} by at least one order of magnitude.
In any case, a satisfactory comparison between data and theory, the
latter including also accurate evaluations of the amounts of cold
dust, would be possible if more and better observations of SN{\ae}
in the FIR/sub-mm become available. The yields based on mixed
models, because of the stronger destruction of dust grains in the
SNRs, better agree with the MIR observations, but considering the
contribution  of cold dust they fail to match  the FIR/sub-mm data
unless the SNa explosion takes place in a low density environment.\\
\textsf{Condensation efficiencies}. Once  the original total yields
from the SN{\ae} models are known, one can derive the condensation
efficiencies of the various elements. One could refer to the study
by  \citet{Nozawa03} who used ad-hoc hydrodynamic models and results
of nucleosynthesis calculations that were based on   the models by
\citet{Umeda02}, but with different properties, like the mass-cut,
progenitor mass, Y$_{e}$ and explosion energy (T. Nozawa, private
communication). Unfortunately these SNa models are not publicly
available. To cope with this, we follow the suggestion by Umeda
(2011, private communication) and make use of the up-to-date
nucleosynthesis calculations by \citet{Nomoto06,Tominaga07} and the
original models by \citet{Umeda02} to  get the final dust-to-gas
ratios from the gaseous yields (the correct correspondence between
the parameters of the SNa models and those used to derive the yield
of \citet{Nozawa03} is secured).

In Figs. \ref{UnmixedDelta} and \ref{MixedDelta} we show the
condensation efficiencies of C, O, Mg, Si, S and Fe for
the unmixed model and O, Mg, Si, Ca and Fe for the mixed
one, both at varying the ambient density n$_{\textrm{H}}$. Obviously for the
highest densities, more grains are destroyed before being injected
into the ISM and therefore the condensation efficiencies are lower.
In our chemical model we consider also the evolution of Ca. This
element  is not considered in the nucleation models by
\citet{Nozawa03}, but is included in the SNa yields by
\citet{Portinari98} and with other refractory elements contributes
to various pyroxene and olivine minerals. For the condensation
efficiency of Ca  we adopt the mean value of the other refractory
elements (Mg, Si, S and Fe). The result for Ca is shown in
Figs. \ref{UnmixedDelta} and \ref{MixedDelta} as a continuous line
with empty squares.

\indent \textsf{How these yields of dust and corresponding
condensation efficiencies compare with the amount of dust that is
estimated to explain the obscured objects at high redshift?}  The
question is still open and vividly debated, see for instance
\citet{Dwek07,Dwek09}, \citet{Draine09} and \citet{Nozawa08}, and in
particular \citet{Maiolino04}, \citet{Wang08a,Wang09},
\citet{Wagg09} and \citet{Michalowski10a,Michalowski10b} for high-z
observations of obscured quasars and LAEs. It is not clear whether
SN{\ae} play a major role as dust producers in high-z, very young
galaxies, when AGB stars still have not yet started contributing to
the total budget \citep{Sugerman06,Bianchi07,Nozawa08}, or grain
accretion in the ISM dominate leaving to SNRs the role of seed
producers over which accretion should take place
\citep{Dwek09,Draine09}. \citet{Dwek09} argue that
$0.1-1$M$_{\odot}$ of dust is produced by every SNa to fully explain
high-z obscured objects, the dust being originating in SNRs. In Fig.
\ref{TheoryVSObservations} we indicate with the hatched area the
$0.1-1$M$_{\odot}$ region. Our theoretical yields agree with the
values falling into the hatched region, in particular for the
unmixed case. They may differ by about one order of magnitude from
the MIR estimates which, as shown in Fig. \ref{SNeYieldObserved},
indicate about $0.01\, \textrm{M}_\odot$ of dust per SNa. Because of
it,  \citet{Dwek09,Draine09} favoured the accretion in the ISM as
the dominant source of dust in high-z quasars. However, more
detailed observations of cold dust in some SNRs
\citep{Gomez09,Dunne09}, and in particular the recent
\citet{Matsuura11} estimate, significantly increase the dust
contribution by  SN{\ae}, that becomes high enough to overwhelm the
ISM accretion in the early stages. This is also what is
theoretically  predicted and modelled in the high SFR inner regions
of the MW Disk by \citet{Piovan11b,Piovan11c}. The dust accretion in
the ISM requires that some enrichment in metals has already occurred
so that some delay is  unavoidable. For very high SFR such as in
QSOs, the delay  can be very short \citep{Gall11a,Gall11b}, thus
further complicating the whole picture. The issue is still debated.\\
 \indent \textsf{Which kind of SN{\ae} produce dust?}  From the entries
of Table \ref{tabella1} we note that nearly all SNa types   are dust
producers. The only exception are type Ia SN{\ae} in which no dust
has been detected  \citep{Borkowski06,Draine09}. In addition to
this,  in meteorites no pre-solar grains formed in type Ia
thermonuclear SN{\ae} explosions have been found \citep{Clayton04}.
Therefore, it is most likely that  type Ia SN{\ae} have almost zero
condensation efficiency. Recently, \citet{Kozasa09} calculated some
models of dust formation in CCSN{\ae}, based upon the same formalism
of \citet{Nozawa03}, but using different underlying models of
SN{\ae}. The aim was  to investigate the effects on dust nucleation
of a different amount of hydrogen  in the envelope at the onset of
the collapse. In Fig. \ref{YieldsBOTH} (left panel) we show the
total  yields of dust by
\citet{Kozasa09} for their unmixed 15,
 18 and 20M$_{\odot}$ models (filled circles), and compare them
with the \citet{Nozawa03} yields. When the hydrogen-rich envelope at
the onset of the collapse is thick, the models agree each other, and
the total amount of dust produced is about the same. On the
contrary, as indicated by the arrow in Fig. \ref{YieldsBOTH} for the
18M$_{\odot}$ star, the effect of the hydrogen-rich envelope on
the amount of dust produced is significant: for type IIb SN{\ae} it
drops  by a factor of about three, from $\thicksim$0.45M$_{\odot}$
to 0.167M$_{\odot}$. This finding   for the 18M$_{\odot}$ model
suggests that in chemical models of galaxies it would be interesting
to distinguish the contribution by different types of CCSN{\ae},
e.g. because of different mass loss histories a different onion-like
structures of the progenitor \citep{Kozasa09,Gall11a}. The
effect of the varying hydrogen envelope could modify our view of the types of SN{\ae} able to
produce significant amounts of dust.

 \indent \textsf{Mixed or unmixed. Which is more consistent with
observations?}  Basing on observations of the Cas A remnant
\citep{Ennis06}, \citet{Kozasa09} prefer to use unmixed models.
Furthermore: (i) the unmixed model better reproduces  the extinction
curves observed in high-z quasars \citet{Hirashita05}; (ii) they are
exactly in the range suggested by \citet{Dwek09} to cope with the
high-z obscured universe and, finally, (iii) SN{\ae} have to produce
some amount of carbonaceous grains according to
 observations of pre-solar dust, whereas the mixed model is not able to
produce C-based dust. It seems therefore that the unmixed model
condensation efficiencies should be preferred. We will present in
our tables only the condensation efficiencies for the unmixed
case.\footnotemark[1]

\footnotetext[1]{The tables of condensation efficiencies for the mixed model will be anyway
available upon request.}

 \indent \textsf{Other condensation efficiencies}.
For the sake of comparison and completeness, we take into account other prescriptions
for the  efficiency of dust condensation  in SNRs. First of all the simple
formulation by \citet{Dwek98} and \citet{Calura08} who  for type II
SN{\ae} adopt  a set of condensation efficiencies \textit{independent} from
the mass/metallicity of the star or the density of the parental environment:

\vspace{-7pt}
\begin{equation}
M_{D,A}\left(M,Z\right)=\delta^{II}_{c}\left(A\right)M_{A}\left(M,Z
\right)
\end{equation}\vspace{-7pt}
\begin{equation}
M_{D,C}\left(M,Z\right)=\delta^{II}_{c}\left(C\right)M_{C}\left(M,Z
\right)
\end{equation}
\vspace{-7pt}

\noindent where $M_{A}\left(M,Z \right)$ is the mass of the generic
element $A$ ejected by the star of mass $M$ and metallicity $Z$, and
A$=$\,Mg,\,Si,\,S,\,Ca,\,Fe (all the refractory elements included
into the model), but carbon. $M_{C}\left(M,Z \right)$ is the ejected
mass of carbon. $\delta^{II}_{c}\left(C\right)$ and
$\delta^{II}_{c}\left(A\right)$ are the condensation efficiencies of
Carbon and refractory elements, finally $M_{D,C}\left(M,Z\right)$
and $M_{D,A}\left(M,Z\right)$ are the ejected mass of dust for
carbon and element $A$. For the oxygen an average between the
refractory elements is used, where every element is weighted for its
mass number $\mathcal{A}_A$:

\vspace{-7pt}
\begin{small}
\begin{equation}
M_{D,O}\left(M,Z\right)=16\sum_{A}\delta^{II}_{c}\left(A\right)\frac{M_{A}\left(M,Z
\right)}{\mathcal{A}_A}
\end{equation}
\end{small}
\vspace{-7pt}

\noindent where A=Mg,Si,S,Ca,Fe. A similar set of equation is used
for type Ia SN{\ae} with the condensation factors  indicated by
$\delta^{I}_{c}$. The adopted values are
$\delta^{II}_{c}\left(A\right)= \delta^{I}_{c}\left(A\right)=0.8$
for $A={Mg,Si,S,Ca,Fe}$ and $\delta^{II}_{c}\left(A\right)=
\delta^{I}_{c}\left(A\right)=0.5$ for C. No distinction is made
for the condensation efficiencies between type II CCSN{\ae} and
thermonuclear type Ia SN{\ae}. The values for the condensation
efficiencies are somewhat arbitrary \citep{Dwek98}; they are simply
meant to indicate the effect of condensation with some destruction.
One of the most controversial issues is the assumption made by
\citet{Dwek98} and \citet{Calura08} about Type Ia SN{\ae}: the condensation
efficiencies are assumed to be high despite the fact that no  dust
formation has been observed in Type Ia  SNRs \citep{Draine09}. This contradictory assumption
has been recently corrected in \citet{Pipino11}.
\citet{Calura08}  also assume condensation efficiencies all equal to
0.1 and compare the results with those of \citet{Dwek98}. They
find that the fraction of newly formed dust is nearly independent
from the condensation efficiencies if dust accretion and destruction
balance each other as it seems to be the case of the MW at the
present age \citep{Dwek98,Zhukovska08}. However, this could not be
true for different ages in the history of MW or for other galaxies
with different SFHs. Finally, the destruction-accretion balance  may
heavily depend on subtle details of the two  processes and their
uncertainties errors. For all these considerations, we suggest that
the more detailed set of condensation efficiencies based on
\citet{Nozawa03,Nozawa06,Nozawa07} that we analyzed above, is more
safe and of general use to be adopted into theoretical models. It
relies on detailed models that are still the most handy available in
literature thanks to the number of modelled masses, to the included
effects of the reverse shock and environmental density on the
surviving mass of dust grains. \\
\indent However, we are still far away from a satisfactory picture.
Indeed, an important point to consider is that the classical
nucleation theory (CNT), widely used to model the formation of dust
in SN{\ae}, has been improperly applied.  The founding hypotheses of
this theory do not hold in the SN{\ae} environment, that is neither
at equilibrium nor at steady state. As recently shown by
\citet{Cherchneff08}, \citet{Cherchneff09a,Cherchneff09b}, and
\citep{Cherchneff10},  the steady state is not  reached,
complicating the problem since molecules would act as a bottleneck
against dust formation \citep{Nozawa08}. Molecules  affect dust
formation by depleting the gas from metals and  cooling the
environment. A kinetically driven approach should be therefore used
and the formation of molecules, as dust precursors, should be
properly treated, thus influencing the nucleation models for dust
formation in SN{\ae}. As shown by \citet{Cherchneff10}, a proper
stochastic, kinetic approach leads to masses of dust that can be 2-5
times less than the amount predicted by \citet{Nozawa03}.
Interestingly, this reduction in the dust mass would produce a worst
agreement between the observational data in Fig.
\ref{TheoryVSObservations} and the predictions of the unmixed model.
Furthermore, as outlined by \citep{Cherchneff10}, the dust mixture
produced with the kinetic description is different. Standing on
these considerations, it is clear that detailed databases of dust
yields by SN{\ae}, taking into account different progenitor masses
and metallicities, hydrogen envelopes and molecules formation, are
needed. Doing this would greatly improve upon the equilibrium and
steady state approximation. Nevertheless, as long as  models at
varying those parameters are not available, the current estimates of
the condensation efficiencies by \citet{Nozawa03,Nozawa06,Nozawa07}
can be safely used in chemical models.\\
\indent Another prescription for dust condensation in SN{\ae} worth
being examined  is the one by \citet{Zhukovska08}. They assume that
SN{\ae} are poor producers of dust.  This hypothesis is likely
contradicted by the recent estimates of dust content  in Cas A,
Kepler SNa, and SN 1987A. Anyway, according to \citet{Zhukovska08},
the uncertainties on dust formation in SN{\ae} are still so large
that purely theoretical yields cannot be safely used. Therefore,
they adopt the same scheme of \citet{Dwek98}, introduce condensation
factors independent from both mass of the progenitor and/or
metallicity, and assume that Type II SN{\ae} produce all types of
dust, while Type Ia produce only
small amounts of iron.\\
\indent To adapt their scheme and use it into our model \citep[see
below Sect. \ref{Yieldseffect} and][for more details about the
chemical model]{Piovan11b}, we must switch from their description
limited to some typical dust grains as a whole to ours in which
single elements are followed both in gas and dust and as a whole in
the ISM. Let $M_{i,j}\left(M,Z\right)$ be the ejecta for the
key-element $i$-th of the $j$-type of grain \citep{Zhukovska08},
coming from a SNa of mass $M$ and metallicity $Z$.  If we divide by
the mass of the key-element $A_{i,j}m_{H}$ we get the number of
atoms of the key-element $i$-th available for dust formation.
Dividing again by the number of atoms of the key-element $\nu_{i,j}$
for one unit of dust $j$ (we can simply assume $\nu_{i,j}=1$) and
multiplying by the mass of one grain we get the maximum mass of dust
that can be formed. A multiplicative factor can be the introduced to
take into account the higher or lower efficiency of the condensation
process

\vspace{-7pt}
\begin{flalign}
M_{D,j}\left(M,Z\right) = \delta_{c,j}^{Ia,II}\cdot
\frac{A_{j,d}}{A_{i,j}}\cdot M_{i,j}\left(M,Z\right)
\end{flalign}\label{DeltaZhukovskaJ}
\vspace{-7pt}

\noindent where $i$ refers to the key-element and $j$ to the type of
dust grain. The ejecta is simply scaled by means of the ratio
between the atomic weights and multiplied for the condensation factor
to get the $j$-th type of dust injected into the ISM. If we want the
amount of the element $k$-th ejected in form of dust we have:

\vspace{-7pt}
\begin{flalign}
M_{D,k}\left(M,Z\right) &= \sum_{j=1}^{n}
M_{D,j}\left(M,Z\right)\frac{A_{k,j}}{A_{j,d}} = \nonumber \\
&=\sum_{j=1}^{n} \delta_{c,j}^{Ia,II}\cdot
\frac{A_{k,j}}{A_{i,j}}\cdot M_{i,j}\left(M,Z\right)
\end{flalign}\label{DeltaZhukovskaI}
\vspace{-7pt}

\noindent where $A_{k,j}=A_{i,j}$ if the element of interest
coincides with the key element. The condensation factors in use here
are much lower than for instance those of \citet{Dwek98,Calura08}:
$\delta_{c,j}^{II}$ is 0.00035, 0.15, 0.001, 0.0003 respectively for
silicates, carbonaceous grains, $\textrm{SiC}$ and iron grains,
while $\delta_{c,j}^{Ia}$  is always zero except for iron where it
is 0.005 (in order to agree with the observations). In particular,
the dust condensation efficiencies for the refractory elements
involved in the formation of  silicates  are very low. Indeed, they
are calibrated on the observational hints  from meteorites and
interplanetary dust particles, where the number of detected
silicates  is small up to now. According to \citet{Zhukovska08},
these low values could be also explained if we take into account all
the \textit{local} destructive processes affecting the SNa grains in
the ISM (mostly  the reverse shock and also shocks inside the star
cluster itself).

\section{Yields of dust from AGB stars} \label{AGByield}

\indent While  SNRs  from massive stars  eject newly formed dust in
amounts that are largely uncertain, low and intermediate mass stars
in  the asymptotic giant branch (AGB) phase are long known to safely
be strong injectors of dust in the ISM. It is worth noting that
previous evolutionary phases of low and intermediate mass-stars are
not so important: dust formation in the red giant branch (RGB) and
even early asymptotic giant branch (E-AGB) stars can be ignored
because the physical properties of their stellar winds do not favour
dust formation and the rates of mass loss are very low
\citep{Gail09}. Only the thermally pulsing AGB (TP-AGB) stars are
expected to form dust in significant amounts.\\
\indent TP-AGB stars have been the subject of an impressive number
of studies based on the theory of stellar evolution and going from
synthetic models \citep[see for
instance][]{Groenewegen93,Marigo96,Wagenhuber98,Marigo02,Izzard06,Marigo07}
to full calculations of evolutionary, even hydrodynamical, models
\citep[see for
instance][]{Herwig97,Karakas02,Ventura02,Herwig04,Weiss09}.
Moreover, dust formation in AGB stars has been the subject of more
and more refined and detailed models
\citep{Gail84,Gail85,Gail87,Dominik93,Gail99,Ferrarotti02,Ferrarotti06,Gail09},
able to calculate the amount of newly formed dust in M-stars,
S-stars and C-stars, along a sequence of growing C/O ratio. This
ratio determines the dust mixtures formed in the outflows
\citep{Piovan03,Ferrarotti06,Gail09}. AGB stars with C/O$<$1 are
oxygen-rich stars, that produce dust grains mainly formed by
refractory elements, generically defined as silicates, like
pyroxenes and olivines, oxides like alumina and maybe iron dust.
When the C/O ratio is higher than one, we have carbon-rich stars in
outflows of which  carbon dust, SiC, and even iron dust can
condensate. SiC is detected by means of the typical MIR feature.
When C/O $\approx$ 1 then we get S-stars where quartz and iron dust
should form \citep{Ferrarotti02}. However the C-rich or O-rich
phases dominate, so that for example the contribution of  SiC
produced during the S phase can be neglected compared to the SiC
produced during the carbon-star phase.\\
\indent Depending on the initial mass, the metallicity and the
complex interplay between the third dredge-up and  mass-loss, the
star will become or not carbon-rich. Typically, low mass stars are
not able to become C-rich, because they loose the envelope before
that carbon overcomes the oxygen abundance, while intermediate mass
stars are able to reach C/O$>$1, even if in some cases only for a
short part of the TP-AGB (as it happens for the most massive AGB
stars). Recently, \citet{Ferrarotti06} presented a detailed database
of dust yields from AGB stars, where many compounds are taken into
account. This database has been later extended  by
\citet{Zhukovska08}. Even if, as pointed out by \citet{Draine09},
the  dust yields from  AGB stars are not known at the same level of
confidence  as  the purely gaseous ones, they are surely much better
known than those from  SN{\ae} ejecta.\\
\indent \citet{Ferrarotti06} models are obtained applying the
schemes for dust formation to synthetic AGB models standing on
\citet{Groenewegen93} and \citet{Marigo96}. Similar recipes have
been proposed  by \citet{vandenHoek97} and have been adopted as
reference scheme for the total gaseous yields \citep[see ][for more
details]{Zhukovska08}. Also, for some elements the results by
\citet{Karakas03} have been used. Since we want to obtain the dust
condensation coefficients $\delta_{c,i}^{w}$ for every element
$i$-th of our set, first of all we must  calculate the amount of
each element embedded in newly formed dust. The details of these
calculations are given Appendix \ref{AppendixB}.

\begin{figure*}
\centerline{\hspace{-39pt}
\includegraphics[height=5.0cm,width=8.0truecm]{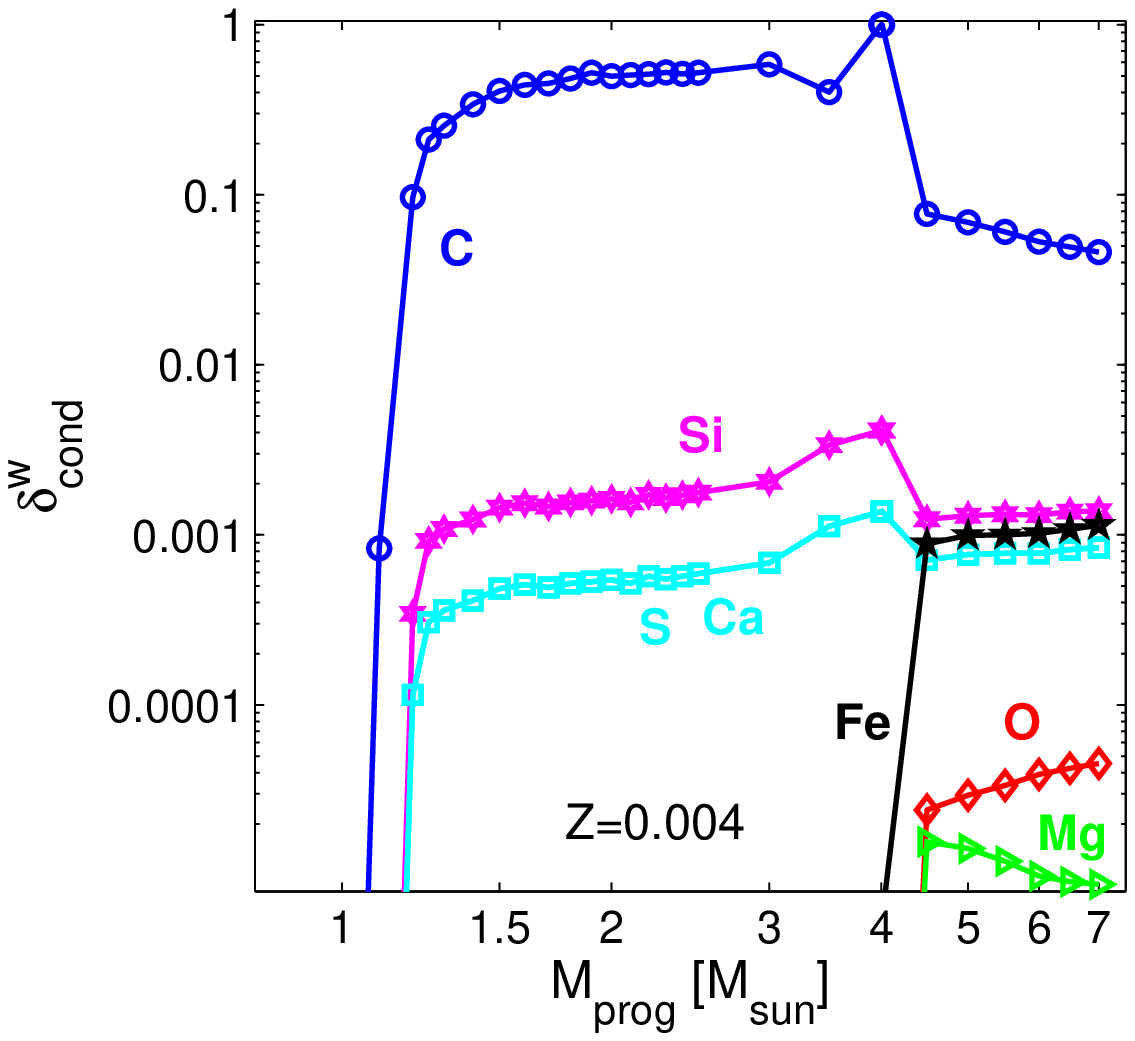}
\includegraphics[height=5.0cm,width=8.0truecm]{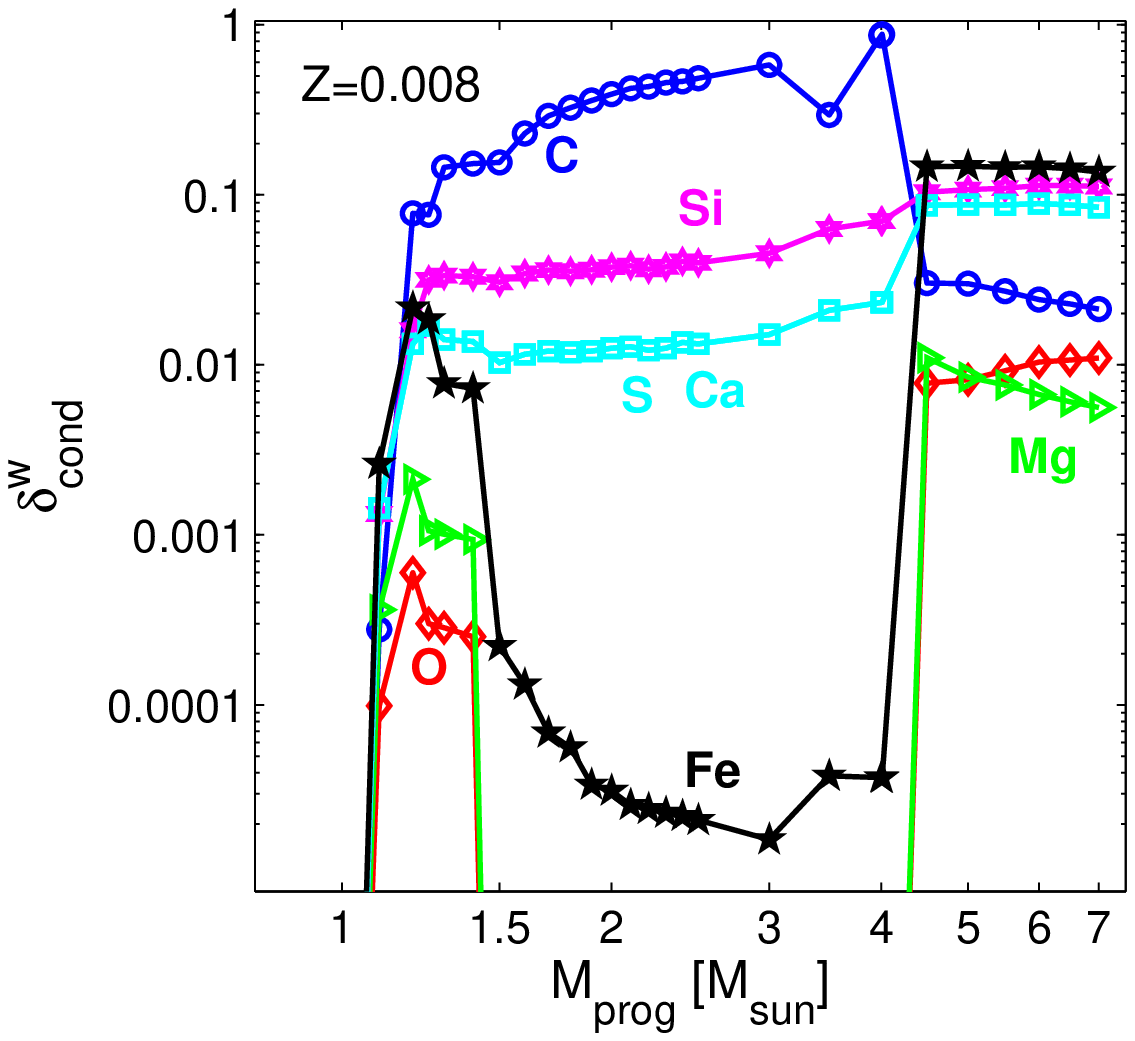}
\hspace{-39pt}} \centerline{\hspace{-39pt}
\includegraphics[height=5.0cm,width=8.0truecm]{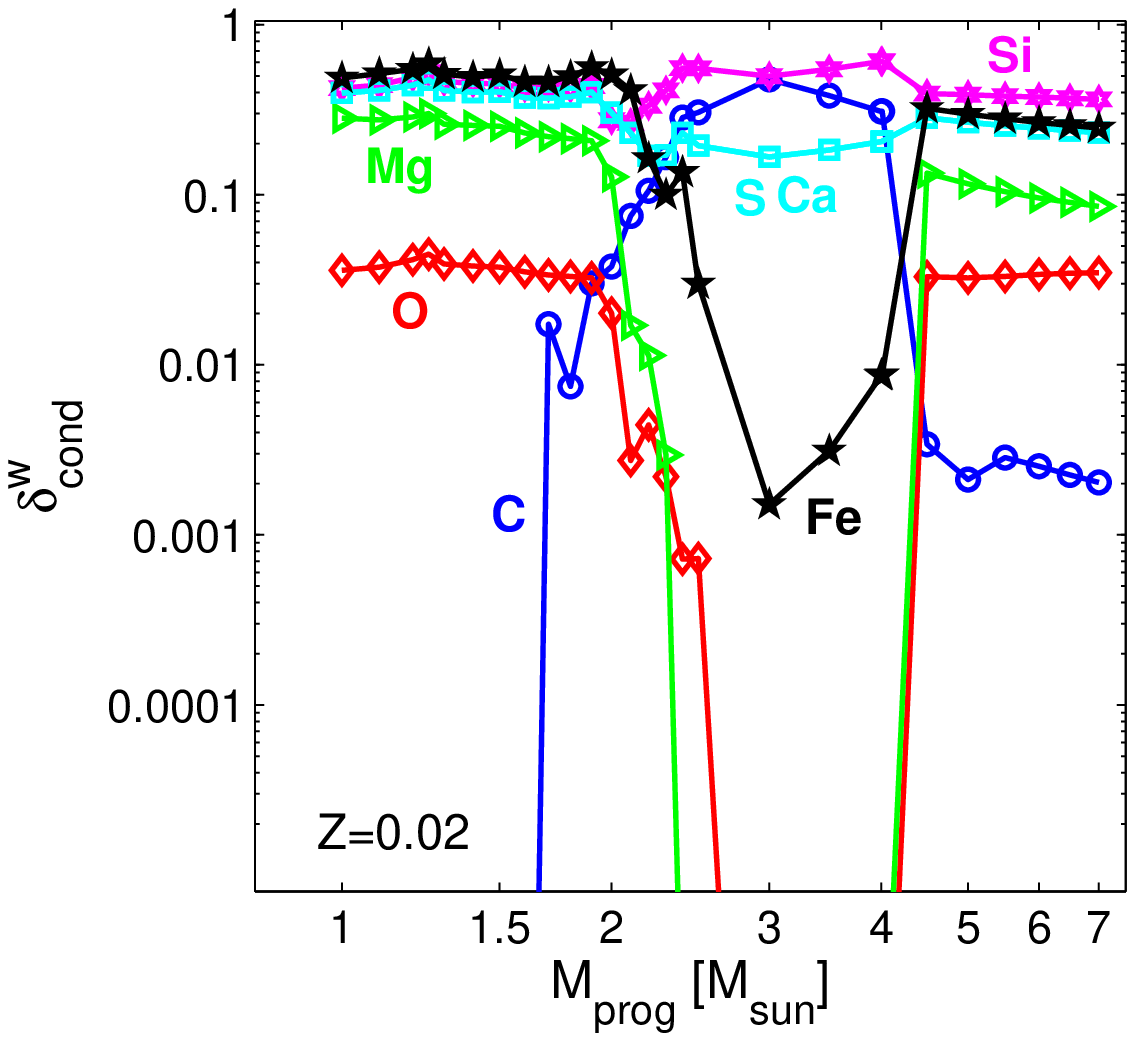}
\includegraphics[height=5.0cm,width=8.0truecm]{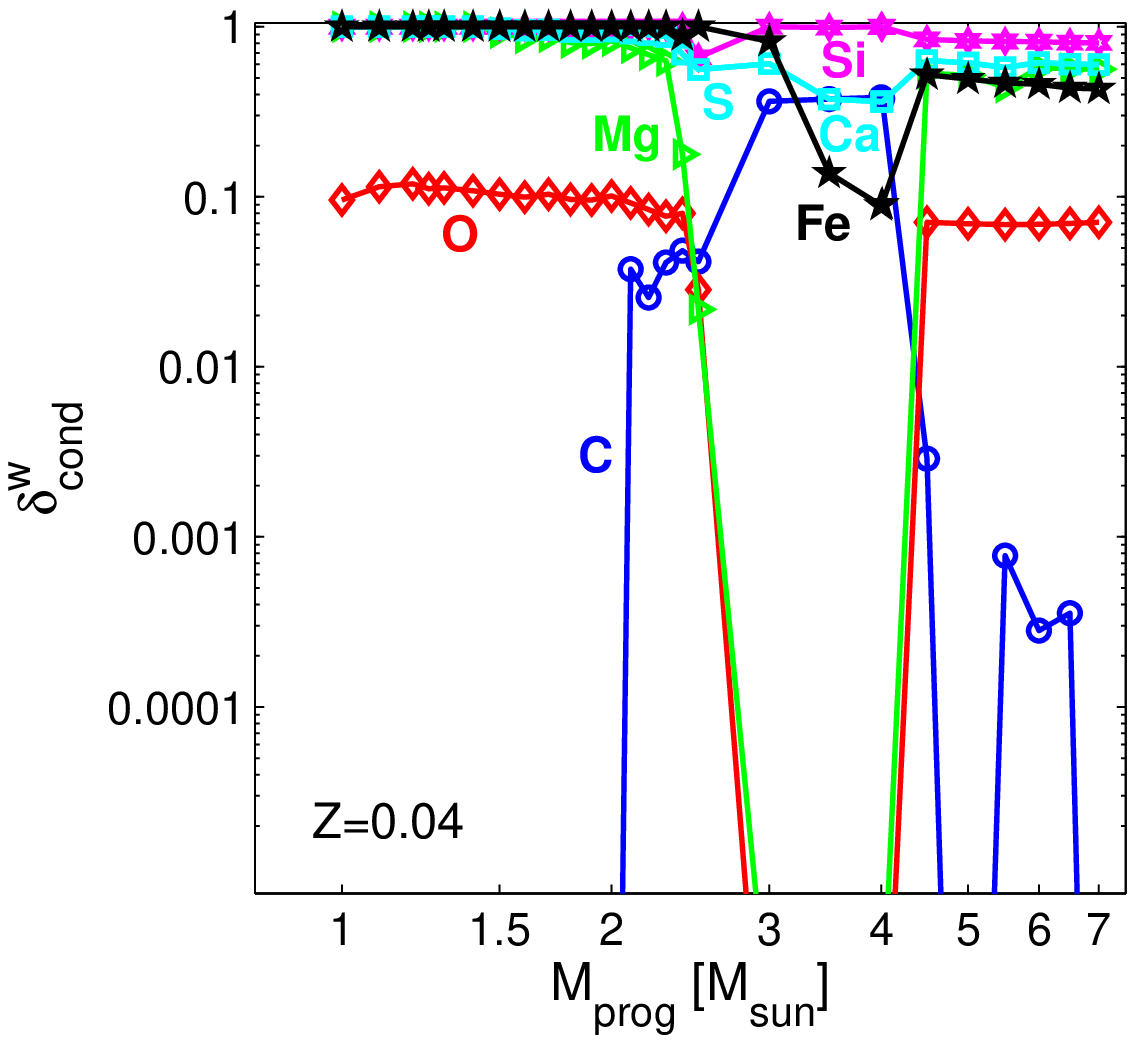}
\hspace{-39pt}} \caption{Condensation efficiencies for the elements
C, O, Mg, Si, Ca, S and Fe in SNRs as a function of
the progenitor mass at varying the metallicity Z. We plot: C
(empty circles and continuous line); O (diamonds and dashed line);
Mg (triangles and dotted line); Si (six-pointed stars and
dot-dashed line); S and Ca (squares and continuous line) and
Fe (five-pointed stars and dashed line).}\label{DeltaAGB}
\end{figure*}

\begin{figure*}
\centerline{\hspace{-39pt}
\includegraphics[height=5.0cm,width=8.0truecm]{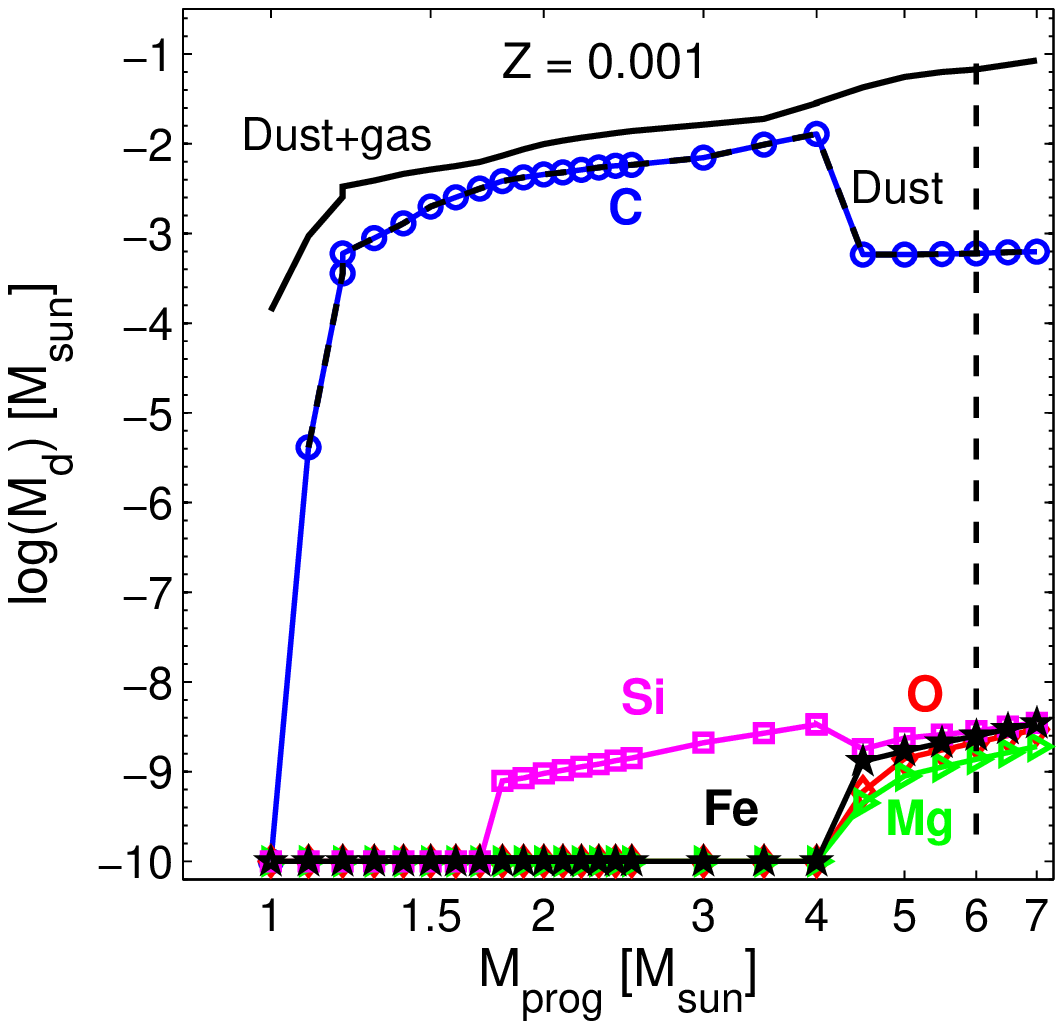}
\includegraphics[height=5.0cm,width=8.0truecm]{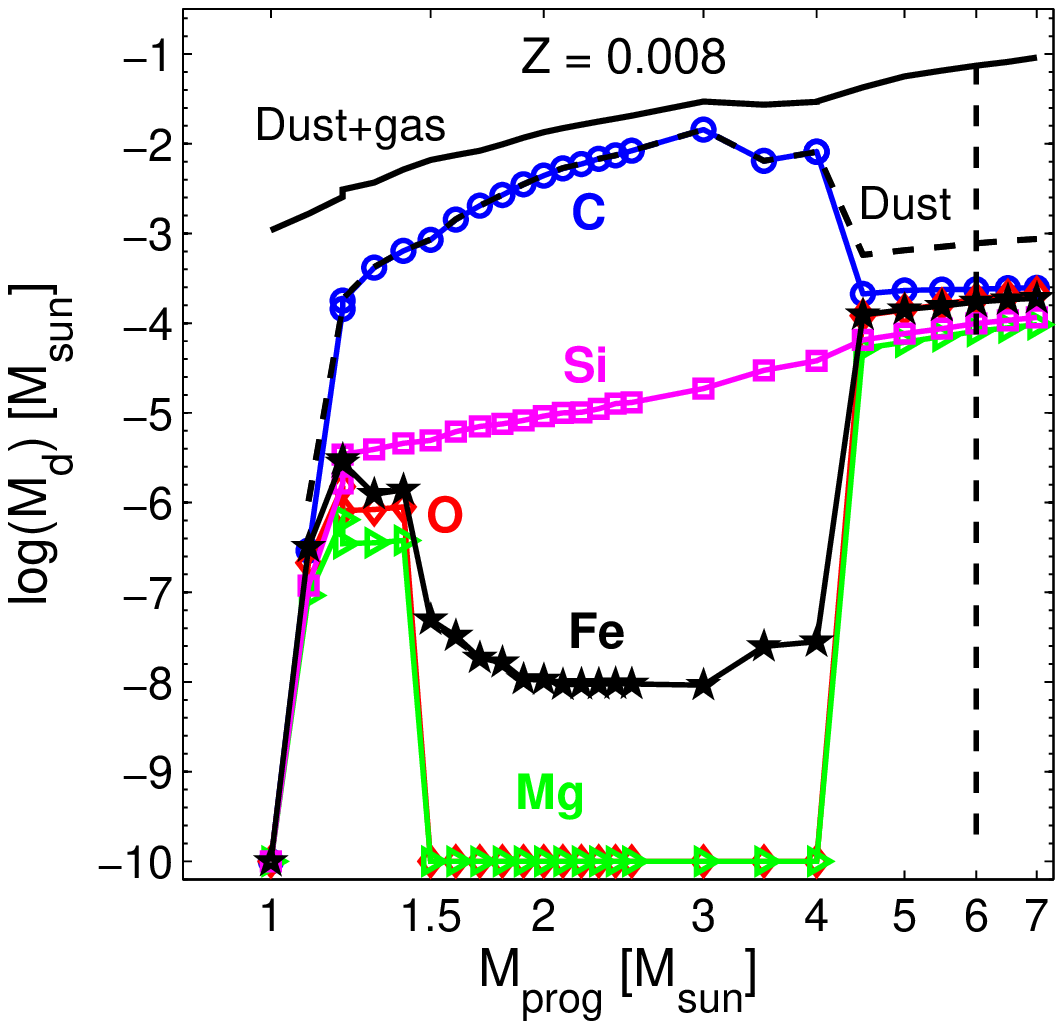}
\hspace{-39pt}} \centerline{\hspace{-39pt}
\includegraphics[height=5.0cm,width=8.0truecm]{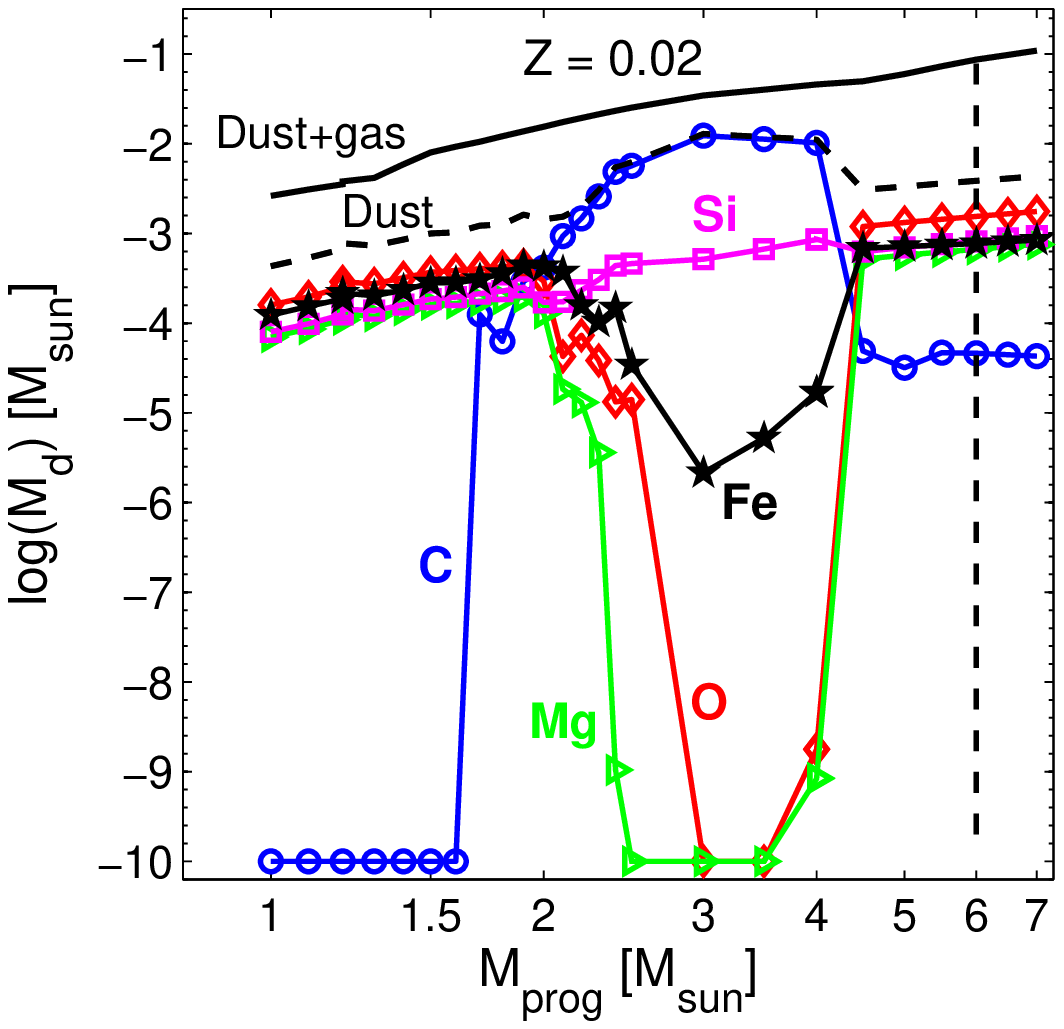}
\includegraphics[height=5.0cm,width=8.0truecm]{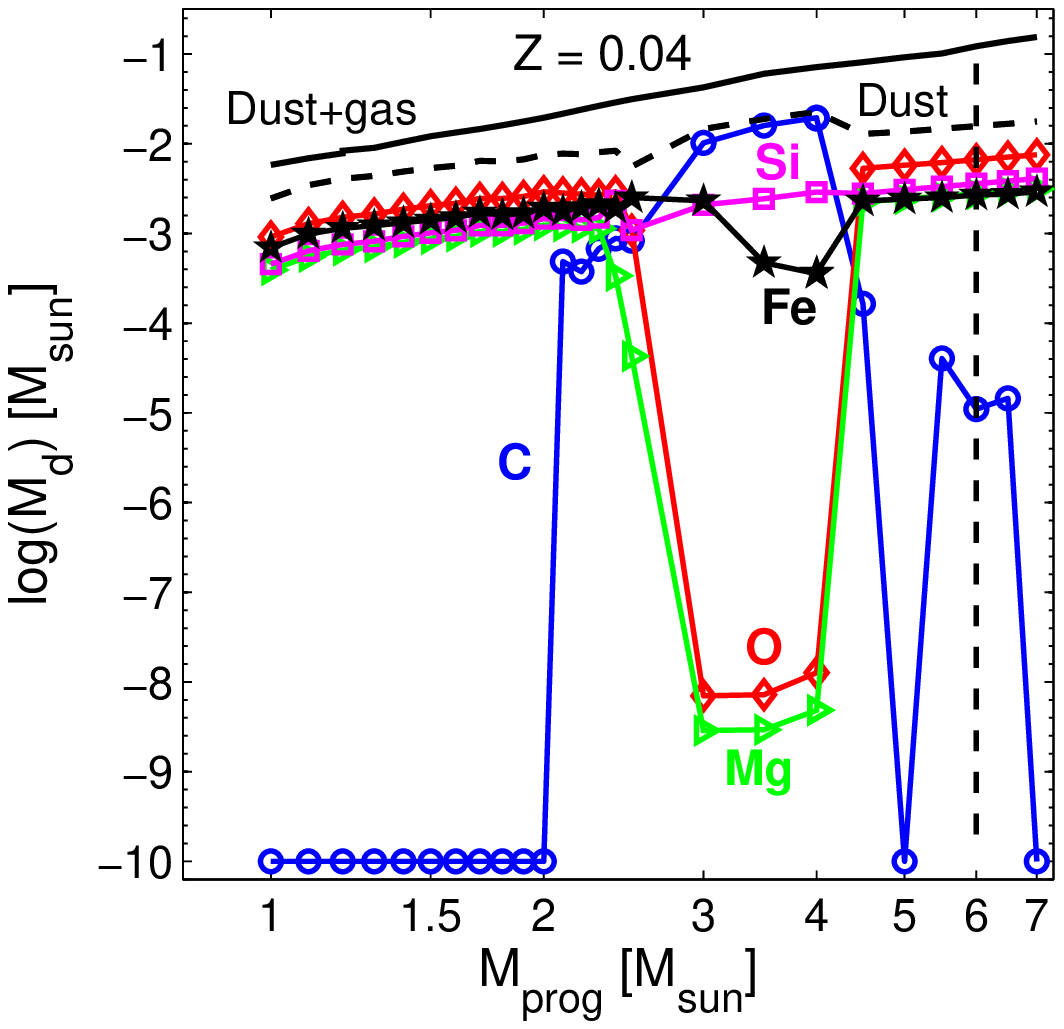}
\hspace{-39pt}} \caption{Dust ejecta for C, O, Mg, Si, and
Fe, calculated by means of dust compounds of
\citet{Ferrarotti06,Zhukovska08} for AGB stars as a  function of the
progenitor mass. Four metallicities are considered. The legend is as
follows: C (empty circles and continuous line); O (diamonds and
dashed line); Mg (triangles and dotted line); Si (six-pointed
stars and dot-dashed line); S (squares and continuous line) and
 Fe (five-pointed stars and dashed line). The dashed line without
markers is the total amount of dust  in the ejecta. The vertical
dashed line represents the 6 M$_{\odot}$ upper AGB limit according
to our set of yields \citet{Marigo96,Marigo98,Portinari98}. Finally,
the continuous line shows the total ejected mass in dust and gas for
what concerns the five elements we have considered in the plot
according to \citet{vandenHoek97}.}\label{YieldsAGB}
\end{figure*}

In Fig. \ref{YieldsAGB}, for the elements C, O, Mg, Si, Fe, we show
the total amount $M_{i,d}\left(M,Z\right)$ of dust formed in the
outflows of AGB stars, according to the models by
\citet{vandenHoek97,Ferrarotti06,Zhukovska08}. Moreover, for each
AGB star, we show the total amount of ejected dust   and compare it
to the total amount of lost material. The following remarks can be
made. First, as shown in the top panels, the dust produced by low
metallicity AGB stars of any mass is carbon dominated. This point
seems to agree with the suggested high-z scenario in which the
appearance of the PAH features and graphite extinction bump in the
UV are both connected to the delayed injection of carbon by AGB
stars, as shown by observations of galaxies of different metallicity
\citep{Dwek05,Galliano08a,Dwek09}. In the early universe, before the
most massive AGB stars start contributing to the dust content of the
ISM, only SN{\ae} are injecting dust. It is worth noticing that this
scenario, based upon the dust production by different stellar
sources, needs deeper investigations. The works by \citet{Dwek09}
and \citet{Draine09} have shown that a significant amount of dust
should be produced also by accretion in the ISM (therefore dust
would not be only of stellar origin). In the extreme case, the role
played by SN{\ae} could be even limited to only injecting metals and
seeds for grain growth. However, the recent observations of large
amounts of cold dust in SN{\ae} \citep{Dunne09,Gomez09,Matsuura11}
reshuffled the problem putting again SN{\ae} on the table as
possible major dust factories. Second, at growing metallicity,
silicate dust starts to be formed in significant amount, however
there is always some injection of carbon dust from stars around
3M$_{\odot}$. Third, the following questions arise: At which mass do
AGB stars disappear? What is the role played by Super-AGB (SAGB)
stars (if any)? Is there any mass range for the existence of
thermonuclear SN{\ae} from single stars igniting carbon? What is the
lower mass limit for core collapse SN{\ae} (CCSN{\ae})? Our recipe
is strictly classical and follows the one adopted by
\citet{Portinari98} to calculate the chemical yields, since we will
base our simulations upon the latest release of those yields. Below
6 solar masses we have stars ending as WDs through the AGB channel;
for masses between 6 and 8M$_{\odot}$ \citet{Portinari98} assumed
1.3M$\odot$ of remnant, either WD or NS, and the overlying layers
expelled either by an explosion or a TP-AGB phase; for masses
M$>8$M$_{\odot}$ we have stars developing an iron core and exploding
as CCSN{\ae}. However, other mass limits would be possible
considering  all the uncertainties affecting the evolution of stars
in the mass range 6 to 12M$_{\odot}$. For instance
\citet{Zhukovska08} and \citet{Gail09}  extend the AGB stars to
stars with initial mass of 8M$_{\odot}$ and neglect the possibility
of SAGB stars; \citet{Calura08} assume also quiescent outflows until
8M$_{\odot}$. Although  the investigation of this point is beyond
the aims of this study, it is worth keeping in mind that stellar
models in this particular mass range are still far from being fully
understood, and therefore  different mixtures of dust could be
formed and ejected.

 \indent Once the masses
of dust ejecta $M_{i,d}\left(M,Z\right)$ are defined, we can obtain
the condensation efficiencies for the element $i$-th during the
TP-AGB phase as:

\vspace{-7pt}
\begin{equation}\label{condensationAGB}
\delta_{i}^{w}=\frac{M_{i,d}\left(M,Z\right)}{\Delta
M^{AGB}\left(M,Z\right)\cdot X_{i,0}\left(M,Z\right)+M\cdot
p_{i}\left(M,Z\right)}
\end{equation}
\vspace{-7pt}

where $\Delta M^{AGB}\left(M,Z\right)\cdot X_{i,0}\left(M,Z\right)$
is the mass of the $i$-th element ejected according to the initial
abundance of that element in the stellar model and $M\cdot
p_{i}\left(M,Z\right)$ is the newly formed and ejected amount of the
same element. In Fig. \ref{DeltaAGB} we show the condensation
efficiencies at growing metallicity. While the condensation
efficiency of carbon keeps quite high, for  oxygen and  other
refractory elements it grows at increasing metallicity. For
metallicity two times solar, for some elements like Si and some
stellar masses, almost all the material is condensed into dust. For
the sake of comparison we consider also another possibility for the condensation
efficiencies, widely adopted in literature, and proposed by \citet{Dwek98} for AGB
stars. They depend on the final C/O ratio in the ejecta, without
following the evolution of the star along the AGB as in the complex
dust nucleation model by \citet{Ferrarotti06} and they are also
\textit{independent} from the metallicity of the stars. If C/O$>$1
then:

\vspace{-7pt}
\begin{equation}
M_{D}\left(C,M\right)=\delta^{w}_{c}\left(C\right)\left[M_{ej}\left(C,M
\right)-\frac{3}{4}M_{ej}\left(O,M \right)\right]
\end{equation}
\vspace{-7pt}
\begin{equation}
M_{D}\left(A,M\right)=\delta^{w}_{c}\left(A\right) \cdot
M_{ej}\left(A,M \right)=0
\end{equation}
\vspace{-7pt}

\noindent where $M_{D}\left(C,M\right)$ and $M_{D}\left(A,M\right)$
are the ejected masses of C and the generic refractory element $A$
embedded into dust; $M_{ej}\left(C,M \right)$, $M_{ej}\left(O,M
\right)$ and $M_{ej}\left(A,M \right)$ are the ejected masses of
carbon, oxygen and refractory element $A$ from the star of mass $M$.
In our case,
$\textrm{A}=\,\textrm{O},\,\textrm{Mg},\,\textrm{Si},\,\textrm{S},\,\textrm{Ca},\,\textrm{Fe}$.
$M_{ej}\left(C,M \right)$ is the ejected mass of $C$.
$\delta^{w}_{c}\left(C\right)$ and $\delta^{w}_{c}\left(A\right)=0$
are the condensation efficiencies of C and refractory elements. When
instead C/O$<$1 then:

\vspace{-7pt}
\begin{equation}
M_{D}\left(C,M\right)=\delta^{w}_{c}\left(C\right) \cdot
M_{ej}\left(C,M \right)=0
\end{equation}
\vspace{-7pt}
\begin{equation}
M_{D}\left(A,M\right)=\delta^{w}_{c}\left(A\right)M_{ej}\left(A,M
\right)
\end{equation}
\vspace{-7pt}
\begin{small}
\begin{equation}
M_{D}\left(O,M\right)=16\sum_A
  \delta^{w}_{c}\left(A\right)\frac{M_{ej}\left(A,M
\right)}{\mathcal{A}_A}
\end{equation}
\end{small}
\vspace{-7pt}

\noindent where A=Mg,Si,S,Ca,Fe.  For the oxygen, the average value
of the refractory elements is introduced, where each element is
weighted by its mass number $\mathcal{A}_A$. In practice
$\delta^{w}_{c}\left(A\right)=1$ is chosen, assuming complete condensation of
the element.

\section{The effect of different yields of dust}\label{Yieldseffect}

We analyzed the most popular recipes adopted in literature to
describe the condensation efficiency of dust in the two main dust
factories, AGB and SN{\ae}, thus deriving theoretical sets of
condensation efficiencies for single refractory elements that can be
generally applied to any set of stellar yields. We want now to test
these different recipes in a full dust formation and evolution
model. At this purpose we introduced the various possibilities for
dust condensation efficiencies into the model by
\citet{Piovan11b,Piovan11c}. This model stands on the classical
formulation for the chemical enrichment of a galaxy
\citep{Chiosi80}, in his latest multi-ring formulation by
\citet{Portinari00} and \citet{Portinari04a} for disk galaxies, with
the introduction of radial flows of matter and galactic bar. The
adopted stellar yields are the original ones by \citet{Portinari98}
in its latest revised version (Portinari 2006 - private
communication). The model is able to describe the evolution of the
abundances of the different refractory elements into dust, properly
simulating the process of injection of the stardust into the ISM and
the accretion/destruction processes to which the dust is subjected.
We trace the evolution of the abundance of both some main typical
grain families usually adopted for a general description of a dusty
ISM and the single elements into dust.\\
\indent The complete description of the equations governing the
model can be found in \citet{Piovan11b,Piovan11c}. In the following
we briefly introduce only the the equations governing the evolution
of the dust component. The evolution of the generic elemental
species $i$-th in the dust at the radial distance $r_{k}$ from the
centre of the galaxy and at the time $t$, is described according to
the extension by \citet{Dwek98} of the formulation for gas only to a
two-component ISM made by gas and dust. Let $\sigma(r,t)$ be the
total surface mass density (gas, dust and stars) of the galaxy at
the radial distance $r$ and time $t$, and $\sigma(r,t_G)$ the same
but at the galactic age $t_G$. The fractionary surface mass density
of dust $D(r,t)$ and of the generic element $i$ in form of dust
$D_i(r,t)$ are given by the following relations

\vspace{-7pt}
\begin{equation}
D\left(r_{k},t\right)=\frac{\sigma^{D}\left(r_{k},t\right)}{\sigma\left(r_{k},t_{G}\right)}
\end{equation}
\vspace{-6pt}

\noindent and

\vspace{-7pt}
\begin{equation}
D_{i}\left(r_{k},t\right)=
\frac{\chi_{i}^{D}\left(r_{k},t\right)\sigma_{D}\left(r_{k},t\right)}
{\sigma\left(r_{k},t_{G}\right)}=
\chi_{i}^{D}\left(r_{k},t\right)D\left(r_{k},t\right)
\end{equation}
\vspace{-7pt}

\noindent where $\chi_i^D(r,t)$ is the fractionary mass abundance of the element $i$ trapped in the dust,
and all surface mass densities are  normalized to $\sigma(r,t_G)$. Identical expressions can be
written for the gas component and the corresponding mass abundance of the generic element
$i$ trapped in the gas is $\chi_i^G(r,t)$. By definition the following companion relationship applies
$\sum_{i}\left[\chi_{i}^{D}\left(r_{k},t\right)+\chi_{i}^{G}\left(r_{k},t\right)\right]=1$,
from which
$\sum_{i}\chi_{i}^{D}\left(r_{k},t\right)\neq 1$ and
$\sum_{i}\chi_{i}^{G}\left(r_{k},t\right)\neq 1$.
Finally, the equations governing the temporal variation of $D_i(r,t)$ is

\begin{small}
\begin{flalign} \label{DUST}
\frac{d}{dt}&D_{i}\left(r_{k},t \right)  = -\chi_{i}^{D}\psi+ \nonumber \\
&+ \int_{0}^{t-\tau_{M_{B,l}}}\psi\left[\phi
\delta^{w}_{c,i}R_{i}\cdot\left(-\frac{dM}{d\tau_{M}}\right)\right]
_{M\left(\tau\right)}dt^{\prime} +  \nonumber  \\
&+ \left(1-A
\right)\int_{t-\tau_{M_{B,l}}}^{t-\tau_{M_{SN{\ae}}}}\psi\left[\phi
\delta^{w}_{c,i}R_{i}\cdot\left(-\frac{dM}{d\tau_{M}}\right)\right]
_{M\left(\tau\right)}dt^{\prime} + \nonumber \\
&+ \left(1-A
\right)\int_{t-\tau_{M_{SN{\ae}}}}^{t-\tau_{M_{B,u}}}\psi\left[\phi
\delta^{II}_{c,i}R_{i}\cdot\left(-\frac{dM}{d\tau_{M}}\right)\right]
_{M\left(\tau\right)}dt^{\prime} +  \nonumber \\
&+ \int_{t-\tau_{M_{B,u}}}^{t-\tau_{M_{u}}}\psi\left[\phi
\delta^{II}_{c,i}R_{i}\cdot\left(-\frac{dM}{d\tau_{M}}\right)\right]
_{M\left(\tau\right)}dt^{\prime} +  \nonumber  \\
&+
A\int_{t-\tau_{M_{SN{\ae}}}}^{t-\tau_{M_{1},max}}\psi\left[\textit{f}
\left(M_{1}\right)\delta^{II}_{c,i}R_{i,1}\cdot\left(-\frac{dM_{1}}{d\tau_{M_{1}}}\right)\right]
_{M\left(\tau\right)}dt^{\prime} +  \nonumber   \\
&+
A\int_{t-\tau_{M_{1},min}}^{t-\tau_{M_{SN{\ae}}}}\psi\left[\textit{f}
\left(M_{1}\right)\delta^{w}_{c,i}R_{i,1}\cdot\left(-\frac{dM_{1}}{d\tau_{M_{1}}}\right)\right]
_{M\left(\tau\right)}dt^{\prime} +  \nonumber   \\
&+ R_{SNI}E_{SNI,i}\delta^{I}_{c,i}+ \nonumber  \\
&-\left[\frac{d}{dt}D_{i}\left(r_{k},t \right) \right]_{out}
+\left[\frac{d}{dt}D_{i}\left(r_{k},t \right) \right]_{rf}+
\nonumber \\
&+\left[\frac{d}{dt}D_{i}\left(r_{k},t \right) \right]_{accr}
-\left[\frac{d}{dt}D_{i}\left(r_{k},t \right) \right]_{SN}
\end{flalign}
\end{small}

\noindent where $\psi(r,t)$ is the star formation rate and
$\phi=\phi(M)$ the initial mass function (IMF). The first term at
the r.h.s. of eqn. (\ref{DUST}) is the depletion of dust because of
the star formation that consumes both gas and dust (assumed
uniformly mixed in the ISM). The second term is the contribution by
stellar winds from low mass stars to the enrichment of the $i$-th
component of the dust. Respect to the classical gas-only
formulation, the so-called condensation coefficients
$\delta^{w}_{c,i}$ determines the fraction of material in stellar
winds that goes into dust with respect to that in gas (local
condensation of dust in the stellar outflow of low-intermediate mass
stars). For these coefficients we can adopt the recipes presented in
Sect. \ref{AGByield}. The third term is the contribution by stars
not belonging to binary systems and not going into type II SN{\ae}
(the same coefficients $\delta^{w}_{c,i}$ are used). The fourth term
is the contribution by stars not belonging to binary systems, but
going into type II SN{\ae}. The condensation efficiency in the
ejecta of type II SN{\ae} are named as $\delta^{II}_{c,i}$. For
these efficiencies the recipes analyzed in Sect.
\ref{SNEyield_details} will be adopted. The fifth term is the
contribution of massive stars going into type II SN{\ae}. The sixth
and seventh term represent the contribution by the primary star of a
binary system, distinguishing between those becoming  type II
SN{\ae} from those failing this stage and using in each situation
the correct coefficients. The eighth term is the contribution of
type Ia SN{\ae}, where the condensation coefficients are named as
$\delta^{I}_{c,i}$ to describe the mass fraction of the ejecta going
into dust. The last four terms describe: (1) the outflow of dust due
to galactic winds (in the case of disk galaxies this term can be set
to zero); (2) the radial flows of matter between contiguous shells;
(3) the accretion term describing the accretion of grain onto bigger
particles in cold clouds; (4) the destruction term taking into
account the effect of the shocks of SN{\ae} on grains, obviously
giving a negative contribution. The infall term in the case of dust
can be neglected because we can assume that the primordial material
entering the galaxy is made by gas only without a solid dust
component mixed to it. \\
\indent In the model by \citet{Piovan11b}, many choices for the
various terms of the eqns. (\ref{DUST}) related to dust have been
considered and tested together with  other  prescriptions that are
needed to solve the companion equations  for gas and total ISM
\citep[See][for the systems of equations describing the evolution of
the gas and ISM]{Piovan11b}. In Table \ref{Parameters} we summarize
the list of assumptions/parameters specifying  a given model. A
detailed description of each possible combination of the parameters
is in \citet{Piovan11b}. Let us shortly  describe here the
parameters in use:

\begin{itemize}
\item The IMF determines the relative number of AGB stars,
SN{\ae}, and very low mass stars (not contributing to the enrichment
of the ISM), thus driving the amounts of  star-dust injected by the
different sources above. Many IMFs are possible. They are examined
in detail by  \citet{Piovan11b}. For the purposes of this paper, to
analyze the effect of different condensation efficiencies, we
consider the IMF by  \citet{Kroupa07} (here indicated by
$\mathcal{G}$ according to our identification code). The IMF is kept
constant throughout this paper.

\item Several laws of star formation can be found in literature
\citep[see][\, for more details]{Piovan11b}.  We choose here the
star formation law by \citet{Dopita94} (indicated by $\mathcal{D}$)
with an intrinsic efficiency given by  $\nu=0.55$ \citep[see][\, for
all details]{Piovan11b}. In this study, the law of star formation is
kept constant.

\item Dust accretion in the cold
regions of the ISM strongly depends on the fraction of molecular
clouds (MCs) with respect to the gas mass of the ISM. this fraction
is named here $\chi_{MC}$. This fraction can be kept constant or
varied in time and space. For this latter case, \citep{Piovan11c}
develop a simple model based on Artificial Neural Networks (ANN) and
observational data on the SFR, content of molecular hydrogen
H$_{2}$, and  total gas mass in the SN and MW Disk. With the aid of
the ANN we  derive from the local values of the SFR and gas mass,
the corresponding H$_{2}$ mass and it is the one adopted here (case
$\mathcal{A}$).

\item Two models for the dust accretion in the ISM are available.
Here we adopt the most complex one (case $\mathcal{B}$), based upon
the work by \citet{Zhukovska08} and taking into account the
numerical densities in the ISM (thus allowing variable time-scales
of accretion), the lifetime and the mass of MCs as cold regions
where the accretion happens and, finally, a set of dust grains
representative of the ISM.

\item The condensation efficiencies of type II SN{\ae}, type Ia SN{\ae}
and AGB stars are the target of this paper and will be varied
and discussed below.

\item The galactic bar and the radial flows mechanism of matter
exchange between contiguous shells is fixed
according to the considerations and models by \citep{Portinari00}.
\end{itemize}

\renewcommand{\arraystretch}{1.3}
\setlength{\tabcolsep}{2.8pt}
\begin{table*}
\scriptsize
\begin{center}
\caption[]{\footnotesize Parameters of the models. Column (1) is the
parameter number, column (2) the associated physical quantity, and
column (3)  the source and the italic symbols are the identification
code we have adopted. See the text for some more details and
\citet{Piovan11b} for a detailed description.}
\begin{tabular}{ccl}
\hline \noalign{\smallskip}
n$^{o}$         & \small Parameter     &\small Source and identification label  \\
\hline \noalign{\smallskip} \small 1 & \small IMF           &\small Salpeter\footnotesize$^{1}$
$(\mathcal{A})$,  Larson\footnotesize$^{2}$ $(\mathcal{B}$), Kennicutt\footnotesize$^{3}$ $(\mathcal{C})$
Kroupa orig.\footnotesize$^{4}$ $(\mathcal{D})$, \\
&&\small  Chabrier\footnotesize$^{5}$ $(\mathcal{E})$, Arimoto\footnotesize$^{6}$ $(\mathcal{F})$, Kroupa
2007\footnotesize$^{7}$ $(\mathcal{G})$,
Scalo\footnotesize$^{8}$ $(\mathcal{H})$, Larson SN\footnotesize$^{9}$ $(\mathcal{I})$ \\

\small 2 &\small SFR law              &\small Constant SFR
$(\mathcal{A})$, Schmidt\footnotesize$^{10}$ $(\mathcal{B})$,
Talbot \& Arnett\footnotesize$^{11}$ $(\mathcal{C})$, Dopita \& Ryder\footnotesize$^{12}$ $(\mathcal{D})$, Wyse \& Silk\footnotesize$^{13}$ $(\mathcal{E})$ \\
\small 3 & \small $\chi_{MC}$ model    &\small Artificial Neural
Networks model\footnotesize$^{14}$ $(\mathcal{A})$, Constant $\chi_{MC}$
as in the Solar Neigh.\footnotesize$^{15}$ $(\mathcal{B})$               \\

\small 4 &\small Accr. model          &\small Modified
\citet{Dwek98} and \citet{Calura08} $(\mathcal{A})$;
adapted \citet{Zhukovska08} model $(\mathcal{B})$         \\

\small 5 &\small SN{\ae} Ia model &\small Dust injection adapted
from: \citet{Dwek98}, \citet{Calura08} $(\mathcal{A})$,
\citet{Zhukovska08}  $(\mathcal{B})$                   \\

\small 6 &\small SN{\ae} II model     &\small Dust injection adapted
from: \citet{Dwek98} $(\mathcal{A})$, \citet{Zhukovska08}
$(\mathcal{B})$, \\
&& \small \citet{Nozawa03,Nozawa06,Nozawa07}
 $(\mathcal{C})$\\

\small 7 &\small AGB model            &\small Dust injection adapted
from: \citet{Dwek98}
$(\mathcal{A})$, \citet{Ferrarotti06}  $(\mathcal{B})$  \\

\small 8 &\small Galactic Bar\footnotesize$^{16}$   &\small No onset
$(\mathcal{A})$, onset at $t_{G}-4$ Gyr
$(\mathcal{B})$, onset at $t_{G}-1$ Gyr $(\mathcal{C})$ \\

\small 9 &\small Efficiency SFR\footnotesize$^{17}$       &\small Low efficiency
$(\mathcal{A})$, medium efficiency $(\mathcal{B})$
, high efficiency $(\mathcal{C})$  \\
\hline \label{Parameters}
\end{tabular}
\end{center}
\renewcommand{\arraystretch}{1}
\footnotesize$^{1}${\citet{Salpeter55}}.\, \footnotesize$^{2}${\citet{Larson86,Larson98}}.\,
\footnotesize$^{3}${\citet{Kennicutt83,Kennicutt94}}.\,\footnotesize$^{4}${\citet{Kroupa98}}.\,
\footnotesize$^{5}${\citet{Chabrier01}}.\,\footnotesize$^{6}${\citet{Arimoto87}}.\,
\footnotesize$^{7}${\citet{Kroupa02a,Kroupa07}}.\,\footnotesize$^{8}${\citet{Scalo86}}.\,
\footnotesize$^{9}${\citet{Larson86,Scalo86,Portinari04a}}.\,\footnotesize$^{10}${\citet{Schmidt59}}.\,
\footnotesize$^{10}${\citet{Talbot75}}.\,\footnotesize$^{11}${\citet{Talbot75}}.\,
\footnotesize$^{12}${\citet{Dopita94}}.\,\footnotesize$^{13}${\citet{Wyse89}}.\,
\footnotesize$^{14}${\citet{Piovan11c}}.\,\footnotesize$^{15}${\citet{Zhukovska08}}.\,
\footnotesize$^{16}${\citet{Portinari00}}.\,\footnotesize$^{16}${\citet{Piovan11b}}.\,
\end{table*}
\renewcommand{\arraystretch}{1}

\noindent Each model can be therefore identified, for the sake of concision, by a
string of nine letters (the number of parameters) in italic face
whose position in the string and the alphabet corresponds to a
particular parameter and choice for it. The sequence should be read
from top to bottom. Just for example, the string
$\mathcal{DBAABABAB}$ corresponds to Kroupa 1998 IMF, Schmidt SFR,
ANN model for $\chi_{MC}$, \citet{Dwek98} accretion model,
\citet{Zhukovska08} SN Ia recipe for dusty yields, \citet{Dwek98}
condensation efficiencies for type II SN{\ae}, \citet{Ferrarotti06}
yields for AGB stars, no bar effect on the inner regions and average
efficiency $\nu$ of the SFR. If not otherwise specified radial flows
will always be included by default. We do not enter the detail of
the different accretion models or IMFs and SF laws \citep[See][for this point]{Piovan11b}, but we want to
focus here on the recipes for dust condensation. Are they reliable?
When the difference between one condensation set and another is more
striking? How the fully theoretical set that we compiled behaves? To
try to face and discuss these issues we define a "standard model", according to the choices
just described above for the IMF and the other parameters,
upon which apply different sets of condensation efficiencies. This
reference model is identified by the string $\mathcal{GDABBCBBB}$.

\begin{figure}
\centerline{\hspace{-3mm}
\includegraphics[height=8.0cm,width=8.7cm]{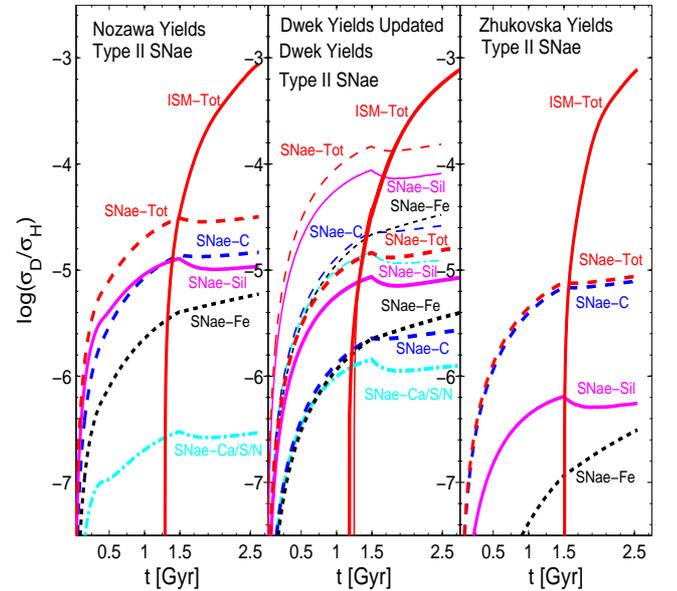}}
\caption{Temporal evolution of the contribution to the dust budget
in the Solar Neighborhood during  the \textit{first 1.5 Gyr-2.5 Gyr}
by type II SN{\ae} according to three different prescriptions for
the dust condensation efficiencies. All the contributions are
already corrected for the effect of dust destruction. We show: the
contribution by accretion of dust grains in the ISM (ISM-tot,
continuous line); the total contribution by SN{\ae} dust yields
(SN{\ae}-tot, dashed line); the contributions by the different kind
of SNa dust grains, that is Silicates (SN{\ae}-Sil, continuous
line), carbonaceous grains (SN{\ae}-C, dashed line), iron dust
grains (SN{\ae}-Fe, dotted line) and Ca/S/N generic dust grains
(SN{\ae}-Ca/S/N, dot-dashed line). \textbf{Left panel}: the results
based on the \citet{Nozawa03,Nozawa06,Nozawa07} condensation
efficiencies. \textbf{Central panel}: the same as in the left panel
but for the \citet{Dwek98} condensation efficiencies and for the
\citet{Pipino11} efficiencies. \textbf{Right panel}: the same as in
the left panel but for the \citet{Calura08} condensation
efficiencies.} \label{SNaeII1Gyr}
\end{figure}

Let us start our analysis of different yields of dust by evaluating
the effect of varying the amount of star-dust injected from type II
SN{\ae}. In the reference model identified by the parameter string
$\mathcal{GDABBCBBB}$ we vary the recipe for dust yields (the sixth
parameter of the string) choosing among  the three possible
solutions presented in Sect. \ref{SNEyield_details}, namely the one
with high condensation efficiencies by \citet{Dwek98} and
\citet{Calura08}, the set we built based upon the most CNT models
for Type II SN{\ae} by \citet{Nozawa03,Nozawa06,Nozawa07}, and the
low SN{\ae} efficiencies proposed by \citet{Zhukovska08}. We also
included in the discussion the recently revised \citet{Dwek98}
efficiencies, as proposed by \citet{Pipino11} in order to reproduce
the observational constraints: the coefficients for type II SN{\ae}
are lowered by a factor of 10. The results of the corresponding
chemical models, with every parameter fixed except the use of
different CCSN{\ae} efficiencies, are presented in Fig.
\ref{SNaeII1Gyr} where for simplicity we divided the contribution by
SN{\ae} grouping the elements in some typical grain families,
silicates (like olivines, pyroxenes and quartz, depleting the gas of
magnesium, silicon, iron and oxygen), carbonaceous grains, iron dust
and other grains involving S/Ca/N \citep[see][for more
details]{Piovan11b}. The simulated region is the Solar Neighbourhood
(SoNe) ring. Even if the SoNe region is not interested by an intense
star formation activity, nevertheless in the early phases of the
evolution, before that dust accretion in the ISM becomes
significant, different efficiencies have a strong impact on the
early evolution of the dust content. The CNT models present a
contribution that is in the middle between low efficiencies by
\citet{Zhukovska08} and the high efficiencies by \citet{Dwek98}. It
is interesting to observe that the corrected contribution by
\citet{Pipino11}, scaled in such a way to agree with the
observations, also shown in the middle panel of Fig.
\ref{SNaeII1Gyr} (thin lines), tends to produce similar total amount
of dust as the CNT models. The relative contribution to the total
mass budget of the elements embedded into dust grains is however
quite different. It is interesting to underline the following point:
\citet{Zhukovska08} type II SN{\ae} condensation efficiencies are
chosen relying upon clues coming from pre-solar dust grains that
seem to suggest a not negligible contribution of carbon based dust
grains coming from SN{\ae}. The level of carbonaceous grains formed
adopting the \citet{Zhukovska08} coefficient for carbon is similar
to the one predicted by the revised ad-hoc efficiencies by
\citet{Pipino11} and the CNT models by \citet{Nozawa03}. This
agreement seems satisfactory and it suggests some confidence in the
carbon coefficient by \citet{Nozawa03}. However, it must be
underlined that the recent 20M$_{\odot}$ and 170M$_{\odot}$ models
by \citet{Cherchneff10} with kinetic approach and following
molecules evolution, do not form carbon based grains. The formation
of carbon is hampered unless C-rich and He$^{+}$ deprived regions
are introduced, thus implying a strong dependence from the mixing
induced by the explosion. Not even SiC is formed, however it is
present in pre-solar dust grains \citep{Hoppe00}. More detailed
models are then probably required as well as more precise abundances
of pre-solar dust grains, still highly uncertain. \\
\indent For the refractory elements involved into the formation of
silicates, the small number of detections of pre-solar grains does
not allow a safe constraint. \citet{Zhukovska08} assume a very low
efficiency of condensation as working hypothesis, while in the other
recipes that we included, a higher efficiency is chosen or comes out
from CNT models. According to the results by \citet{Matsuura11},
this latter choice seems to be  the most realistic one. They try to
fit the emission spectrum of SN 1987A in the FIR/sub-mm observed
with the PACS, using  different combinations of typical dust grains.
They find that a single dust species implies unrealistic dust
masses. Therefore,  a combinations of different dust grains is
necessary. In this mixture, silicates are present as the most
abundant species, thus suggesting that they condense in significant
amounts. These recipes \citep{Nozawa03,Pipino11} do not differ too
much each other except from the original highly efficient
\citet{Dwek98} recipe that produces a lot of dust. The kind of
mixture of silicates anyway is strongly model dependent, and quite
different partitions of grains are formed if we adopt the CNT
hypothesis or the stochastic/kinetic models. We can therefore
conclude that it is not so safe to follow the specific grain
compounds that are injected into the ISM, due to the uncertainties
on the specific mixture of dust produced. A bit more safe is to rely
on a database like the one we calculated with the condensation
efficiencies of the single elements, that in some way represents an
average of the amount of an element involved into dust.\\

\begin{figure}
\centerline{\hspace{-3mm}
\includegraphics[height=7.5cm,width=8.3cm]{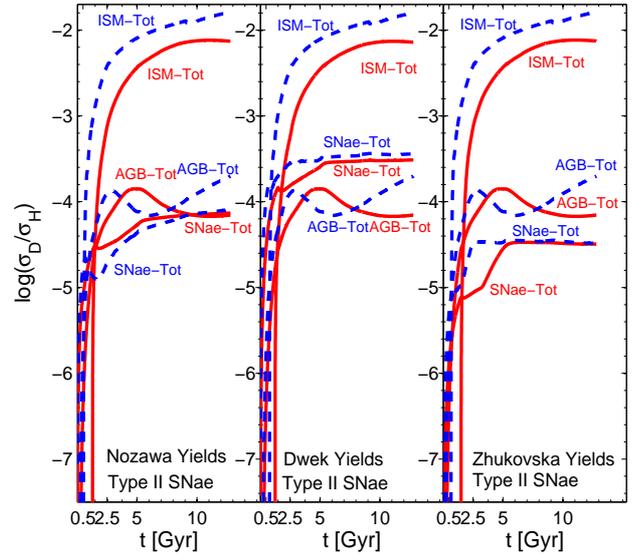}}
\caption{Temporal evolution of the contribution to the dust budget
by type II SN{\ae} according to three different prescriptions for
the dust condensation efficiencies. The data refer to  the Solar
Neighbourhood  (continuous lines) and the innermost region of the MW
(dashed lines). All the contributions are already corrected for the
effect of dust destruction. We show: the contribution by accretion
of dust grain in the ISM (ISM-tot) and the total contributions by
SN{\ae} (SN{\ae}-tot) and AGB stars (AGB-tot). \textbf{Left panel}:
the results for the \citet{Nozawa03,Nozawa06,Nozawa07} condensation
efficiencies. \textbf{Central panel}: the same as in the left panel
but for the \citet{Dwek98} and \citet{Calura08} condensation
efficiencies. \textbf{Right panel}: the same as in the left panel
but for the \citet{Zhukovska08} condensation efficiencies.}
\label{SNeII13Gyr}
\end{figure}

While the influence of different yields of dust by SN{\ae}  can be
very important  in the early stages of the evolution, once the
accretion process in the ISM becomes the dominant dust factory and
the SFR declines, there is in practice no influence on the relative
dust budget compared to the gas amount at the current age, as
clearly shown in Fig. \ref{SNeII13Gyr}. In this figure we present
the evolution until the current time of three region of the MW, that
are an inner ring around 2.3 Kpc (left panel), the Solar
Neighbourhood at 8.5 Kpc (middle panel) and an outer ring at 15.1
Kpc (right panel). The three regions,from left to right, can be
taken in some way as representative of three environments where a
high/average/low star formation, respectively, occurred
\citep{Piovan11b}. The poor influence of the different recipes on
the current total mass budget is true even for the innermost regions
of the MW with higher SFR: once the accretion in MCs is the
dominating process, the stardust injection becomes a secondary
issue. The final dust content is controlled by dust accretion in the
cold region of the ISM. In the same figure we also show for
comparison the AGB reference contribution of the
$\mathcal{GDABBCBBB}$ model based on the \citet{Ferrarotti06} and
\citet{Zhukovska08} estimate  of the AGB efficiencies. Finally, it
is interesting to note that different  prescriptions for the dust
yields by Type II SN{\ae}  cannot affect the amounts  of the various
elements embedded into dust in a low-star-forming environment in
which the  ISM is already enriched in atoms of metals and dust
seeds. In these conditions the amount of free metals available in
the ISM does not appreciably vary  as a consequence  of changes in
the SN{\ae} condensations, even in the case of  SN{\ae} yields
dominating over those by AGB stars as in the model presented in the
middle panel.

\begin{figure}
\centerline{\hspace{-3mm}
\includegraphics[height=7.5cm,width=8.3cm]{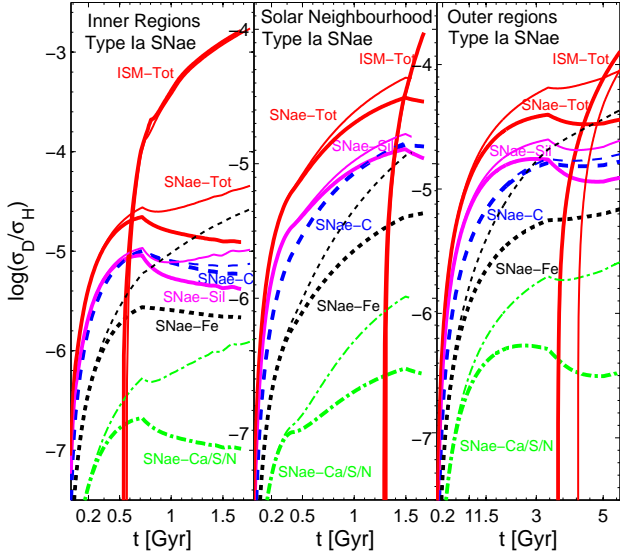}}
\caption{Temporal evolution of the contribution to the dust budget
during  the \textit{first 1.5 Gyr-2.5 Gyr} by type Ia SN{\ae} for
two prescriptions of dust ejecta and three  different regions of the
MW: an inner ring at  2.3 Kpc, the SoNe, and an outer ring at 15.1
Kpc. The results for prescription  $\mathcal{A}$ based on
\citet{Dwek98,Calura08} are represented with thin lines, whereas
those for case $\mathcal{B}$  based on \citet{Zhukovska08} are shown
with thick lines. All the contributions are already corrected for
the effect of dust destruction. We show: the contribution by
accretion of dust grain in the ISM (ISM-tot, continuous lines); the
total contribution by SN{\ae} dust yields (SN{\ae}-tot, type Ia +
type II SN{\ae} - continuous line); the contributions by the
different kind of SNa dust grains, that is silicates (SN{\ae}-Sil,
continuous lines), carbonaceous grains (SN{\ae}-C, dashed lines),
iron dust grains (SN{\ae}-Fe, dotted lines) and Ca/S/N based dust
grains (SN{\ae}-Ca/S/N, dot-dashed lines). \textbf{Left panel}: the
results for the inner ring. \textbf{Central panel}: the same as in
the left panel but for the SoNe. \textbf{Right panel}: the same as
in the left panel but for the outer ring.} \label{SNIa1Gyr}
\end{figure}

\indent We turn now to examine the  contribution to the dust content
by Type Ia SN{\ae}. Two different evaluations have been included
into the model by \citet{Piovan11b}, namely one with high efficiency
of dust formation by type Ia SN{\ae} \citep[see][- our parameter
n$^{o}$ 5 in Tab. \ref{Parameters}, model $\mathcal{A}$]{Dwek98} and
one with very low efficiency, in closer  agreement with the clues
from observations (model $\mathcal{B}$). It must be underlined that
in \citet{Pipino11} the efficiency of condensation in type Ia
SN{\ae} by \citet{Calura08} has been corrected from the high
original value taken from \citet{Dwek98} to a much lower one, in
order to satisfy what the observations actually tell us, that is no
dust condensation is still observed in type Ia SN{\ae}. In any case
we want to explore the whole range of possibilities and for this
reason we included two different solution. In Fig. \ref{SNIa1Gyr} we
present the evolution in the first Gyrs of the star-dust injected by
the SN{\ae}  in three different regions of the MW Disk as before
(left panel for the innermost regions, middle panel for the solar
vicinity, right panel for the outermost one). The contribution by
SN{\ae} is in turn split according to the different types of grains
that have been considered in \citet{Piovan11a} and on which the
single elements have been suitably grouped together. Finally, two
models are displayed, one with high efficiency (thin lines) and one
with low  efficiency (thick lines). As expected, Type Ia SN{\ae} do
not affect the dust evolution  during the first 0.5-1 Gyr, simply
according to the current scenario for the origin of these objects
since they still have to come in significant numbers. In brief, in
the double-degenerate picture they start to contribute only after a
certain amount of time has elapsed and  their effect depends on the
SFR history of  the region. Furthermore, in the innermost regions,
with high SFR and fast ISM enrichment, the dust accretion process
starts to dominate early on the dust production and when type Ia
SN{\ae} come into play, it is already too late to have a significant
role. However, in regions with low SFR such as the solar vicinity
or, even more significantly, the outermost regions, it may happen
that Type Ia SN{\ae} significantly affect the dust enrichment,
simply because the ISM accretion mechanism is delayed because of the
poor enrichment in metals. This effect gets clearly stronger at
lowering the SFR.

\begin{figure}
\centerline{\hspace{-3mm}
\includegraphics[height=7.5cm,width=8.3cm]{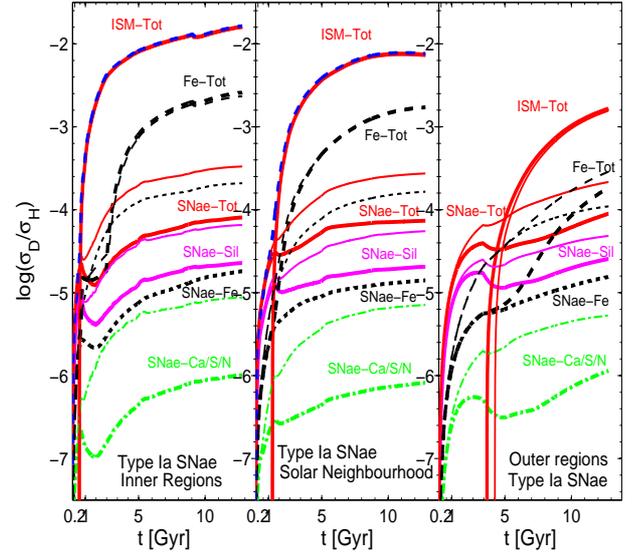}}
\caption{Temporal evolution up to  13 Gyr of the contribution to the
dust budget by type Ia SN{\ae} for two prescriptions of dust ejecta
(models $\mathcal{A}$ and $\mathcal{B}$ of Fig. \ref{SNIa1Gyr}) and
three  different regions of the MW. The results for model
$\mathcal{A}$  are represented by thin lines, whereas those of model
$\mathcal{B}$ are indicated by  thick lines. We show: the
contribution by accretion of dust grain in the ISM (ISM-tot,
continuous lines); the total evolution of the iron-dust (Fe-tot,
dashed lines); the contributions by some of the SNa dust grains,
that is silicates (SN{\ae}-Sil, continuous lines), iron dust grains
(SN{\ae}-Fe, dotted lines) and Ca/S/N based dust grains
(SN{\ae}-Ca/S/N, dot-dashed lines). \textbf{Left panel}: the results
for the inner ring. \textbf{Central panel}: the same but for the
SoNe. \textbf{Right panel}: the same but for the outer ring.}
\label{SNIa13Gyr}
\end{figure}

\indent Do different choices for the Type Ia SN{\ae} condensation
efficiencies influence the final depletion of the elements into dust
at the current age? We show in Fig. \ref{SNIa13Gyr} the evolution of
the dust budget for our two different recipes for Type Ia SN{\ae}:
there is no difference in the ISM accretion process even for the low
SFR environment. Even if the two recipes for the dust condensation
in Type Ia SN{\ae} produce very different amount of dust (see the
thin and thick lines) this has no influence on  the ISM process that
mainly depends on the global amount of metals available in the ISM.
We also checked what is the effect on the iron-dust budget: since
Type Ia SN{\ae} are the main iron polluters we would expect if not
an effect on the global dust budget, at least some influence on the
iron dust. Indeed, we find some differences, but only when we
consider a low-star-forming environment. In that case the iron dust
from SN{\ae} can play a role in the final depletion. In the solar
vicinity, on the contrary, at the present time  no effect can be
seen even if varying the condensation coefficients in Type Ia
SN{\ae}.


\begin{figure} \centerline{\hspace{-3mm}
\includegraphics[height=7.5cm,width=8.3cm]{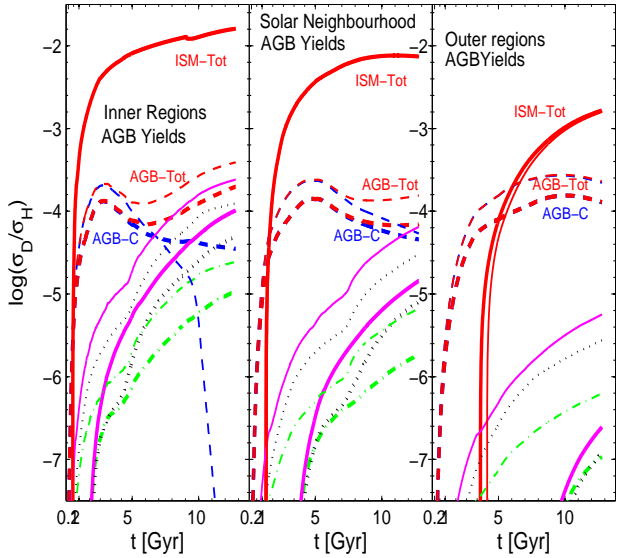}}
\caption{Temporal evolution of the contribution from AGB stars to
the dust budget calculated until the present age. All the
contributions have been properly corrected for the effect of dust
destruction. Three regions of the MW Disk are considered as usual:
the  inner ring (left panel), the SoNe (middle panel), and the outer
region (right panel). Thick lines represent our model $\mathcal{A}$
based upon \citet{Dwek98}, whereas the  thin lines represent model
$\mathcal{B}$ based upon \citet{Ferrarotti06}. We show: the total
contribution by accretion of dust grain in the ISM (continuous
lines) and the total contribution by AGB stars (dashed lines); the
contributions by the various kinds of AGBs dust grains, in the
specific: the iron-stardust (dotted lines); the silicates
(continuous lines), the carbonaceous grains (dashed lines) and
Ca/S/N based dust grains (dot-dashed lines). \textbf{Left panel}:
The results for the  inner ring. \textbf{Central panel}: the same as
in the left panel but for the SoNe. \textbf{Right panel}: the same
as in the left panel but for the outer ring.} \label{AGB13Gyr}
\end{figure}

Finally, we examine the role played by AGB stars that are long known
to pollute the ISM with metals and dust of various kinds depending
on the C/O ratio. In the very early stages of the evolution the
contribution by AGB stars is negligible: since the most massive AGB
star that we included in our models has 6 M$_{\odot}$, it takes some
time before  AGB stars start polluting the ISM. Depending on the
condensation efficiency of the SN{\ae} dust, if this latter is low
it may happen that there could be a possibility for AGB stars to
contribute significantly before  the ISM accretion process starts
dominating. In Fig. \ref{AGB13Gyr} we show the evolution of the AGB
dust budget for the  three  regions of the MW disk for our two
models $\mathcal{A}$ and $\mathcal{B}$ (see Table \ref{Parameters}).
The difference between the models is less striking than for SN{\ae},
with some exception like carbon evolution at high metallicities
(left panel). Once more,  dust production by ISM accretion
overwhelms that by stars (the AGB stars in this  case) and for low
SFR (right panel - outer regions) we see a temporal window where AGB
could produce some effect before the ISM accretion process  becomes
predominant. In any case we consider more reliable the use of the
\citet{Ferrarotti06} models, that includes the effect of the
metallicity on the development of the Carbon rich phase in the AGB,
while the simple recipe by \citet{Dwek98,Calura08,Pipino11} do not
take into account this point. The approach by which dust formation
is simulated in the circumstellar envelope of AGB stars, in spite of
its limitations, is more reliable than the hypotheses assumed for
dust formation in SN{\ae} \citep{Cherchneff10}, and we can rely on
\citet{Ferrarotti06} results. The inclusion of the metallicity
effect allows to respect an important characteristic of the dust
mixture: we have, as expected from AGB models, that low metallicity
stars more easily enter the carbon rich phase and produce more
carbon dust, while high metallicity stars, mainly avoiding or
briefly entering the C-rich part of the evolution mainly contribute
with silicates.\\
\noindent As we said, we fixed at 6 M$_{\odot}$  the transition
between AGB stars and SN{\ae}. In classical chemical models this
limit typically varies from 5 to 8 M$_{\odot}$: this would have only
a very small effect on the relative proportion of dust generated by
SN{\ae} and AGB stars, but it could affect the way by which AGB
stars contribute to the early evolution \citep{Valiante09}.
Interestingly enough, recently a complete  set of yields from  S-AGB
stars spanning the mass interval from  7.5 and 10.5 M$_{\odot}$ has
been calculated \citep{Siess10a,Siess10b}. Even if the number of
stars in this mass range nearly parallels the number of stars more
massive than this limit  \citep{Siess10a}, because of their short
lifetime  there are no firm evaluations of the role played by S-AGB
stars in the chemical evolution of galaxies and even more important
what could be their effect on the early evolution of the dust
content in galaxies.

\begin{figure}
\centerline{\hspace{-3mm}
\includegraphics[height=7.5cm,width=8.3cm]{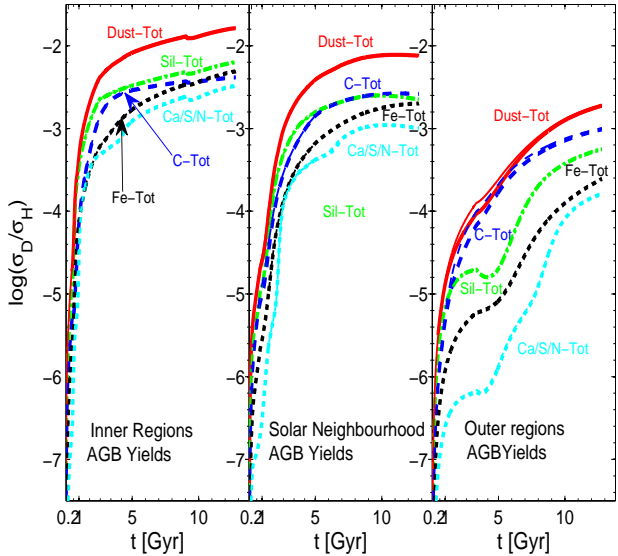}}
\caption{Temporal evolution of the contribution from AGB stars to
the dust budget calculated until the present age.  Three regions of
the MW Disk are considered as usual: the  inner ring (left panel),
the SoNe (middle panel), and the outer region (right panel). Thick
lines represent our model $\mathcal{A}$ based upon \citet{Dwek98},
while thin lines represent model $\mathcal{B}$, based upon
\citet{Ferrarotti06}. We show: the total amount of dust grains in
the ISM (continuous lines) and the total amount of the various grain
families, where with total we mean AGB stardust plus SN{\ae}
stardust plus accreted dust in the ISM. In detail we show: the
iron-stardust (dotted lines); the silicates (dot-dashed lines), the
carbonaceous grains (dashed lines) and Ca/S/N based dust grains
(dotted lines). \textbf{Left panel}: The results for the inner
region of the MW Disk.   \textbf{Central panel}: the same as in the
left panel but for the SoNe. \textbf{Right panel}: the same as in
the left panel but for the outer ring. } \label{AGB13GyrTOT}
\end{figure}

Does the different possible recipes for AGB star-dust lead to some
differences in the evolution of the dust content in the ISM? In Fig.
\ref{AGB13GyrTOT} we show (always for the three regions of the disk)
the evolution of the \emph{total} normalized abundance of the
various dust grains families for the prescriptions  for the AGB
stars. Only in the outer regions with very low star formation rates
and low metallicities we can notice a difference between one model
of dust nucleation and injection by AGB stars and another. In the
other  regions, there is in practice no difference in the total
budget at varying the AGB condensation efficiencies. This simply
means that in the early stages, with the typical SF laws for the MW,
AGB stars   dust factory is dominated by SN{\ae} (unless we assume a
very low condensation efficiencies by SN{\ae}), and later by the
accretion process in the ISM. Obviously AGB stars play a fundamental
role in refueling the ISM with metals and seeds, but the dust
factory of the ISM accretion mechanism is the starring actor. Of
course,  if some peculiar dust grains (like SiC) are injected by
AGBs and not formed by accretion in the ISM, in that case AGB stars
play a crucial role in determining the evolution of that kind of
dust.

\section{Discussion and conclusions}\label{Discus_Concl}

In this paper we have described the database of condensation
efficiencies for the main elements that form the dust emitted by the
most important dust factories in nature, i.e. SN{\ae} and AGB stars.
The results are organized in tables containing the condensation
efficiencies for the refractory elements C, O, Mg, Si, S, Ca (for
this latter and partially for S, the average values between the
condensation efficiencies of the other refractory elements has been
adopted) and Fe in AGB stars and SN{\ae}. The condensation
efficiencies, multiplied by the gaseous ejecta provide an estimate
of the amount of each element trapped  into dust that is injected
into the ISM. Our compilation stands upon the \textit{theoretical}
work by \citet{Ferrarotti06,Zhukovska08} and
\citet{Nozawa03,Nozawa06,Nozawa07} and allows us  to take into
account: (1) the different metallicity of the AGB stars, thus
simulating the growing number of C-rich stars forming carbonaceous
grains at lowering the metallicity; (2) the different density of the
environment in which the SNa explosion occurs which crucially
determines the final amount of dust surviving to the
shocks. \\
\indent There is an important aspect of the results to note. The
mixture of dust grains emitted by stars, in particular the one by
SN{\ae}, is still very model dependent: for instance the kinetic
models by \citet{Cherchneff10}  not only predict different total
amounts of dust as compared to the classical CNT models by
\citet{Nozawa03}, but also different  composition for the dust. To
somehow cure this point of uncertainty we calculated (and tested)
\textit{average} condensation coefficients for \textit{each
elements}. To this aim, we made use of the best prescriptions for
the production of dust by stars, and compared the results one would
obtain by introducing them in the classical chemical model for the
MW Disk and SoNe \citep{Piovan11b,Piovan11c}. As already mentioned,
the largest uncertainty is due to SN{\ae}: to highlight the point we
compare the model results with the available data about the amounts
of dust and its composition, taking into account the observational
hints by pre-solar grains and IR emission from SNa remnants. In
particular we look for the best solution to be adopted for SN{\ae}.
We find that the CNT approximation widely adopted for dust formation
in SN{\ae}, despite its limitations as a first order approximation
of a complex situation neither at equilibrium nor at steady state,
still gives good results. Indeed, (i) it produces amounts of
carbonaceous grains that agree with the estimate required by the
pre-solar dust grains; (ii) it produces significant amounts of
silicates that seem to be necessary to reproduce very recent cold
dust emission spectra in the SN 1987A; (iii) the amount of dust
produced by SN{\ae} is not negligible, as indicated by the most
recent in observations at long wavelengths and, (iv) the CNT unmixed
model agrees well with both the FIR/sub-mm estimates of newly formed
dust in SN{\ae} and the amount of dust required to explain high-z
obscured objects with the SN{\ae} playing a decisive role.\\
\indent The whole picture we have just outlined could be considered
as satisfactory.  However, it is worth noticing: (i) the reduction
in the amount of dust predicted by the more physically correct
kinetic models, goes in the opposite direction with respect to what
indicated by observations of cold dust; (ii) the poor data available
on pre-solar silicates does not support an efficient condensation of
silicates
as indicated  by estimates of cold dust  in SN{\ae}.\\
\indent This simply means that our understanding of dust formation
in SN{\ae} is still very uncertain, both theoretically and
observationally. More data on  cold dust for a wider sample of
SN{\ae} is needed. A new generation of models of dust formation in
SN{\ae} is therefore required to highlight the many points of
uncertainty and contradiction. They go from the obvious uncertainty
in the correct approach to use, the complicated description of the
dust formation and the the limited number of accurate models in
literature to poor statistics, lack of high quality observations of
the cold dust in the FIR/sub-mm for
many SN{\ae}, and  uncertainties on estimating the amount of fresh dust).\\

\textit{Acknowledgements}. L. Piovan acknowledges A. Weiss and the
Max Planck Institut Fur AstroPhysik (Garching - Germany) for the
very warm and friendly hospitality and providing unlimited
computational support during the visits as EARA fellow when a
significant part of this study has been carried out.
 The authors are also deeply grateful to  S. Zhukovska and H. P. Gail
for many explanations and clarifications about their  yields of dust from AGB stars,  T. Nozawa
and H. Umeda for many fruitful discussions about SNa dust yields.
This work has been financed by the University of Padua with the dedicated
fellowship "Numerical Simulations of galaxies (dynamical, chemical and
spectrophotometric models), strategies of parallelization in
dynamical lagrangian approach, communication cell-to-cell into
hierarchical tree codes, algorithms and optimization techniques" as  part of the AACSE
Strategic Research Project.

\appendix
\section*{Appendix A} \label{AppendixA}

The models of dust formation in the ejecta of Pop III CCSN{\ae}
(core collapse) and PISN{\ae} (pair-instability) by \citet{Nozawa03}
are obtained taking into account two extreme cases, i.e. no mixing
and full mixing in the He core. In the un-mixed case the typical
onion-like structure is considered. In the uniformly mixed model all
the elements are mixed in the He core. In both cases many types of
dust grain are considered: Fe, FeS, Si, V, T, Cr, Co, Ni, Cu, C,
SiC, TiC, Al$_{2}$O$_{3}$, MgSiO$_{3}$, Mg$_{2}$SiO$_{4}$,
SiO$_{2}$, MgO, Fe$_{3}$O$_{4}$, FeO. In the final yields of dust,
only some of these species are present. In the un-mixed case we have
Fe, FeS, Si, C, Al$_{2}$O$_{3}$, MgSiO$_{3}$, Mg$_{2}$SiO$_{4}$,
SiO$_{2}$ and MgO, whereas for the mixed one we have SiO$_{2}$,
Al$_{2}$O$_{3}$, MgSiO$_{3}$, Mg$_{2}$SiO$_{4}$ and Fe$_{3}$O$_{4}$.
For the abundances of the single elements in the yields of dust  of
the $i$-th progenitor we used the following equations. For the
unmixed case:

\begin{small}
\begin{align}\label{UnmixedElements}
M_{i}&\left(C\right) = M_{i}\left(C\right) \\
M_{i}&\left(O\right) = A\left( O
\right)\left[\frac{4M_{i}\left(MgSiO_{4}\right)}{A\left(MgSiO_{4}\right)}+
\frac{3M_{i}\left(MgSiO_{3}\right)}{A\left(MgSiO_{3}\right)}+\right. \nonumber \\
&+\left.\frac{3M_{i}\left(Al_{2}O_{3}\right)}{A\left(Al_{2}O_{3}\right)}+
\frac{M_{i}\left(2SiO_{2}\right)}{A\left(SiO_{2}\right)}+
\frac{M_{i}\left(MgO\right)}{A\left(MgO\right)}\right]\\
M_{i}&\left(Mg\right) =A\left( Mg
\right)\left[\frac{M_{i}\left(MgSiO_{3}\right)}{A\left(MgSiO_{3}\right)}+
\frac{M_{i}\left(MgSiO_{4}\right)}{A\left(MgSiO_{4}\right)}+\right.
\nonumber \\
&+\left.\frac{M_{i}\left(MgO\right)}{A\left(MgO\right)}\right]\\
M_{i}&\left(Si\right) =A\left( Si
\right)\left[\frac{M_{i}\left(MgSiO_{3}\right)}{A\left(MgSiO_{3}\right)}+
\frac{M_{i}\left(MgSiO_{4}\right)}{A\left(MgSiO_{4}\right)}+\right.
\nonumber \\
&+\left.\frac{M_{i}\left(Si\right)}{A\left(Si\right)}+
\frac{M_{i}\left(SiO_{2}\right)}{A\left(SiO_{2}\right)}\right]\\
M_{i}&\left(S\right) = A\left(S\right)\frac{M_{i}\left(FeS\right)}{A\left(FeS\right)} \\
M_{i}&\left(Fe\right) =A\left( Fe
\right)\left[\frac{M_{i}\left(Fe\right)}{A\left(Fe\right)}+
\frac{M_{i}\left(FeS\right)}{A\left(FeS\right)}\right]
\end{align}
\end{small}

\noindent where $A\left(...\right)$ are the mass numbers of the
considered elements or molecules and $M_{i}\left(...\right)$ are the
dust yields of a given element or molecule for the $i$-th
progenitor. For the mixed case the equations in use  are:

\begin{small}
\begin{align}\label{UnmixedElements}
M_{i}&\left(O\right) = A\left( O
\right)\left[\frac{4M_{i}\left(MgSiO_{4}\right)}{A\left(MgSiO_{4}\right)}+
\frac{3M_{i}\left(MgSiO_{3}\right)}{A\left(MgSiO_{3}\right)}+\right. \nonumber \\
&+\left.\frac{3M_{i}\left(Al_{2}O_{3}\right)}{A\left(Al_{2}O_{3}\right)}+
\frac{M_{i}\left(2SiO_{2}\right)}{A\left(SiO_{2}\right)}+
\frac{3M_{i}\left(Al_{2}O_{3}\right)}{A\left(Al_{2}O_{3}\right)}\right]\\
M_{i}&\left(Mg\right) =A\left( Mg
\right)\left[\frac{M_{i}\left(MgSiO_{3}\right)}{A\left(MgSiO_{3}\right)}+
\frac{M_{i}\left(MgSiO_{4}\right)}{A\left(MgSiO_{4}\right)}\right]\\
M_{i}&\left(Si\right) =A\left( Si
\right)\left[\frac{M_{i}\left(MgSiO_{3}\right)}{A\left(MgSiO_{3}\right)}+
\frac{M_{i}\left(MgSiO_{4}\right)}{A\left(MgSiO_{4}\right)}+\right.
\nonumber \\
&+\left.\frac{M_{i}\left(SiO_{2}\right)}{A\left(SiO_{2}\right)}\right]\\
M_{i}&\left(Fe\right) =A\left( Fe
\right)\left[\frac{3M_{i}\left(Fe_{3}O_{4}\right)}{A\left(Fe_{3}O_{4}\right)}\right]
\end{align}
\end{small}

In  this latter case,  Aluminium $M_{i}\left(Al\right)$ is not taken
into account, because our chemical model does not include the
evolution of this element. \\

\section*{Appendix B} \label{AppendixB}

According to  \citet{Zhukovska08}, dust formation in AGB stars  has
been interpolated and extrapolated from the data and models by
\citet{Ferrarotti06} to eight metallicities, going from $Z=0.001$ to
$Z=0.04$ and to a set of twenty-seven masses from $1$M$_{\odot}$ to
$7$M$_{\odot}$. Three types of mixtures are considered depending on
the C/O ratio, i.e. for M-stars, S-stars and C-stars. For
M-stars they consider the formation of the following dust
molecules: (1) magnesium iron silicates, like the olivines
forsterite (Mg$_{2}$SiO$_{4}$, with Mg as the end-member) and
fayalite (Fe$_{2}$SiO$_{4}$,  with Fe as the end-member), the
pyroxenes enstatite (MgSiO$_{3}$, with Mg as the end-member), and
ferrosilite (FeSiO$_{3}$, with Fe as the end-member); (2) quartz
SiO$_{2}$ and (3) iron dust grains Fe. For S-stars they only
consider (1) quartz SiO$_{2}$ and (2) iron dust grains Fe, while
for carbon rich C-stars they include (1) carbonaceous grains made
by C and H, again (2) quartz SiO$_{2}$ and (3) iron dust grains
Fe.

For the abundances of the single elements in the dust yields of the
$i$-th progenitor we used the following equations:

\noindent For $M$-stars

\begin{small}
\begin{align}\label{M-stars}
M_{i}&\left(O\right) = A\left( O
\right)\left[\frac{4M_{i}\left(MgSiO_{4}\right)}{A\left(MgSiO_{4}\right)}+
\frac{3M_{i}\left(MgSiO_{3}\right)}{A\left(MgSiO_{3}\right)}+\right. \nonumber \\
&+\left.\frac{4M_{i}\left(FeSiO_{4}\right)}{A\left(FeSiO_{4}\right)}+
\frac{3M_{i}\left(FeSiO_{3}\right)}{A\left(FeSiO_{3}\right)}+
\frac{M_{i}\left(2SiO_{2}\right)}{A\left(SiO_{2}\right)}
\right]\\
M_{i}&\left(Mg\right) =A\left( Mg
\right)\left[\frac{M_{i}\left(MgSiO_{3}\right)}{A\left(MgSiO_{3}\right)}+
\frac{M_{i}\left(MgSiO_{4}\right)}{A\left(MgSiO_{4}\right)}\right]\\
M_{i}&\left(Si\right) =A\left( Si
\right)\left[\frac{M_{i}\left(MgSiO_{3}\right)}{A\left(MgSiO_{3}\right)}+
\frac{M_{i}\left(MgSiO_{4}\right)}{A\left(MgSiO_{4}\right)}+\right.
\nonumber \\
&+\left.\frac{M_{i}\left(SiO_{2}\right)}{A\left(SiO_{2}\right)}\right]\\
M_{i}&\left(Fe\right) =A\left( Fe
\right)\left[\frac{M_{i}\left(FeSiO_{4}\right)}{A\left(FeSiO_{4}\right)}+
\frac{M_{i}\left(FeSiO_{3}\right)}{A\left(FeSiO_{3}\right)}+\right.
\nonumber \\
&+\left.\frac{M_{i}\left(Fe\right)}{A\left(Fe\right)} \right]
\end{align}
\end{small}

\noindent where $A\left(...\right)$ are the mass numbers of the
considered elements or molecules and $M_{i}\left(...\right)$ are the
dust yields of a given element or molecule for the $i$-th
progenitor.

\noindent For $S$-stars

\begin{small}
\begin{align}\label{S-stars}
M_{i}&\left(O\right) = A\left( O \right)\left[
\frac{2M_{i}\left(SiO_{2}\right)}{A\left(SiO_{2}\right)}
\right]\\
M_{i}&\left(Si\right) =A\left( Si
\right)\left[\frac{M_{i}\left(SiO_{2}\right)}{A\left(SiO_{2}\right)}\right]\\
M_{i}&\left(Fe\right) =A\left( Fe
\right)\left[\frac{M_{i}\left(Fe\right)}{A\left(Fe\right)} \right]
\end{align}
\end{small}

\noindent Finally, for C-stars

\begin{small}
\begin{align}\label{S-stars}
M_{i}&\left(Si\right) =A\left( Si \right)\left[
\frac{M_{i}\left(SiC\right)}{A\left(SiC\right)}\right]\\
M_{i}&\left(C\right) =A\left( C
\right)\left[\frac{M_{i}\left(C\right)}{A\left(C\right)}+
\frac{M_{i}\left(SiC\right)}{A\left(SiC\right)}\right]\\
M_{i}&\left(Fe\right) =A\left( Fe
\right)\left[\frac{M_{i}\left(Fe\right)}{A\left(Fe\right)} \right]
\end{align}
\end{small}

\noindent with the usual meaning of the symbols.\\

\begin{scriptsize}
\bibliographystyle{apj}                          
\bibliography{mnemonic,PiovanI}    
\end{scriptsize}

\renewcommand{\arraystretch}{1.05}
\begin{table*}
\footnotesize
\begin{center}
\caption[]{Dust condensation efficiencies for AGB stars of metallicity Z=0.001.} \begin{tabular}{ccccccccc}
\hline \hline \vspace{0.1cm} M$_{\odot}$ & $\delta_{C}$ & $\delta_{O}$
& $\delta_{Mg} $ & $\delta_{Si}$& $\delta_{Fe} $ & $\delta_{Ca}$& $\delta_{S} $ & $\delta_{Al}$ \\   \hline
  1.00  &    0.00000000 &    0.00000000  &   0.00000000 &   0.00000000 &   0.00000000 &   0.00000000 &   0.00000000 &   0.00000000   \\
  1.10  &    0.00481670 &    0.00000000  &   0.00000000 &   0.00000000 &   0.00000000 &   0.00000000 &   0.00000000 &   0.00000000   \\
  1.20  &    0.14639919 &    0.00000000  &   0.00000000 &   0.00000000 &   0.00000000 &   0.00000000 &   0.00000000 &   0.00000000   \\
  1.25  &    0.18576120 &    0.00000000  &   0.00000000 &   0.00000000 &   0.00000000 &   0.00000000 &   0.00000000 &   0.00000000   \\
  1.30  &    0.23425915 &    0.00000000  &   0.00000000 &   0.00000000 &   0.00000000 &   0.00000000 &   0.00000000 &   0.00000000   \\
  1.40  &    0.29019752 &    0.00000000  &   0.00000000 &   0.00000000 &   0.00000000 &   0.00000000 &   0.00000000 &   0.00000000   \\
  1.50  &    0.39619540 &    0.00000000  &   0.00000000 &   0.00000000 &   0.00000000 &   0.00000000 &   0.00000000 &   0.00000000   \\
  1.60  &    0.45922137 &    0.00000000  &   0.00000000 &   0.00000000 &   0.00000000 &   0.00000000 &   0.00000000 &   0.00000000   \\
  1.70  &    0.52030200 &    0.00000000  &   0.00000000 &   0.00000000 &   0.00000000 &   0.00000000 &   0.00000000 &   0.00000000   \\
  1.80  &    0.53833388 &    0.00000000  &   0.00000000 &   0.00003080 &   0.00000000 &   0.00001026 &   0.00001026 &   0.00001026   \\
  1.90  &    0.50111438 &    0.00000000  &   0.00000000 &   0.00003117 &   0.00000000 &   0.00001039 &   0.00001039 &   0.00001039   \\
  2.00  &    0.47470917 &    0.00000000  &   0.00000000 &   0.00003262 &   0.00000000 &   0.00001087 &   0.00001087 &   0.00001087   \\
  2.10  &    0.46005052 &    0.00000000  &   0.00000000 &   0.00003321 &   0.00000000 &   0.00001107 &   0.00001107 &   0.00001107   \\
  2.20  &    0.45734066 &    0.00000000  &   0.00000000 &   0.00003437 &   0.00000000 &   0.00001145 &   0.00001145 &   0.00001145   \\
  2.30  &    0.46136134 &    0.00000000  &   0.00000000 &   0.00003487 &   0.00000000 &   0.00001162 &   0.00001162 &   0.00001162   \\
  2.40  &    0.46360417 &    0.00000000  &   0.00000000 &   0.00003586 &   0.00000000 &   0.00001195 &   0.00001195 &   0.00001195   \\
  2.50  &    0.45318091 &    0.00000000  &   0.00000000 &   0.00003663 &   0.00000000 &   0.00001221 &   0.00001221 &   0.00001221   \\
  3.00  &    0.47763680 &    0.00000000  &   0.00000000 &   0.00004363 &   0.00000000 &   0.00001454 &   0.00001454 &   0.00001454   \\
  3.50  &    0.53036135 &    0.00000000  &   0.00000000 &   0.00004637 &   0.00000000 &   0.00001545 &   0.00001545 &   0.00001545   \\
  4.00  &    1.00000000 &    0.00000000  &   0.00000000 &   0.00004927 &   0.00000000 &   0.00001642 &   0.00001642 &   0.00001642   \\
  4.50  &    0.15586482 &    0.00000014  &   0.00000006 &   0.00002279 &   0.00001202 &   0.00001162 &   0.00001162 &   0.00001162   \\
  5.00  &    0.14262594 &    0.00000033  &   0.00000008 &   0.00002682 &   0.00001382 &   0.00001357 &   0.00001357 &   0.00001357   \\
  5.50  &    0.12204322 &    0.00000049  &   0.00000008 &   0.00002615 &   0.00001530 &   0.00001384 &   0.00001384 &   0.00001384   \\
  6.00  &    0.10409485 &    0.00000073  &   0.00000007 &   0.00002593 &   0.00001666 &   0.00001422 &   0.00001422 &   0.00001422   \\
  6.50  &    0.09531371 &    0.00000081  &   0.00000006 &   0.00002658 &   0.00001798 &   0.00001487 &   0.00001487 &   0.00001487   \\
  7.00  &    0.08813571 &    0.00000083  &   0.00000006 &   0.00002702 &   0.00001928 &   0.00001545 &   0.00001545 &   0.00001545   \\
  \hline
\end{tabular}
\end{center}
\end{table*}
\renewcommand{\arraystretch}{1}

\renewcommand{\arraystretch}{1.05}
\begin{table*}
\footnotesize
\begin{center}
\caption[]{Dust condensation efficiencies for AGB stars of metallicity Z=0.002.} \begin{tabular}{ccccccccc}
\hline \hline \vspace{0.1cm} M$_{\odot}$ & $\delta_{C}$ & $\delta_{O}$
& $\delta_{Mg} $ & $\delta_{Si}$& $\delta_{Fe} $ & $\delta_{Ca}$& $\delta_{S} $ & $\delta_{Al}$ \\   \hline
  1.00  &   0.00000000  &   0.00000000  &   0.00000000  &  0.00000000  &  0.00000000  &  0.00000000  &  0.00000000  &  0.00000000  \\
  1.10  &   0.00265956  &   0.00000000  &   0.00000000  &  0.00000000  &  0.00000000  &  0.00000000  &  0.00000000  &  0.00000000  \\
  1.20  &   0.15026726  &   0.00000000  &   0.00000000  &  0.00003450  &  0.00000000  &  0.00001150  &  0.00001150  &  0.00001150  \\
  1.25  &   0.15487339  &   0.00000000  &   0.00000000  &  0.00004420  &  0.00000000  &  0.00001473  &  0.00001473  &  0.00001473  \\
  1.30  &   0.21380446  &   0.00000000  &   0.00000000  &  0.00006667  &  0.00000000  &  0.00002222  &  0.00002222  &  0.00002222  \\
  1.40  &   0.29515964  &   0.00000000  &   0.00000000  &  0.00010309  &  0.00000000  &  0.00003436  &  0.00003436  &  0.00003436  \\
  1.50  &   0.36691218  &   0.00000102  &   0.00000000  &  0.00015487  &  0.00000000  &  0.00005162  &  0.00005162  &  0.00005162  \\
  1.60  &   0.39952639  &   0.00000000  &   0.00000000  &  0.00012928  &  0.00000000  &  0.00004309  &  0.00004309  &  0.00004309  \\
  1.70  &   0.41162551  &   0.00000000  &   0.00000000  &  0.00012490  &  0.00000000  &  0.00004163  &  0.00004163  &  0.00004163  \\
  1.80  &   0.42151132  &   0.00000000  &   0.00000000  &  0.00013551  &  0.00000000  &  0.00004517  &  0.00004517  &  0.00004517  \\
  1.90  &   0.41303055  &   0.00000000  &   0.00000000  &  0.00014318  &  0.00000000  &  0.00004772  &  0.00004772  &  0.00004772  \\
  2.00  &   0.40014155  &   0.00000000  &   0.00000000  &  0.00015340  &  0.00000000  &  0.00005113  &  0.00005113  &  0.00005113  \\
  2.10  &   0.37498383  &   0.00000000  &   0.00000000  &  0.00015437  &  0.00000000  &  0.00005145  &  0.00005145  &  0.00005145  \\
  2.20  &   0.38755967  &   0.00000000  &   0.00000000  &  0.00015997  &  0.00000000  &  0.00005332  &  0.00005332  &  0.00005332  \\
  2.30  &   0.37049519  &   0.00000000  &   0.00000000  &  0.00016498  &  0.00000000  &  0.00005499  &  0.00005499  &  0.00005499  \\
  2.40  &   0.37186737  &   0.00000000  &   0.00000000  &  0.00016891  &  0.00000000  &  0.00005630  &  0.00005630  &  0.00005630  \\
  2.50  &   0.36282426  &   0.00000000  &   0.00000000  &  0.00017613  &  0.00000000  &  0.00005871  &  0.00005871  &  0.00005871  \\
  3.00  &   0.34320933  &   0.00000000  &   0.00000000  &  0.00021219  &  0.00000000  &  0.00007073  &  0.00007073  &  0.00007073  \\
  3.50  &   0.46943942  &   0.00000000  &   0.00000000  &  0.00024423  &  0.00000000  &  0.00008141  &  0.00008141  &  0.00008141  \\
  4.00  &   1.00000000  &   0.00000000  &   0.00000000  &  0.00027761  &  0.00000000  &  0.00009253  &  0.00009253  &  0.00009253  \\
  4.50  &   0.12055474  &   0.00000142  &   0.00000059  &  0.00011654  &  0.00006326  &  0.00006013  &  0.00006013  &  0.00006013  \\
  5.00  &   0.10926296  &   0.00000177  &   0.00000053  &  0.00012074  &  0.00007193  &  0.00006440  &  0.00006440  &  0.00006440  \\
  5.50  &   0.09486079  &   0.00000242  &   0.00000052  &  0.00012001  &  0.00007762  &  0.00006605  &  0.00006605  &  0.00006605  \\
  6.00  &   0.08336151  &   0.00000323  &   0.00000044  &  0.00011962  &  0.00008168  &  0.00006725  &  0.00006725  &  0.00006725  \\
  6.50  &   0.07595158  &   0.00000366  &   0.00000043  &  0.00012366  &  0.00008886  &  0.00007098  &  0.00007098  &  0.00007098  \\
  7.00  &   0.07009158  &   0.00000393  &   0.00000043  &  0.00012864  &  0.00009646  &  0.00007518  &  0.00007518  &  0.00007518  \\
  \hline
\end{tabular}
\end{center}
\end{table*}
\renewcommand{\arraystretch}{1}

\renewcommand{\arraystretch}{1.05}
\begin{table*}
\footnotesize
\begin{center}
\caption[]{Dust condensation efficiencies for AGB stars of metallicity Z=0.004.} \begin{tabular}{ccccccccc}
\hline \hline \vspace{0.1cm} M$_{\odot}$ & $\delta_{C}$ & $\delta_{O}$
& $\delta_{Mg} $ & $\delta_{Si}$& $\delta_{Fe} $ & $\delta_{Ca}$& $\delta_{S} $ & $\delta_{Al}$ \\   \hline
  1.00 &   0.00000000 &   0.00000000 &   0.00000000 &   0.00000000 &   0.00000000 &   0.00000000 &   0.00000000 &   0.00000000 \\
  1.10 &   0.00083222 &   0.00000000 &   0.00000000 &   0.00000000 &   0.00000000 &   0.00000000 &   0.00000000 &   0.00000000 \\
  1.20 &   0.09697428 &   0.00000000 &   0.00000000 &   0.00034345 &   0.00000000 &   0.00011448 &   0.00011448 &   0.00011448 \\
  1.25 &   0.21074678 &   0.00000000 &   0.00000000 &   0.00092058 &   0.00000000 &   0.00030686 &   0.00030686 &   0.00030686 \\
  1.30 &   0.25507001 &   0.00000000 &   0.00000000 &   0.00108098 &   0.00000000 &   0.00036032 &   0.00036032 &   0.00036032 \\
  1.40 &   0.33996781 &   0.00000000 &   0.00000000 &   0.00123257 &   0.00000000 &   0.00041085 &   0.00041085 &   0.00041085 \\
  1.50 &   0.40811000 &   0.00000000 &   0.00000000 &   0.00144564 &   0.00000000 &   0.00048188 &   0.00048188 &   0.00048188 \\
  1.60 &   0.44345154 &   0.00000000 &   0.00000000 &   0.00153059 &   0.00000000 &   0.00051019 &   0.00051019 &   0.00051019 \\
  1.70 &   0.45103152 &   0.00000000 &   0.00000000 &   0.00146990 &   0.00000000 &   0.00048996 &   0.00048996 &   0.00048996 \\
  1.80 &   0.48278305 &   0.00000000 &   0.00000000 &   0.00155159 &   0.00000000 &   0.00051719 &   0.00051719 &   0.00051719 \\
  1.90 &   0.52098757 &   0.00000000 &   0.00000000 &   0.00159094 &   0.00000000 &   0.00053031 &   0.00053031 &   0.00053031 \\
  2.00 &   0.49801535 &   0.00000000 &   0.00000000 &   0.00162689 &   0.00000000 &   0.00054229 &   0.00054229 &   0.00054229 \\
  2.10 &   0.50604017 &   0.00000000 &   0.00000000 &   0.00155156 &   0.00000000 &   0.00051718 &   0.00051718 &   0.00051718 \\
  2.20 &   0.51122053 &   0.00000000 &   0.00000000 &   0.00171615 &   0.00000000 &   0.00057205 &   0.00057205 &   0.00057205 \\
  2.30 &   0.52421529 &   0.00000000 &   0.00000000 &   0.00164786 &   0.00000000 &   0.00054928 &   0.00054928 &   0.00054928 \\
  2.40 &   0.51358635 &   0.00000000 &   0.00000000 &   0.00171132 &   0.00000000 &   0.00057044 &   0.00057044 &   0.00057044 \\
  2.50 &   0.52089403 &   0.00000258 &   0.00000000 &   0.00177490 &   0.00000000 &   0.00059163 &   0.00059163 &   0.00059163 \\
  3.00 &   0.58487881 &   0.00000000 &   0.00000000 &   0.00204975 &   0.00000385 &   0.00068453 &   0.00068453 &   0.00068453 \\
  3.50 &   0.40179017 &   0.00000000 &   0.00000000 &   0.00336510 &   0.00000554 &   0.00112355 &   0.00112355 &   0.00112355 \\
  4.00 &   1.00000000 &   0.00000000 &   0.00000000 &   0.00410518 &   0.00000553 &   0.00137024 &   0.00137024 &   0.00137024 \\
  4.50 &   0.07722777 &   0.00002405 &   0.00001569 &   0.00123602 &   0.00088501 &   0.00071224 &   0.00071224 &   0.00071224 \\
  5.00 &   0.06873504 &   0.00002940 &   0.00001430 &   0.00129496 &   0.00099224 &   0.00076717 &   0.00076717 &   0.00076717 \\
  5.50 &   0.06051039 &   0.00003360 &   0.00001203 &   0.00132337 &   0.00099961 &   0.00077834 &   0.00077834 &   0.00077834 \\
  6.00 &   0.05301306 &   0.00003907 &   0.00000986 &   0.00130422 &   0.00102121 &   0.00077843 &   0.00077843 &   0.00077843 \\
  6.50 &   0.04938816 &   0.00004243 &   0.00000911 &   0.00136929 &   0.00107905 &   0.00081915 &   0.00081915 &   0.00081915 \\
  7.00 &   0.04591880 &   0.00004523 &   0.00000880 &   0.00138212 &   0.00113674 &   0.00084255 &   0.00084255 &   0.00084255 \\
\hline
\end{tabular}
\end{center}
\end{table*}
\renewcommand{\arraystretch}{1}

\renewcommand{\arraystretch}{1.05}
\begin{table*}
\footnotesize
\begin{center}
\caption[]{Dust condensation efficiencies for AGB stars of metallicity Z=0.008.} \begin{tabular}{ccccccccc}
\hline \hline \vspace{0.1cm} M$_{\odot}$ & $\delta_{C}$ & $\delta_{O}$
& $\delta_{Mg} $ & $\delta_{Si}$& $\delta_{Fe} $ & $\delta_{Ca}$& $\delta_{S} $ & $\delta_{Al}$ \\   \hline
   1.00 &  0.00000000 &  0.00000000 &  0.00000000 &   0.00000000 &   0.00000000 &   0.00000000 &   0.00000000 &   0.00000000 \\
   1.10 &  0.00027758 &  0.00009887 &  0.00036252 &   0.00132490 &   0.00261710 &   0.00143484 &   0.00143484 &   0.00143484 \\
   1.20 &  0.07760180 &  0.00059906 &  0.00212571 &   0.01598957 &   0.02168562 &   0.01326697 &   0.01326697 &   0.01326697 \\
   1.25 &  0.07606858 &  0.00030102 &  0.00105843 &   0.03153497 &   0.01847338 &   0.01702226 &   0.01702226 &   0.01702226 \\
   1.30 &  0.14477275 &  0.00028493 &  0.00100876 &   0.03348390 &   0.00776547 &   0.01408604 &   0.01408604 &   0.01408604 \\
   1.40 &  0.15255537 &  0.00025173 &  0.00093851 &   0.03289920 &   0.00726190 &   0.01369987 &   0.01369987 &   0.01369987 \\
   1.50 &  0.15482117 &  0.00000000 &  0.00000000 &   0.03063335 &   0.00022130 &   0.01028488 &   0.01028488 &   0.01028488 \\
   1.60 &  0.22931682 &  0.00000000 &  0.00000000 &   0.03446039 &   0.00013123 &   0.01153054 &   0.01153054 &   0.01153054 \\
   1.70 &  0.29105982 &  0.00000000 &  0.00000000 &   0.03601382 &   0.00006852 &   0.01202745 &   0.01202745 &   0.01202745 \\
   1.80 &  0.32492563 &  0.00000000 &  0.00000000 &   0.03546248 &   0.00005612 &   0.01183953 &   0.01183953 &   0.01183953 \\
   1.90 &  0.35826111 &  0.00000000 &  0.00000000 &   0.03614184 &   0.00003385 &   0.01205856 &   0.01205856 &   0.01205856 \\
   2.00 &  0.39031755 &  0.00000000 &  0.00000000 &   0.03749559 &   0.00003115 &   0.01250891 &   0.01250891 &   0.01250891 \\
   2.10 &  0.42299855 &  0.00000000 &  0.00000000 &   0.03821450 &   0.00002576 &   0.01274675 &   0.01274675 &   0.01274675 \\
   2.20 &  0.43290204 &  0.00000000 &  0.00000000 &   0.03648384 &   0.00002441 &   0.01216942 &   0.01216942 &   0.01216942 \\
   2.30 &  0.45596038 &  0.00000000 &  0.00000000 &   0.03759902 &   0.00002318 &   0.01254073 &   0.01254073 &   0.01254073 \\
   2.40 &  0.46422409 &  0.00000000 &  0.00000000 &   0.04062868 &   0.00002216 &   0.01355028 &   0.01355028 &   0.01355028 \\
   2.50 &  0.48453652 &  0.00000000 &  0.00000000 &   0.03974079 &   0.00002097 &   0.01325392 &   0.01325392 &   0.01325392 \\
   3.00 &  0.57992976 &  0.00000000 &  0.00000000 &   0.04517592 &   0.00001616 &   0.01506402 &   0.01506402 &   0.01506402 \\
   3.50 &  0.29496188 &  0.00000000 &  0.00000000 &   0.06281384 &   0.00003823 &   0.02095069 &   0.02095069 &   0.02095069 \\
   4.00 &  0.86459902 &  0.00000000 &  0.00000000 &   0.06980109 &   0.00003735 &   0.02327948 &   0.02327948 &   0.02327948 \\
   4.50 &  0.03021447 &  0.00779002 &  0.01101550 &   0.10334552 &   0.14612063 &   0.08682722 &   0.08682722 &   0.08682722 \\
   5.00 &  0.03000158 &  0.00816375 &  0.00846570 &   0.10708606 &   0.14715856 &   0.08757011 &   0.08757011 &   0.08757011 \\
   5.50 &  0.02708609 &  0.00923751 &  0.00761757 &   0.10935401 &   0.14469343 &   0.08722167 &   0.08722167 &   0.08722167 \\
   6.00 &  0.02422391 &  0.01037559 &  0.00670013 &   0.11334449 &   0.14625306 &   0.08876589 &   0.08876589 &   0.08876589 \\
   6.50 &  0.02282019 &  0.01068092 &  0.00606410 &   0.11320376 &   0.14183756 &   0.08703514 &   0.08703514 &   0.08703514 \\
   7.00 &  0.02128779 &  0.01095849 &  0.00561900 &   0.11206287 &   0.13593213 &   0.08453800 &   0.08453800 &   0.08453800 \\
\hline
\end{tabular}                                                                                                      \end{center}
\end{table*}
\renewcommand{\arraystretch}{1}

\renewcommand{\arraystretch}{1.05}
\begin{table*}
\footnotesize
\begin{center}
\caption[]{Dust condensation efficiencies for AGB stars of metallicity Z=0.015.} \begin{tabular}{ccccccccc}
\hline \hline \vspace{0.1cm} M$_{\odot}$ & $\delta_{C}$ & $\delta_{O}$
& $\delta_{Mg} $ & $\delta_{Si}$& $\delta_{Fe} $ & $\delta_{Ca}$& $\delta_{S} $ & $\delta_{Al}$ \\   \hline
   1.00 &  0.00000000 &  0.01622151 &  0.10337936 &   0.19067569 &   0.26824351 &   0.18743286 &   0.18743286 &   0.18743286 \\
   1.10 &  0.00000000 &  0.01989686 &  0.11994899 &   0.23451746 &   0.31372527 &   0.22273057 &   0.22273057 &   0.22273057 \\
   1.20 &  0.00000000 &  0.01897280 &  0.10806979 &   0.22449623 &   0.29882128 &   0.21046243 &   0.21046243 &   0.21046243 \\
   1.25 &  0.00000000 &  0.01940105 &  0.10893929 &   0.23035203 &   0.29258844 &   0.21062658 &   0.21062658 &   0.21062658 \\
   1.30 &  0.00000000 &  0.02117347 &  0.11832043 &   0.25223618 &   0.31781088 &   0.22945583 &   0.22945583 &   0.22945583 \\
   1.40 &  0.00000000 &  0.01880107 &  0.10771517 &   0.22584695 &   0.30122424 &   0.21159545 &   0.21159545 &   0.21159545 \\
   1.50 &  0.01612098 &  0.01844121 &  0.10805736 &   0.24602971 &   0.31441874 &   0.22283527 &   0.22283527 &   0.22283527 \\
   1.60 &  0.00187487 &  0.01829934 &  0.10631659 &   0.22389462 &   0.33054988 &   0.22025369 &   0.22025369 &   0.22025369 \\
   1.70 &  0.03372210 &  0.01095275 &  0.06322135 &   0.18120554 &   0.55583957 &   0.26675549 &   0.26675549 &   0.26675549 \\
   1.80 &  0.05378203 &  0.00771053 &  0.00727006 &   0.24215832 &   0.11816384 &   0.12253074 &   0.12253074 &   0.12253074 \\
   1.90 &  0.10033417 &  0.00123331 &  0.00688251 &   0.17667234 &   0.04297937 &   0.07551141 &   0.07551141 &   0.07551141 \\
   2.00 &  0.17177962 &  0.00117904 &  0.00650155 &   0.33888422 &   0.15096066 &   0.16544881 &   0.16544881 &   0.16544881 \\
   2.10 &  0.22003572 &  0.00084726 &  0.00461739 &   0.31107965 &   0.02916475 &   0.11495393 &   0.11495393 &   0.11495393 \\
   2.20 &  0.26818177 &  0.00062284 &  0.00344337 &   0.29958994 &   0.02299745 &   0.10867692 &   0.10867692 &   0.10867692 \\
   2.30 &  0.31008388 &  0.00141144 &  0.00000089 &   0.32679560 &   0.00348235 &   0.11009295 &   0.11009295 &   0.11009295 \\
   2.40 &  0.34559179 &  0.00000010 &  0.00000059 &   0.28613619 &   0.00427951 &   0.09680543 &   0.09680543 &   0.09680543 \\
   2.50 &  0.36478837 &  0.00000000 &  0.00000000 &   0.29653113 &   0.00712501 &   0.10121871 &   0.10121871 &   0.10121871 \\
   3.00 &  0.53051299 &  0.00000000 &  0.00000000 &   0.29465242 &   0.00022888 &   0.09829376 &   0.09829376 &   0.09829376 \\
   3.50 &  0.34012190 &  0.00000000 &  0.00000000 &   0.33390767 &   0.00067675 &   0.11152814 &   0.11152814 &   0.11152814 \\
   4.00 &  0.28373240 &  0.00000000 &  0.00000000 &   0.40260222 &   0.00445742 &   0.13568654 &   0.13568654 &   0.13568654 \\
   4.50 &  0.00883125 &  0.02274328 &  0.06481755 &   0.27767137 &   0.26289180 &   0.20179357 &   0.20179357 &   0.20179357 \\
   5.00 &  0.00763527 &  0.02265625 &  0.05189337 &   0.27938663 &   0.24193878 &   0.19107293 &   0.19107293 &   0.19107293 \\
   5.50 &  0.00685113 &  0.02352466 &  0.04558380 &   0.27452172 &   0.23402431 &   0.18470995 &   0.18470995 &   0.18470995 \\
   6.00 &  0.00577691 &  0.02453750 &  0.03980196 &   0.26985066 &   0.22267650 &   0.17744304 &   0.17744304 &   0.17744304 \\
   6.50 &  0.00555197 &  0.02492647 &  0.03636159 &   0.26652971 &   0.21356319 &   0.17215149 &   0.17215149 &   0.17215149 \\
   7.00 &  0.00502738 &  0.02528332 &  0.03375876 &   0.26303478 &   0.20576146 &   0.16751834 &   0.16751834 &   0.16751834 \\
\hline
\end{tabular}
\end{center}
\end{table*}
\renewcommand{\arraystretch}{1}

\begin{table*}
\footnotesize
\begin{center}
\caption[]{Dust condensation efficiencies for AGB stars of metallicity Z=0.02.} \begin{tabular}{ccccccccc}
\hline \hline \vspace{0.1cm} M$_{\odot}$ & $\delta_{C}$ & $\delta_{O}$
& $\delta_{Mg} $ & $\delta_{Si}$& $\delta_{Fe} $ & $\delta_{Ca}$& $\delta_{S} $ & $\delta_{Al}$ \\   \hline
   1.00 &  0.00000000 &  0.03588087 &  0.28188633 & 0.42317978 & 0.48590908 & 0.39699173 &  0.39699173 &   0.39699173 \\
   1.10 &  0.00000000 &  0.03744050 &  0.27525320 & 0.44111016 & 0.51543185 & 0.41059841 &  0.41059841 &   0.41059841 \\
   1.20 &  0.00000000 &  0.04159802 &  0.28644563 & 0.48998219 & 0.54661994 & 0.44101592 &  0.44101592 &   0.44101592 \\
   1.25 &  0.00000000 &  0.04468728 &  0.30066679 & 0.52663114 & 0.58942348 & 0.47224047 &  0.47224047 &   0.47224047 \\
   1.30 &  0.00000000 &  0.03918550 &  0.26010042 & 0.46278830 & 0.51112469 & 0.41133780 &  0.41133780 &   0.41133780 \\
   1.40 &  0.00000000 &  0.03815650 &  0.25531810 & 0.45374694 & 0.49276080 & 0.40060861 &  0.40060861 &   0.40060861 \\
   1.50 &  0.00000000 &  0.03755734 &  0.25306780 & 0.44863829 & 0.50755628 & 0.40308746 &  0.40308746 &   0.40308746 \\
   1.60 &  0.00000000 &  0.03523968 &  0.23093010 & 0.42641801 & 0.45943436 & 0.37226082 &  0.37226082 &   0.37226082 \\
   1.70 &  0.01731944 &  0.03372272 &  0.21993045 & 0.42843372 & 0.46029551 & 0.36955323 &  0.36955323 &   0.36955323 \\
   1.80 &  0.00747658 &  0.03313635 &  0.21398305 & 0.42163598 & 0.49490421 & 0.37684108 &  0.37684108 &   0.37684108 \\
   1.90 &  0.03012721 &  0.03244870 &  0.20786083 & 0.43504426 & 0.55978612 & 0.40089707 &  0.40089707 &   0.40089707 \\
   2.00 &  0.03783436 &  0.02008472 &  0.12747051 & 0.27485541 & 0.50827910 & 0.30353501 &  0.30353501 &   0.30353501 \\
   2.10 &  0.07491230 &  0.00273484 &  0.01705400 & 0.26454065 & 0.41019760 & 0.23059742 &  0.23059742 &   0.23059742 \\
   2.20 &  0.10537837 &  0.00442499 &  0.01136481 & 0.33324845 & 0.16352652 & 0.16937993 &  0.16937993 &   0.16937993 \\
   2.30 &  0.16635710 &  0.00219678 &  0.00294874 & 0.41019668 & 0.09997012 & 0.17103852 &  0.17103852 &   0.17103852 \\
   2.40 &  0.28411965 &  0.00072087 &  0.00000080 & 0.55655533 & 0.13558008 & 0.23071207 &  0.23071207 &   0.23071207 \\
   2.50 &  0.30505662 &  0.00072684 &  0.00000000 & 0.55319755 & 0.02987900 & 0.19435885 &  0.19435885 &   0.19435885 \\
   3.00 &  0.47605672 &  0.00000000 &  0.00000000 & 0.49814355 & 0.00150995 & 0.16655116 &  0.16655116 &   0.16655116 \\
   3.50 &  0.38344394 &  0.00000000 &  0.00000000 & 0.54722350 & 0.00311482 & 0.18344611 &  0.18344611 &   0.18344611 \\
   4.00 &  0.30805414 &  0.00000005 &  0.00000028 & 0.60713409 & 0.00867781 & 0.20527073 &  0.20527073 &   0.20527073 \\
   4.50 &  0.00340680 &  0.03293037 &  0.13415468 & 0.39329761 & 0.32212277 & 0.28319169 &  0.28319169 &   0.28319169 \\
   5.00 &  0.00211536 &  0.03241885 &  0.11598430 & 0.38785476 & 0.29799660 & 0.26727855 &  0.26727855 &   0.26727855 \\
   5.50 &  0.00284376 &  0.03312643 &  0.10431749 & 0.37975338 & 0.27933865 & 0.25446984 &  0.25446984 &   0.25446984 \\
   6.00 &  0.00253550 &  0.03402182 &  0.09563033 & 0.37537577 & 0.26596970 & 0.24565860 &  0.24565860 &   0.24565860 \\
   6.50 &  0.00224815 &  0.03448482 &  0.09007775 & 0.36997262 & 0.25614723 & 0.23873253 &  0.23873253 &   0.23873253 \\
   7.00 &  0.00202929 &  0.03486064 &  0.08524557 & 0.36396209 & 0.24701413 & 0.23207393 &  0.23207393 &   0.23207393 \\
\hline
\end{tabular}
\end{center}
\end{table*}
\renewcommand{\arraystretch}{1}

\renewcommand{\arraystretch}{1.05}
\begin{table*}
\footnotesize
\begin{center}
\caption[]{Dust condensation efficiencies for AGB stars of metallicity Z=0.03.} \begin{tabular}{ccccccccc}
\hline \hline \vspace{0.1cm} M$_{\odot}$ & $\delta_{C}$ & $\delta_{O}$
& $\delta_{Mg} $ & $\delta_{Si}$& $\delta_{Fe} $ & $\delta_{Ca}$& $\delta_{S} $ & $\delta_{Al}$ \\   \hline
   1.00 &  0.00000000 &  0.07082836 &  0.75108259 & 0.83103533 & 0.98354473 & 0.85522088 & 0.85522088 & 0.85522088 \\
   1.10 &  0.00000000 &  0.07613366 &  0.75363042 & 0.89351633 & 1.00000000 & 0.88238225 & 0.88238225 & 0.88238225 \\
   1.20 &  0.00000000 &  0.07597349 &  0.69104898 & 0.89162629 & 1.00000000 & 0.86089176 & 0.86089176 & 0.86089176 \\
   1.25 &  0.00000000 &  0.07699561 &  0.67622678 & 0.90447559 & 1.00000000 & 0.86023412 & 0.86023412 & 0.86023412 \\
   1.30 &  0.00000000 &  0.07586636 &  0.64606965 & 0.89227962 & 0.98093727 & 0.83976218 & 0.83976218 & 0.83976218 \\
   1.40 &  0.00000000 &  0.07277666 &  0.59900888 & 0.85792214 & 0.92814101 & 0.79502401 & 0.79502401 & 0.79502401 \\
   1.50 &  0.00000000 &  0.06978718 &  0.56132950 & 0.82496602 & 0.90455843 & 0.76361798 & 0.76361798 & 0.76361798 \\
   1.60 &  0.00000000 &  0.06493798 &  0.51116937 & 0.77327433 & 0.87573021 & 0.72005797 & 0.72005797 & 0.72005797 \\
   1.70 &  0.00750569 &  0.06582141 &  0.50562327 & 0.78475048 & 0.79965006 & 0.69667461 & 0.69667461 & 0.69667461 \\
   1.80 &  0.01899001 &  0.06746586 &  0.51131057 & 0.83609174 & 0.79478610 & 0.71406280 & 0.71406280 & 0.71406280 \\
   1.90 &  0.01806721 &  0.06679816 &  0.50277474 & 0.86088724 & 0.87817759 & 0.74727986 & 0.74727986 & 0.74727986 \\
   2.00 &  0.03160594 &  0.06807771 &  0.50599904 & 0.85074611 & 0.77483299 & 0.71052605 & 0.71052605 & 0.71052605 \\
   2.10 &  0.03794097 &  0.06224528 &  0.46910262 & 0.79211867 & 0.93645211 & 0.73255780 & 0.73255780 & 0.73255780 \\
   2.20 &  0.04378164 &  0.04456768 &  0.33003823 & 0.55757792 & 0.96248021 & 0.61669879 & 0.61669879 & 0.61669879 \\
   2.30 &  0.05061630 &  0.04602102 &  0.04360384 & 0.98405163 & 0.65148339 & 0.55971296 & 0.55971296 & 0.55971296 \\
   2.40 &  0.06320275 &  0.00576956 &  0.01617968 & 0.29795188 & 0.96971374 & 0.42794843 & 0.42794843 & 0.42794843 \\
   2.50 &  0.11939945 &  0.00143085 &  0.01141116 & 0.35561157 & 0.58280965 & 0.31661079 & 0.31661079 & 0.31661079 \\
   3.00 &  0.42346974 &  0.00000002 &  0.00000022 & 0.84328022 & 0.04839004 & 0.29722349 & 0.29722349 & 0.29722349 \\
   3.50 &  0.42175841 &  0.00000006 &  0.00000046 & 0.80959367 & 0.02271603 & 0.27743672 & 0.27743672 & 0.27743672 \\
   4.00 &  0.42738901 &  0.00000008 &  0.00000057 & 0.83127976 & 0.01961049 & 0.28363028 & 0.28363028 & 0.28363028 \\
   4.50 &  0.00227282 &  0.05200890 &  0.32983869 & 0.62278143 & 0.39312278 & 0.44858097 & 0.44858097 & 0.44858097 \\
   5.00 &  0.00158088 &  0.05099778 &  0.26781185 & 0.60590159 & 0.37157343 & 0.41509562 & 0.41509562 & 0.41509562 \\
   5.50 &  0.00000000 &  0.05080716 &  0.29316400 & 0.59229197 & 0.38538367 & 0.42361321 & 0.42361321 & 0.42361321 \\
   6.00 &  0.00058438 &  0.05156336 &  0.29533909 & 0.59156759 & 0.34692253 & 0.41127640 & 0.41127640 & 0.41127640 \\
   6.50 &  0.00059364 &  0.05209715 &  0.29094937 & 0.58501816 & 0.34266797 & 0.40621183 & 0.40621183 & 0.40621183 \\
   7.00 &  0.00039084 &  0.05278038 &  0.28655054 & 0.57970047 & 0.33187034 & 0.39937378 & 0.39937378 & 0.39937378 \\
\hline
\end{tabular}
\end{center}
\end{table*}
\renewcommand{\arraystretch}{1}

\renewcommand{\arraystretch}{1.05}
\begin{table*}
\footnotesize
\begin{center}
\caption[]{Dust condensation efficiencies for AGB stars of metallicity Z=0.04.} \begin{tabular}{ccccccccc}
\hline \hline \vspace{0.1cm} M$_{\odot}$ & $\delta_{C}$ & $\delta_{O}$
& $\delta_{Mg} $ & $\delta_{Si}$& $\delta_{Fe} $ & $\delta_{Ca}$& $\delta_{S} $ & $\delta_{Al}$ \\   \hline
   1.00 &  0.00000000 &  0.09575488 &  1.00000000 & 1.00000000 & 1.00000000 & 1.00000000 & 1.00000000 & 1.00000000  \\
   1.10 &  0.00000000 &  0.11433025 &  1.00000000 & 1.00000000 & 1.00000000 & 1.00000000 & 1.00000000 & 1.00000000  \\
   1.20 &  0.00000000 &  0.11894327 &  1.00000000 & 1.00000000 & 1.00000000 & 1.00000000 & 1.00000000 & 1.00000000  \\
   1.25 &  0.00000000 &  0.11149659 &  1.00000000 & 1.00000000 & 1.00000000 & 1.00000000 & 1.00000000 & 1.00000000  \\
   1.30 &  0.00000000 &  0.11257290 &  1.00000000 & 1.00000000 & 1.00000000 & 1.00000000 & 1.00000000 & 1.00000000  \\
   1.40 &  0.00000000 &  0.10862042 &  1.00000000 & 1.00000000 & 1.00000000 & 1.00000000 & 1.00000000 & 1.00000000  \\
   1.50 &  0.00000000 &  0.10340609 &  0.92197638 & 1.00000000 & 1.00000000 & 0.97399212 & 0.97399212 & 0.97399212  \\
   1.60 &  0.00000000 &  0.09942303 &  0.85732739 & 1.00000000 & 1.00000000 & 0.95244246 & 0.95244246 & 0.95244246  \\
   1.70 &  0.00000000 &  0.10355969 &  0.87022826 & 1.00000000 & 1.00000000 & 0.95674275 & 0.95674275 & 0.95674275  \\
   1.80 &  0.00000000 &  0.09636815 &  0.80451336 & 1.00000000 & 1.00000000 & 0.93483778 & 0.93483778 & 0.93483778  \\
   1.90 &  0.00000000 &  0.09577716 &  0.79577738 & 1.00000000 & 1.00000000 & 0.93192579 & 0.93192579 & 0.93192579  \\
   2.00 &  0.00000000 &  0.10104673 &  0.82643983 & 1.00000000 & 1.00000000 & 0.94214661 & 0.94214661 & 0.94214661  \\
   2.10 &  0.03748761 &  0.09204577 &  0.74974691 & 1.00000000 & 1.00000000 & 0.91658230 & 0.91658230 & 0.91658230  \\
   2.20 &  0.02558362 &  0.08406925 &  0.68316619 & 1.00000000 & 1.00000000 & 0.89438873 & 0.89438873 & 0.89438873  \\
   2.30 &  0.04100302 &  0.07695840 &  0.62984640 & 1.00000000 & 1.00000000 & 0.87661546 & 0.87661546 & 0.87661546  \\
   2.40 &  0.04817184 &  0.07991250 &  0.17760612 & 1.00000000 & 0.88695599 & 0.68818737 & 0.68818737 & 0.68818737  \\
   2.50 &  0.04149373 &  0.02841000 &  0.02179646 & 0.66093798 & 1.00000000 & 0.56091148 & 0.56091148 & 0.56091148  \\
   3.00 &  0.36332013 &  0.00000014 &  0.00000130 & 1.00000000 & 0.81994561 & 0.60664897 & 0.60664897 & 0.60664897  \\
   3.50 &  0.37492885 &  0.00000012 &  0.00000100 & 0.98702456 & 0.13749214 & 0.37483924 & 0.37483924 & 0.37483924  \\
   4.00 &  0.38434334 &  0.00000019 &  0.00000142 & 1.00000000 & 0.08887748 & 0.36295963 & 0.36295963 & 0.36295963  \\
   4.50 &  0.00287962 &  0.07073514 &  0.53140808 & 0.84193033 & 0.52269274 & 0.63201038 & 0.63201038 & 0.63201038  \\
   5.00 &  0.00000000 &  0.06938901 &  0.51852810 & 0.82662466 & 0.49251262 & 0.61255513 & 0.61255513 & 0.61255513  \\
   5.50 &  0.00077662 &  0.06863618 &  0.44163665 & 0.81843120 & 0.46734515 & 0.57580433 & 0.57580433 & 0.57580433  \\
   6.00 &  0.00028089 &  0.06881210 &  0.57364937 & 0.81409916 & 0.45906608 & 0.61560487 & 0.61560487 & 0.61560487  \\
   6.50 &  0.00035566 &  0.06955029 &  0.56709971 & 0.80625341 & 0.43526870 & 0.60287394 & 0.60287394 & 0.60287394  \\
   7.00 &  0.00000000 &  0.07071037 &  0.56273765 & 0.80240491 & 0.43006581 & 0.59840279 & 0.59840279 & 0.59840279  \\
\hline
\end{tabular}
\end{center}
\end{table*}
\renewcommand{\arraystretch}{1}

\renewcommand{\arraystretch}{1.05}
\begin{table*}
\footnotesize
\begin{center}
\caption[]{Dust condensation efficiencies for CCSN{\ae} and PISN{\ae} with environmental density of hydrogen n$_{H}$=0.1cm$^{-3}$. Unmixed model. Since for masses under 10M$_{\odot}$ no models are available we kept
the condensation efficiencies obtained for the 10M$_{\odot}$.} \begin{tabular}{cccccccc}          \hline \hline \vspace{0.1cm} M$_{\odot}$ & $\delta_{C}$ & $\delta_{O}$
& $\delta_{Mg} $ & $\delta_{Si}$& $\delta_{S} $ & $\delta_{Ca}$& $\delta_{Fe}$ \\
\hline
    8.0 &    0.761600 &    0.079599 &   0.240478 &    0.580435 &    0.017779 &    0.352828 &    0.572620  \\
    9.0 &    0.761600 &    0.079599 &   0.240478 &    0.580435 &    0.017779 &    0.352828 &    0.572620  \\
   10.0 &    0.761600 &    0.079599 &   0.240478 &    0.580435 &    0.017779 &    0.352828 &    0.572620  \\
   15.0 &    0.728914 &    0.045431 &   0.383057 &    0.593230 &    0.030652 &    0.355489 &    0.415019  \\
   20.0 &    0.683200 &    0.034797 &   0.440417 &    0.627777 &    0.092563 &    0.359057 &    0.275471  \\
   25.0 &    0.736000 &    0.069298 &   0.548345 &    0.728713 &    0.145156 &    0.459010 &    0.413825  \\
   30.0 &    0.560800 &    0.037068 &   0.401763 &    0.639319 &    0.135128 &    0.294052 &    0.000000  \\
   35.0 &    0.560815 &    0.034143 &   0.394543 &    0.623800 &    0.139907 &    0.289562 &    0.000000  \\
   40.0 &    0.560861 &    0.027818 &   0.385938 &    0.615143 &    0.131841 &    0.283230 &    0.000000  \\
   45.0 &    0.560938 &    0.023009 &   0.328929 &    0.598507 &    0.105834 &    0.258318 &    0.000000  \\
   50.0 &    0.561045 &    0.020911 &   0.300423 &    0.579052 &    0.086477 &    0.241488 &    0.000000  \\
   55.0 &    0.561183 &    0.020168 &   0.290482 &    0.558445 &    0.071386 &    0.230078 &    0.000000  \\
   60.0 &    0.561351 &    0.020142 &   0.289856 &    0.538117 &    0.059253 &    0.312632 &    0.363304  \\
   65.0 &    0.561550 &    0.020501 &   0.293767 &    0.519060 &    0.049295 &    0.311544 &    0.384056  \\
   70.0 &    0.561780 &    0.021062 &   0.299610 &    0.501823 &    0.041013 &    0.311025 &    0.401653  \\
   75.0 &    0.562040 &    0.021721 &   0.305915 &    0.486621 &    0.034067 &    0.310401 &    0.415001  \\
   80.0 &    0.562331 &    0.022416 &   0.311832 &    0.473449 &    0.028213 &    0.309334 &    0.423840  \\
   85.0 &    0.562652 &    0.023109 &   0.306223 &    0.462181 &    0.023270 &    0.305019 &    0.428402  \\
   90.0 &    0.563004 &    0.023778 &   0.299257 &    0.452633 &    0.019094 &    0.300035 &    0.429157  \\
   95.0 &    0.563387 &    0.024408 &   0.292389 &    0.444605 &    0.015570 &    0.294806 &    0.426659  \\
  100.0 &    0.563800 &    0.024992 &   0.285570 &    0.437902 &    0.012604 &    0.289384 &    0.421459  \\
  110.0 &    0.564718 &    0.026004 &   0.271909 &    0.427778 &    0.008036 &    0.278165 &    0.404936  \\
  120.0 &    0.565759 &    0.026800 &   0.257972 &    0.421099 &    0.004879 &    0.266737 &    0.382999  \\
  130.0 &    0.566922 &    0.027390 &   0.243487 &    0.404934 &    0.002754 &    0.252359 &    0.358260  \\
  140.0 &    0.568208 &    0.027807 &   0.228194 &    0.375880 &    0.001381 &    0.234548 &    0.332739  \\
  150.0 &    0.569616 &    0.030000 &   0.211937 &    0.492890 &    0.003355 &    0.237831 &    0.243140  \\
  160.0 &    0.571147 &    0.037363 &   0.194414 &    0.512762 &    0.000679 &    0.227533 &    0.202279  \\
  170.0 &    0.572800 &    0.046971 &   0.175410 &    0.465367 &    0.000000 &    0.204242 &    0.176192  \\
  180.0 &    0.582523 &    0.046626 &   0.155560 &    0.449846 &    0.000000 &    0.191561 &    0.160840  \\
  190.0 &    0.604258 &    0.043842 &   0.135302 &    0.355407 &    0.000000 &    0.158891 &    0.144855  \\
  200.0 &    0.632000 &    0.044636 &   0.114842 &    0.209691 &    0.000000 &    0.114537 &    0.133616  \\
  210.0 &    0.632000 &    0.049919 &   0.114534 &    0.209710 &    0.000000 &    0.114942 &    0.135523  \\
  220.0 &    0.632000 &    0.053117 &   0.114203 &    0.207845 &    0.000000 &    0.113978 &    0.133862  \\
  230.0 &    0.632000 &    0.059726 &   0.106786 &    0.207785 &    0.000000 &    0.111808 &    0.132663  \\
  240.0 &    0.632000 &    0.065658 &   0.100796 &    0.207621 &    0.000000 &    0.109667 &    0.130252  \\
  250.0 &    0.632000 &    0.072806 &   0.096465 &    0.207607 &    0.000000 &    0.108192 &    0.128694  \\
  260.0 &    0.632000 &    0.079860 &   0.093527 &    0.205513 &    0.000000 &    0.106698 &    0.127754  \\
  270.0 &    0.632000 &    0.086956 &   0.088870 &    0.205730 &    0.000000 &    0.105588 &    0.127754  \\ \hline
\end{tabular}
\end{center}
\end{table*}
\renewcommand{\arraystretch}{1}

\renewcommand{\arraystretch}{1.05}
\begin{table*}
\footnotesize
\begin{center}
\caption[]{Dust condensation efficiencies for CCSN{\ae} and PISN{\ae} with environmental density of hydrogen n$_{H}$=1cm$^{-3}$. Unmixed model. Since for masses under 10M$_{\odot}$ no models are available we kept
the condensation efficiencies obtained for the 10M$_{\odot}$.}
\begin{tabular}{cccccccc}          \hline \hline \vspace{0.1cm} M$_{\odot}$ & $\delta_{C}$ & $\delta_{O}$
& $\delta_{Mg} $ & $\delta_{Si}$& $\delta_{S} $ & $\delta_{Ca}$& $\delta_{Fe}$ \\   \hline
    8.0 &    0.384000 &    0.008791 &   0.011565 &    0.222091 &    0.000000 &    0.109540 &    0.204505  \\
    9.0 &    0.384000 &    0.008791 &   0.011565 &    0.222091 &    0.000000 &    0.109540 &    0.204505  \\
   10.0 &    0.384000 &    0.008791 &   0.011565 &    0.222091 &    0.000000 &    0.109540 &    0.204505  \\
   15.0 &    0.393024 &    0.004674 &   0.023645 &    0.250208 &    0.000068 &    0.105031 &    0.146205  \\
   20.0 &    0.442400 &    0.003497 &   0.039002 &    0.318588 &    0.000778 &    0.112494 &    0.091607  \\
   25.0 &    0.572800 &    0.030027 &   0.163564 &    0.502400 &    0.007565 &    0.213708 &    0.181303  \\
   30.0 &    0.126400 &    0.005055 &   0.021344 &    0.372298 &    0.011137 &    0.101195 &    0.000000  \\
   35.0 &    0.118441 &    0.004327 &   0.020441 &    0.346292 &    0.011531 &    0.094566 &    0.000000  \\
   40.0 &    0.110897 &    0.003307 &   0.018818 &    0.328854 &    0.010866 &    0.089635 &    0.000000  \\
   45.0 &    0.103753 &    0.002471 &   0.014699 &    0.302104 &    0.008723 &    0.081381 &    0.000000  \\
   50.0 &    0.096994 &    0.002015 &   0.012299 &    0.273533 &    0.007127 &    0.073240 &    0.000000  \\
   55.0 &    0.090605 &    0.001748 &   0.010922 &    0.245334 &    0.005883 &    0.065535 &    0.000000  \\
   60.0 &    0.084573 &    0.001582 &   0.010032 &    0.219029 &    0.004883 &    0.085645 &    0.108635  \\
   65.0 &    0.078882 &    0.001469 &   0.009365 &    0.195426 &    0.004063 &    0.084041 &    0.127310  \\
   70.0 &    0.073516 &    0.001384 &   0.008789 &    0.174773 &    0.003380 &    0.082018 &    0.141130  \\
   75.0 &    0.068463 &    0.001313 &   0.008237 &    0.156966 &    0.002808 &    0.079383 &    0.149520  \\
   80.0 &    0.063706 &    0.001247 &   0.007677 &    0.141722 &    0.002325 &    0.076115 &    0.152737  \\
   85.0 &    0.059231 &    0.001181 &   0.006860 &    0.128695 &    0.001918 &    0.072232 &    0.151457  \\
   90.0 &    0.055024 &    0.001114 &   0.006063 &    0.117540 &    0.001574 &    0.067920 &    0.146501  \\
   95.0 &    0.051068 &    0.001043 &   0.005321 &    0.107944 &    0.001283 &    0.063308 &    0.138684  \\
  100.0 &    0.047351 &    0.000969 &   0.004631 &    0.099633 &    0.001039 &    0.058512 &    0.128745  \\
  110.0 &    0.040570 &    0.000811 &   0.003398 &    0.085988 &    0.000662 &    0.048756 &    0.104977  \\
  120.0 &    0.034563 &    0.000645 &   0.002357 &    0.075184 &    0.000402 &    0.039298 &    0.079248  \\
  130.0 &    0.029212 &    0.000479 &   0.001507 &    0.064294 &    0.000227 &    0.030121 &    0.054455  \\
  140.0 &    0.024398 &    0.000324 &   0.000847 &    0.052817 &    0.000114 &    0.021625 &    0.032722  \\
  150.0 &    0.020003 &    0.000211 &   0.000376 &    0.091431 &    0.000277 &    0.026143 &    0.012488  \\
  160.0 &    0.015910 &    0.000131 &   0.000094 &    0.089070 &    0.000056 &    0.023149 &    0.003375  \\
  170.0 &    0.012000 &    0.000107 &   0.000000 &    0.066617 &    0.000000 &    0.016778 &    0.000495  \\
  180.0 &    0.008553 &    0.000168 &   0.000000 &    0.049857 &    0.000000 &    0.012765 &    0.001203  \\
  190.0 &    0.005723 &    0.000305 &   0.000000 &    0.027392 &    0.000000 &    0.007572 &    0.002896  \\
  200.0 &    0.003200 &    0.000519 &   0.000000 &    0.001629 &    0.000000 &    0.001683 &    0.005103  \\
  210.0 &    0.003200 &    0.000593 &   0.000000 &    0.001625 &    0.000000 &    0.001700 &    0.005176  \\
  220.0 &    0.003200 &    0.000636 &   0.000000 &    0.001585 &    0.000000 &    0.001674 &    0.005113  \\
  230.0 &    0.003200 &    0.000727 &   0.000000 &    0.001581 &    0.000000 &    0.001662 &    0.005067  \\
  240.0 &    0.003200 &    0.000807 &   0.000000 &    0.001576 &    0.000000 &    0.001638 &    0.004975  \\
  250.0 &    0.003200 &    0.000903 &   0.000000 &    0.001574 &    0.000000 &    0.001622 &    0.004915  \\
  260.0 &    0.003200 &    0.000995 &   0.000000 &    0.001530 &    0.000000 &    0.001602 &    0.004879  \\
  270.0 &    0.003200 &    0.001091 &   0.000000 &    0.001534 &    0.000000 &    0.001603 &    0.004879  \\ \hline
\end{tabular}
\end{center}
\end{table*}
\renewcommand{\arraystretch}{1}

 \renewcommand{\arraystretch}{1.05}
 \begin{table*}
 \footnotesize
 \begin{center}
 \caption[]{Dust condensation efficiencies for CCSN{\ae} and PISN{\ae} with environmental density of hydrogen n$_{H}$=10cm$^{-3}$. Unmixed model. Since for masses under 10M$_{\odot}$ no models are available we kept
the condensation efficiencies obtained for the 10M$_{\odot}$.}
 \begin{tabular}{cccccccc}          \hline \hline \vspace{0.1cm} M$_{\odot}$ & $\delta_{C}$ & $\delta_{O}$
 & $\delta_{Mg} $ & $\delta_{Si}$& $\delta_{S} $ & $\delta_{Ca}$& $\delta_{Fe}$ \\   \hline
    8.0 &    0.085600 &    0.000287 &   0.000000 &    0.064344 &    0.000000 &    0.023620 &    0.030138  \\
    9.0 &    0.085600 &    0.000287 &   0.000000 &    0.064344 &    0.000000 &    0.023620 &    0.030138  \\
   10.0 &    0.085600 &    0.000287 &   0.000000 &    0.064344 &    0.000000 &    0.023620 &    0.030138  \\
   15.0 &    0.084599 &    0.000090 &   0.000000 &    0.071024 &    0.000000 &    0.022832 &    0.020303  \\
   20.0 &    0.083200 &    0.000000 &   0.000000 &    0.090492 &    0.000000 &    0.025242 &    0.010478  \\
   25.0 &    0.218400 &    0.000103 &   0.001033 &    0.166885 &    0.000000 &    0.051658 &    0.038714  \\
   30.0 &    0.008000 &    0.000651 &   0.000000 &    0.158752 &    0.000000 &    0.039688 &    0.000000  \\
   35.0 &    0.007439 &    0.000585 &   0.000000 &    0.152355 &    0.000000 &    0.038089 &    0.000000  \\
   40.0 &    0.006898 &    0.000475 &   0.000000 &    0.148896 &    0.000000 &    0.037224 &    0.000000  \\
   45.0 &    0.006378 &    0.000372 &   0.000000 &    0.139471 &    0.000000 &    0.034868 &    0.000000  \\
   50.0 &    0.005878 &    0.000316 &   0.000000 &    0.127779 &    0.000000 &    0.031945 &    0.000000  \\
   55.0 &    0.005398 &    0.000286 &   0.000000 &    0.115116 &    0.000000 &    0.028779 &    0.000000  \\
   60.0 &    0.004939 &    0.000269 &   0.000000 &    0.102513 &    0.000000 &    0.034057 &    0.033713  \\
   65.0 &    0.004500 &    0.000260 &   0.000000 &    0.090634 &    0.000000 &    0.033253 &    0.042379  \\
   70.0 &    0.004082 &    0.000256 &   0.000000 &    0.079811 &    0.000000 &    0.032318 &    0.049461  \\
   75.0 &    0.003684 &    0.000254 &   0.000000 &    0.070145 &    0.000000 &    0.031175 &    0.054556  \\
   80.0 &    0.003306 &    0.000252 &   0.000000 &    0.061600 &    0.000000 &    0.029801 &    0.057605  \\
   85.0 &    0.002949 &    0.000250 &   0.000000 &    0.054069 &    0.000000 &    0.028206 &    0.058753  \\
   90.0 &    0.002612 &    0.000247 &   0.000000 &    0.047420 &    0.000000 &    0.026416 &    0.058244  \\
   95.0 &    0.002296 &    0.000242 &   0.000000 &    0.041522 &    0.000000 &    0.024468 &    0.056351  \\
  100.0 &    0.002000 &    0.000234 &   0.000000 &    0.036259 &    0.000000 &    0.022402 &    0.053348  \\
  110.0 &    0.001469 &    0.000214 &   0.000000 &    0.027245 &    0.000000 &    0.018061 &    0.044999  \\
  120.0 &    0.001020 &    0.000183 &   0.000000 &    0.019785 &    0.000000 &    0.013680 &    0.034936  \\
  130.0 &    0.000653 &    0.000145 &   0.000000 &    0.013139 &    0.000000 &    0.009422 &    0.024550  \\
  140.0 &    0.000367 &    0.000101 &   0.000000 &    0.007555 &    0.000000 &    0.005631 &    0.014968  \\
  150.0 &    0.000163 &    0.000062 &   0.000000 &    0.007067 &    0.000000 &    0.003155 &    0.005554  \\
  160.0 &    0.000041 &    0.000023 &   0.000000 &    0.002702 &    0.000000 &    0.001010 &    0.001340  \\
  170.0 &    0.000000 &    0.000000 &   0.000000 &    0.000518 &    0.000000 &    0.000130 &    0.000000  \\
  180.0 &    0.000000 &    0.000000 &   0.000000 &    0.000277 &    0.000000 &    0.000069 &    0.000000  \\
  190.0 &    0.000000 &    0.000000 &   0.000000 &    0.000113 &    0.000000 &    0.000028 &    0.000000  \\
  200.0 &    0.000000 &    0.000000 &   0.000000 &    0.000000 &    0.000000 &    0.000000 &    0.000000  \\
  210.0 &    0.000000 &    0.000000 &   0.000000 &    0.000000 &    0.000000 &    0.000000 &    0.000000  \\
  220.0 &    0.000000 &    0.000000 &   0.000000 &    0.000000 &    0.000000 &    0.000000 &    0.000000  \\
  230.0 &    0.000000 &    0.000000 &   0.000000 &    0.000000 &    0.000000 &    0.000000 &    0.000000  \\
  240.0 &    0.000000 &    0.000000 &   0.000000 &    0.000000 &    0.000000 &    0.000000 &    0.000000  \\
  250.0 &    0.000000 &    0.000000 &   0.000000 &    0.000000 &    0.000000 &    0.000000 &    0.000000  \\
  260.0 &    0.000000 &    0.000000 &   0.000000 &    0.000000 &    0.000000 &    0.000000 &    0.000000  \\
  270.0 &    0.000000 &    0.000000 &   0.000000 &    0.000000 &    0.000000 &    0.000000 &    0.000000  \\
\hline
\end{tabular}
\end{center}
\end{table*}
\renewcommand{\arraystretch}{1}

\end{document}